\documentclass[12pt,a4paper]{article}
\pdfoutput=1
\RequirePackage{amsmath}%
\RequirePackage{amsthm}%
\usepackage{amsfonts}%
\usepackage{amssymb}%
\usepackage{graphicx}%
\usepackage{url}%
\usepackage{color}
\usepackage{booktabs}
\usepackage{multirow}
\usepackage{caption}
\usepackage{subcaption}
\usepackage{wrapfig}
\usepackage{booktabs}
\usepackage{algorithmic,algorithm,float}
\usepackage[section]{placeins}
\usepackage{multirow}

\renewcommand{\i}{\mathrm{i}}

\begin{document}
\title{The Motion of a Point Vortex in Multiply Connected Polygonal Domains}
	\author{El Mostafa Kalmoun, Mohamed M S Nasser, Khalifa A. Hazaa}
	
	\date{}
	\maketitle
	
	\vskip-0.8cm %
	\centerline{Department of Mathematics, Statistics and Physics, Qatar University,} %
	\centerline{P.O. Box 2713, Doha, Qatar}%
	\centerline{ekalmoun@qu.edu.qa; mms.nasser@qu.edu.qa; khalifa.alhazaa@qu.edu.qa}

\begin{abstract}
We study the motion of a single point vortex in simply and multiply connected polygonal domains. 
	In case of multiply connected domains, the 	polygonal obstacles can be viewed as the cross-sections of 3D polygonal cylinders.
	First, we utilize conformal mappings to transfer the polygonal domains onto circular domains. Then, we employ the Schottky–Klein prime function to compute the Hamiltonian governing the point vortex motion in circular domains.
	We compare between the topological structures of the contour lines of the Hamiltonian in symmetric and asymmetric domains. Special attention is paid to the interaction of point vortex trajectories with the polygonal obstacles. In this context, we discuss the effect of symmetry breaking, and obstacle location and shape on the behavior of vortex motion.
\end{abstract}

\begin{center}
\begin{quotation}
\end{quotation}
\end{center}

%%%%%%%%%%%%%%%%%%%%%%%%%%%%%%%%%%%%%%%%%%
\section{Introduction}
\label{sc:int}
%-------------------------------------------------------------------
Since the seminal work by Helmholtz~\cite{helmholtz}, 2D point vortex motion has sparked intense research due to its important applications in fluid dynamics and geophysics~\cite{aref2007point,boatto2006point,newton2002n-vortex,saffman1992vortex}. 
A point vortex is a model of potential flow in which vorticity of the surrounding flow is present only at a single point. In addition to the various physical applications, the 2D point vortex model, despite its simplicity, provides a {\it mathematics playground} in the words of Aref~\cite{aref2007point}, that links a variety of mathematical areas such as dynamical systems, ordinary and partial differential equations, Hamiltonian dynamics, theory of polynomials, elliptic functions, to name a few.

The dynamics of point vortices is governed by a Hamiltonian system, which can be derived from the Euler equations describing the velocity field and pressure of a 2D incompressible and inviscid flow. 
If the fluid domain has no boundaries, like the whole complex plane for instance, the Hamiltonian of the governing dynamical system is written in terms of the complex Green function~\cite{saffman1992vortex}. 
The presence of solid boundaries might break symmetries that would otherwise exist and thus the Hamiltonian system is constrained by the fluid non-penetration condition expressed by zero normal velocity on all solid boundaries. For simple geometries containing special symmetries like the upper-half plane, a channel between two infinite walls or the exterior of a circular cylinder, an explicit formula of the Hamiltonian of a point vortex can be obtained by the classical method of images in which solid boundaries are treated as streamlines of a flow field. This is usually combined with conformal mapping ideas as can be seen in~\cite{newton2002n-vortex,saffman1992vortex} for a variety of examples. Similarly, in bounded and simply connected domains, basic results concerning single vortex motion are well established~\cite{flucher1997vortex}. In this case the generalized Hamiltonian is known as the \emph{Kirchhoff-Routh path function}~\cite{kirchhoff,routh}.

As many fluid domains are not simply connected (e.g. in aerodynamics), a growing interest has centered around studying $N$-point vortex dynamics in multiply connected domains.  Lin~\cite{lin1941motion} had  early proved the existence and uniqueness of the Kirchhoff-Routh path function in multiply connected domains but no explicit formula was given. Recently, Johnson \& McDonald~\cite{johnson2004motion} studied the motion of one point vortex in doubly and triply connected domains exterior to either one or two circular cylinders using elliptic functions, provided that zero circulations are imposed around the circles.
Under the same boundary conditions, Crowdy \& Marshall~\cite{cm-ana} presented an analytical formula for the Hamiltonian in multiply connected circular domains by applying methods of complex function theory. In particular, their analysis makes use of a special transcendental function known as the \emph{Schottky-Klein prime function}.
The formula can be employed for a general multiply connected domain as long as the conformal mapping onto the canonical circular domain is known. Indeed, this formula was already applied to study point vortex motion around an arbitrary finite number of circular cylinders~\cite{cm-mot1} and through gaps in walls~\cite{cm-mot2}.
When the multiply connected domain is symmetric with respect to the real axis, Sakajo~\cite{Tak-eq} showed that the two point vortices' motion is reduced to that of a single point vortex in a multiply connected semicircle. Thus the Crowdy \& Marshall analytic formula was again utilized to plot the trajectories of two point vortices for several circular domains. 
%The topological structure that consists of the saddle/centre points, their heteroclinic/homoclinic orbits in the contour lines of the Hamiltonian we presented for several domains.

In this paper, we study the motion of a single point vortex in simply and multiply connected polygonal domains. In case of multiply connected dmains, the polygonal obstacles can be viewed as the cross-sections of polygonal cylinders in 3D. Our method will first make use of conformal mappings to transfer the polygonal domain onto the circular domain where the Crowdy \& Marshall analytic formula can be employed. The objective of our study is to compare between the topological structures of the contour lines of the Hamiltonian in circular and polygonal domains. We aim also to emphasize how these trajectories interact with the polygonal obstacles. In this context, we discuss the effect of symmetry breaking, and obstacle location and shape on the behavior of vortex motion.

\section{Computing the Hamiltonian of a Point Vortex Motion}
\label{sc:ham}
%-------------------------------------------------------------------

We  present in this section a numerical method for computing the Hamiltonian of a point vortex motion in multiply connected polygonal domains. The method is based on using conformal mappings to map polygonal domains onto circular ones, and then appealing to the analytic formula presented in~\cite{cm-ana} for the Hamiltonian in multiply connected circular domains.

Let $G$ denote the multiply connected domain exterior to $m$ non-overlapping polygons $\Gamma_1,\ldots,\Gamma_m$ (clockwise orientated) and interior to a polygon $\Gamma_{m+1}$ (in the counterclockwise orientation). Such a domain will be called a polygonal domain. We assume that no corner of these polygons is a cusp.
Let us suppose that the interior of $G$ is occupied with incompressible fluid. The flow is assumed to undergo irrotational motion except for a single point vortex. Therefore, the circulations around the polygonal obstacles $\Gamma_1,\ldots,\Gamma_m$ are zero, and the point vortex has a given circulation $\chi$.

\subsection{The Conformal Mapping}

  It is known that there exists a conformal mapping  $\zeta=f(z)$ from the polygonal domain $G$ onto a multiply connected circular domain $D$ interior to the unit circle and exterior to $m$ non-overlapping circles~\cite{Krantz}. We denote the inner $m$ circles by $C_k$, $k=1,2,\ldots,m$ and the unit circle by $C_{m+1}$. By Carath\'eodory's theorem~\cite[p.~111]{Krantz}, the mapping function $f$ can be extended to $\overline{G}=G\cup \partial G$ in such a way that the polygon $\Gamma_k$ is mapped onto the circle $C_k$, for $k=1,2,\ldots,m+1$. Note that $\partial G$ denotes the boundary of the domain $G$, which consists of $m+1$ piecewise smooth Jordan curves.
  For the uniqueness of the conformal mapping, we fix a point $\alpha$ in $G$, and we assume that $f(\alpha)=0$ and $f'(\alpha)>0$.

The centers and radii of the inner circles are unknown and uniquely determined by the polygonal domain $G$. Thus, computing the conformal mapping $f$ requires also computing the centers and radii in the circular domain. 
In this paper, the conformal mapping $f$ will be computed using the MATLAB toolbox (PlgCirMap) presented in~\cite{Nas-plg}.

\subsection{The Hamiltonian for Circular Domains}

The trajectories of a point vortex in the multiply connected circular domain $D$ can be written as the contours of the Hamiltonian, aka the Kirchhoff-Routh path function~\cite{cm-ana}. The existence and uniqueness of such a function is established by Lin~\cite{lin1941motion}. 

Let $H_D(\zeta,\overline{\zeta})$ denote the Hamiltonian for the motion of a single point vortex in the circular domain $D$. 
An explicit formula for $H_D(\zeta,\overline{\zeta})$ has been derived by Crowdy \& Marshall~\cite{cm-ana},
\begin{equation}\label{eq:H_D}
H_D(\zeta,\overline{\zeta})=-\frac{\chi^2}{8\pi}\log\left|\frac{1}{\zeta^2}\frac{\hat{\omega}(\zeta,\zeta)\overline{\hat{\omega}}(1/\zeta,1/\zeta)}{\omega(\zeta,1/\overline{\zeta})\overline{\hat{\omega}}(1/\zeta,\overline{\zeta})}\right|,
\end{equation}
where $\overline{\zeta}$ is the complex conjugate of $\zeta$, 
 $\omega(\zeta,\xi)$ is the Schottky-Klein prime function (prime function, for short) associated with the circular domain $D$, and the function $\hat{\omega}(\zeta,\xi)$ is defined through
\[
\omega(\zeta,\xi)=(\zeta-\xi)\hat{\omega}(\zeta,\xi), \quad \zeta,\xi\in D,
\]
and $\overline{\hat\omega}(\zeta,\xi)=\overline{\hat{\omega}(\overline{\zeta},\overline{\xi})}.$

The Schottky-Klein prime function is an important special transcendental function. 
Due to the works of Crowdy with various collaborators~\cite{c-book} in the last two decades, the prime function has become a key tool to construct analytic formulas in solving several problems involving multiply connected circular domains. %However, numerical methods are needed for computing the prime function. 
For more details on the properties of the prime function, its applications and numerical computation, the reader is referred to the review article~\cite{ckgn} and the recent monograph~\cite{c-book}.

A simplified form of~\eqref{eq:H_D} has been proposed by Sakajo~\cite{Tak-eq},
\[
H_D(\zeta,\overline{\zeta})=-\frac{\chi^2}{4\pi}\log\left|\frac{\hat{\omega}(\zeta,\zeta)}{\zeta\omega(\zeta,1/\overline{\zeta})}\right|.
\]
In the two numerical methods presented in~\cite{ckgn}, the prime function is normalized with $w'(\zeta,\zeta)=1$, which yields
\begin{equation}\label{eq:ham-c}
H_D(\zeta,\overline{\zeta})%=-\frac{\chi^2}{4\pi}\log\left|\frac{1}{\zeta\omega(\zeta,1/\overline{\zeta})}\right|
=\frac{\chi^2}{4\pi}\log\left|\zeta\omega(\zeta,1/\overline{\zeta})\right|.
\end{equation}

By making use of an important property of the prime function~\cite{c-book} that states
\begin{equation}\label{eq:z_in_C}
\omega(\zeta,1/\overline{\zeta})=0\; \hbox{ for } \zeta\in \bigcup_{k=1}^{m+1} C_k,
\end{equation}
it follows from~\eqref{eq:ham-c} that $H_D(\zeta,\overline{\zeta})\to-\infty$ as $\zeta$ approaches $\partial D$.

\subsection{The Hamiltonian for Polygonal Domains}

By the help of conformal mappings, the above exact formula~\eqref{eq:ham-c} can be extended to the general multiply connected domains~\cite{cm-ana}. In fact, using the conformal mapping $\zeta=f(z)$ to map the polygonal domain $G$ onto the circular domain $D$, the Hamiltonian $H_G(z,\overline{z})$ of a single vortex motion of circulation $\chi$ in the bounded polygonal domain $G$ is given by~\cite{cm-ana}
\[
H_G(z,\overline{z})=\frac{\chi^2}{4\pi}\log\left|\zeta\omega(\zeta,1/\overline{\zeta})\right|
+\frac{\chi^2}{4\pi}\log\left|(f^{-1})'(\zeta)\right|,
\]
which can be simplified as
\begin{equation}\label{eq:ham-p}
H_G(z,\overline{z})=\frac{\chi^2}{4\pi}\log\left|\frac{|f(z)|\omega\left(f(z),1/\overline{f(z)}\right)}{f'(z)}\right|.
\end{equation}
The formula~\eqref{eq:z_in_C} implies that $H_G(z,\overline{z})\to-\infty$ as $z$ approaches $\partial G$ as well.
In the sequel, we assume that the point vortex has a circulation $\chi = 1$.

\subsection{Numerical Implementation}

Recently, a MATLAB toolbox (PlgCirMap) for computing the conformal mapping $\zeta=f(z)$, from the polygonal domain $G$ onto the circular domain $D$, and its inverse $z=f^{-1}(\zeta)$ has been presented in~\cite{Nas-plg}. 
The method used in~\cite{Nas-plg} is based on a fast implementation of Koebe's iterative
method~\cite{Nas-CMFT} using the boundary integral equation with the generalized Neumann kernel~\cite{Nas-ETNA} and fast multipole method~\cite{Gre-Gim12}.
The toolbox works also if $G$ is simply connected ($m=1$), and hence $D$ is the unit disk. 
On the other hand, to compute the prime function, two numerical methods are presented in~\cite{ckgn} with MATLAB codes available online. Their first procedure is a spectral method based on global relations, while the second depends on the boundary integral equation with the generalized Neumann kernel.

In this paper, the Hamiltonian in~\eqref{eq:ham-p} is computed by a combination of the MATLAB toolbox PlgCirMap and one or the other of the two numerical methods presented in~\cite{ckgn}. The outline of our procedure is summarized in Algorithm~\ref{algo}.

\begin{algorithm}[!htb]
	\setlength\baselineskip{18pt}
	\caption{\label{algo} Computing the Hamiltonian $H_G$ for the polygonal domain $G$.}
	\begin{algorithmic}[1]
		\STATE Define the vertices of the polygons $\Gamma_k,\; k=1,\ldots,m+1$, as a cell array \verb|ver{k}|.
		\STATE Choose an auxiliary point \verb|alpha| in $G$.
		\STATE Use the MATLAB function \verb|plgcirmap| to compute
		\begin{center}
			\verb|f = plgcirmap(ver,alpha)|
		\end{center} 
		where \verb|f| is a MATLAB {\it struct} with several fields.
		\STATE Extract the centers and radii of the inner circles in $D$ from \verb|f| by \verb|f.cent(1:m)| and \verb|f.rad(1:m)|.
		\STATE Discretize the domain $G$ by a matrix of points \verb|Z|.
		\STATE Call the functions \verb|evalu| and \verb|evalud| from the toolbox PlgCirMap to compute the values of the conformal mapping \verb|fZ| and its derivative \verb|dfZ| on the meshgrid~\verb|Z|:
		\begin{center}
			\verb|fZ = evalu(f,Z,'d');  dfZ = evalud(f,Z,'d')|
		\end{center} 
		\STATE Compute the prime function by applying one of the two methods presented in~\cite{ckgn}.
		\STATE Use Equation~\eqref{eq:ham-p} to compute the Hamiltonian.
	\end{algorithmic}
\end{algorithm}

Note that the point $\alpha $ in Step 2 of  Algorithm~\ref{algo} has no effect on the numerical computation of the Hamiltonian as long as it is sufficiently far from the boundary of $G$.

\section{Simply Connected Domains}
\label{sc:sim}
%-------------------------------------------------------------------

When $D$ is the unit disk, the associated prime function is simply
\[
\omega(\zeta,\alpha)=\zeta-\alpha.
\]
Thus, the Hamiltonian~\eqref{eq:ham-c} of a single vortex motion of circulation $\chi$ in the unit disk $D$ reduces to~\cite{cm-ana}
\[
H_D(\zeta,\overline{\zeta})
=\frac{1}{4\pi}\log|1-\zeta\,\overline{\zeta}|
=\frac{1}{4\pi}\log(1-|\zeta|^2).
\]
Hence, the contour lines of the Hamiltonian are circles with a center point at the origin.
It is clear that $0<1-|\zeta|^2\le1$ when $\zeta$ is in the unit disk $D$ and $1-|\zeta|^2\to0^+$ as $\zeta$ approaches $\partial D$. Hence $H_D(\zeta,\overline{\zeta}) \rightarrow -\infty$ as $\zeta$ approaches $\partial D$ and $H_D(0,0)=0$ is the unique global maximum.

Similarly, the Hamiltonian~\eqref{eq:ham-p} in the bounded simply connected polygonal domain $G$ is given by~\cite{cm-ana}
\[
H_G(z,\overline{z})
=\frac{1}{4\pi}\log\left|\frac{|f(z)|\left(f(z)-1/\overline{f(z)}\right)}{f'(z)}\right|
=\frac{1}{4\pi}\log\frac{1-|f(z)|^2}{|f'(z)|},
\]
where $f$ is the above described conformal mapping from $G$ onto $D$.
Note also that $H_G(z,\overline{z}) \rightarrow -\infty$ as $z$ approaches $\partial G$. 

For convex domains, Gustafsson~\cite{gustafsson1979motion} proved that the Hamiltonian is strictly concave, and therefore the point vortex has a unique equilibrium point, which is the global maximum of $H_G$. 
In this case, contour plots of the Hamiltonian $H_G$, which correspond to vortex trajectories span the domain $G$ by encircling the unique critical point in a similar topological way as in the disk $D$. When passing to non convex domains, the uniqueness of critical points does not hold in general as noticed by Gustafsson~\cite{gustafsson1990convexity} who provided a counter-example for star-shaped domains. 
In the next subsection we present a numerical example that illustrates this fact. 

\subsection{A Star-Shaped Domain}

\begin{figure}[!htb] %
	\centering
	\begin{subfigure}[b]{0.32\textwidth}
		\centering
		\includegraphics[scale=0.4,trim=0 0 0 0,clip]{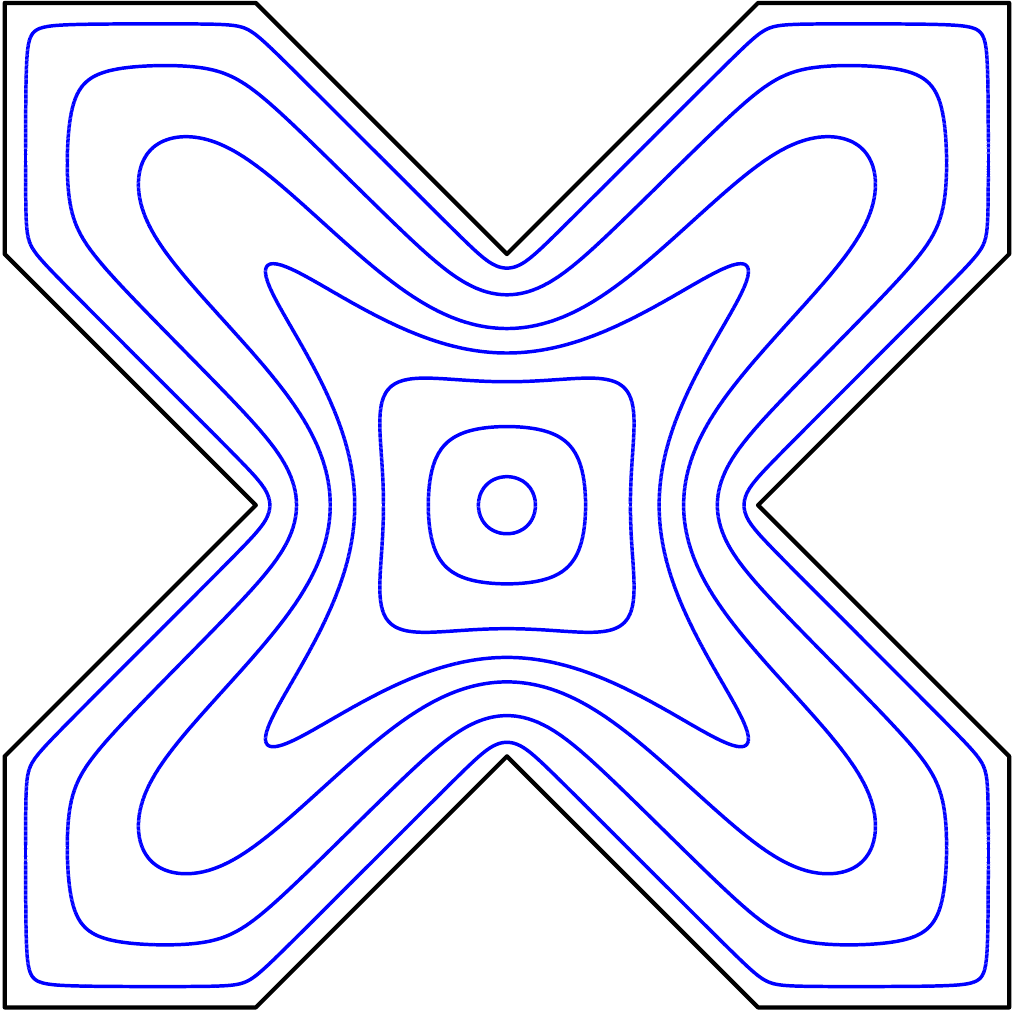}
		\caption{$(a,b)=(0.5,0.5)$}
		\label{fig:simX1}
	\end{subfigure}%
	~
	\begin{subfigure}[b]{0.32\textwidth}
		\centering
		\includegraphics[scale=0.4,trim=0 0 0 0,clip]{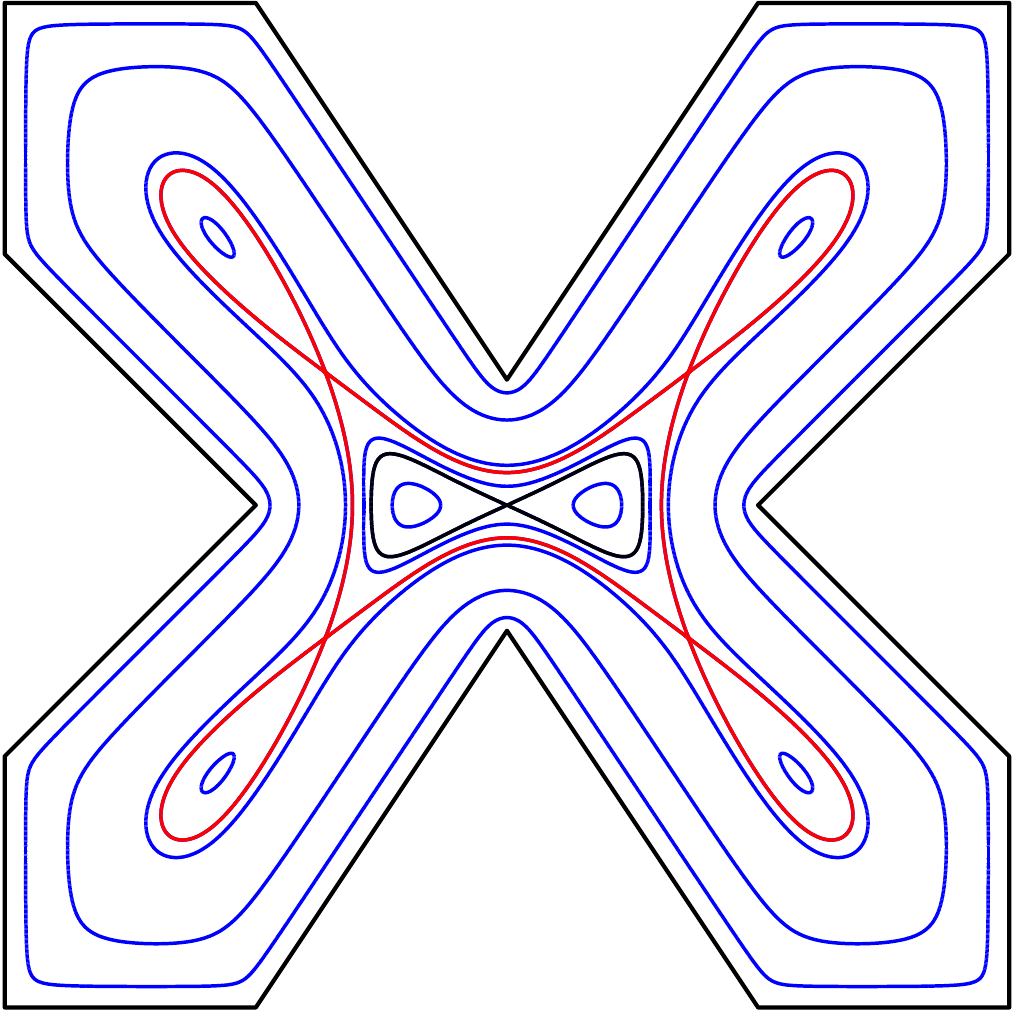}
		\caption{$(a,b)=(0.25,0.5)$}
		\label{fig:simX2}
	\end{subfigure}%
	~
	\begin{subfigure}[b]{0.32\textwidth}
		\centering
		\includegraphics[scale=0.4,trim=0 0 0 0,clip]{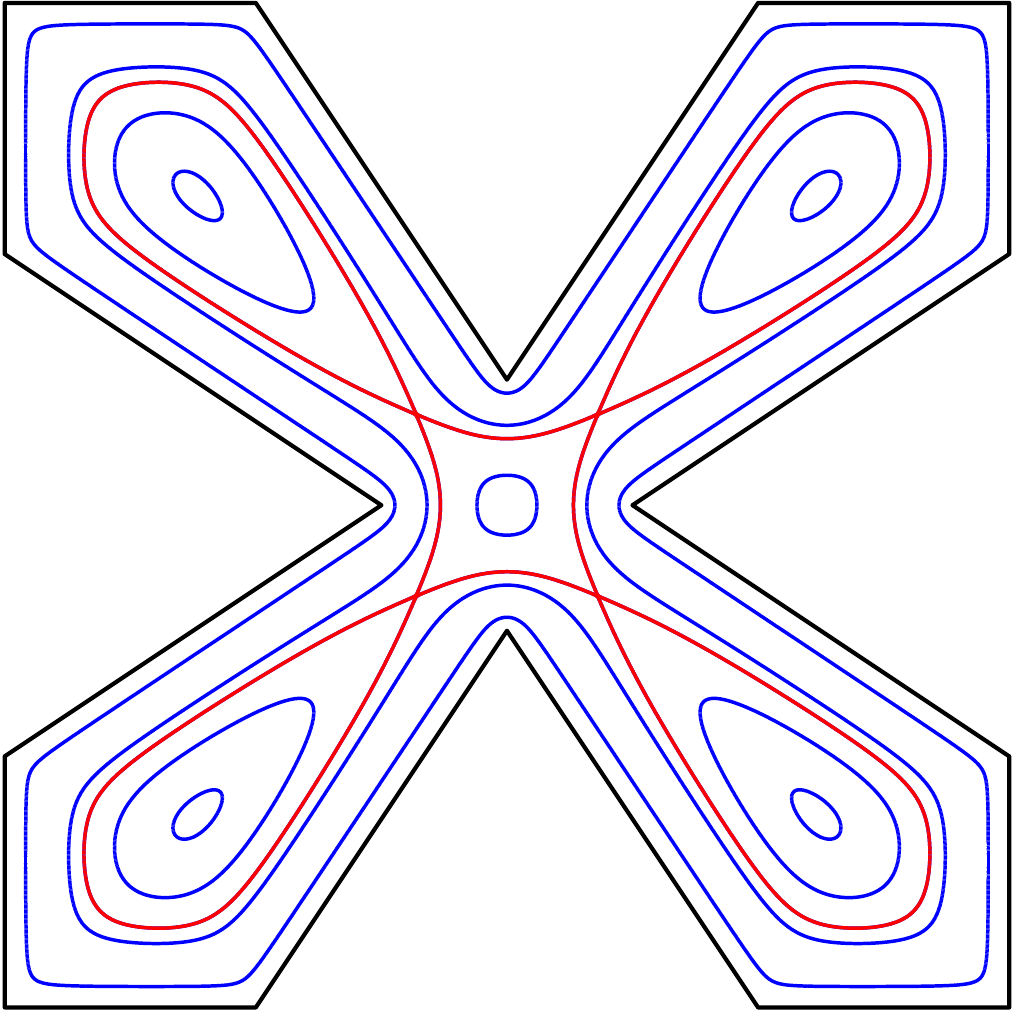}
		\caption{$(a,b)=(0.25,0.25)$}
		\label{fig:simX3}
	\end{subfigure}%
	\caption{Vortex trajectories in  a star-shaped domain. The vertices of the polygon are: $1+0.5\i,1+\i,0.5+\i,a\i,-0.5+\i,-1+\i,-1+0.5\i,-b,-1-0.5\i,-1-\i,-0.5-\i,-a\i,0.5-\i,1-\i,1-0.5\i$ and $b$.}
	\label{fig:simX}
\end{figure}

The vertices of the star-shaped domain in Figure~\ref{fig:simX} are: $1+0.5\i,1+\i,0.5+\i,a\i,-0.5+\i,-1+\i,-1+0.5\i,-b,-1-0.5\i,-1-\i,-0.5-\i,-a\i,0.5-\i,1-\i,1-0.5\i$ and $b$ for three different values of $(a,b)$. The  X-shaped polygon in Figure~\ref{fig:simX1} has $(a,b)=(0.5,0.5)$, which makes the four concave corners located on the same circle of center zero and radius $0.5$. Similar to the unit disk case, the Hamiltonian in this polygonal domain has a unique critical point, which is a center at the origin surrounded by closed orbits. 

If the two vertices on the imaginary axis are displaced by $0.25$ unit toward each other, so that they are not anymore located on the same circle with the other two vertices on the real axis, then we have more than one critical point; see Figure~\ref{fig:simX2}. Indeed, the neutrally stable center point becomes a saddle, which is connected to itself by a pair of symmetric homoclinic orbits forming together a separatrix. 
Two new vortex centers appear on the real axis at the same distance from the saddle point zero. 
The Hamiltonian has four other centers where each point is surrounded by a homoclinic orbit connecting a saddle point. These four hyperbolic points are themselves connected by a heteroclinic loop. Observe also how all equilibria of this system are  distributed in a way that reflects line symmetries in the domain.  

Another worth commenting dynamical feature in this example is the reappearance of the center point at the origin instead of the homoclinic structure when the four concave corners are placed again on a circle of center zero and radius $0.25$. However, we do not recover the uniqueness of critical points since the previous outer homoclinic-heteroclinic pattern still occurs; see Figure~\ref{fig:simX3}.

\subsection{Non Convex Polygonal Domains}

\begin{figure}[!htb] %
	\centering
	\begin{subfigure}[b]{0.32\textwidth}
		\centering
		\includegraphics[scale=0.4,trim=0 0 0 0,clip]{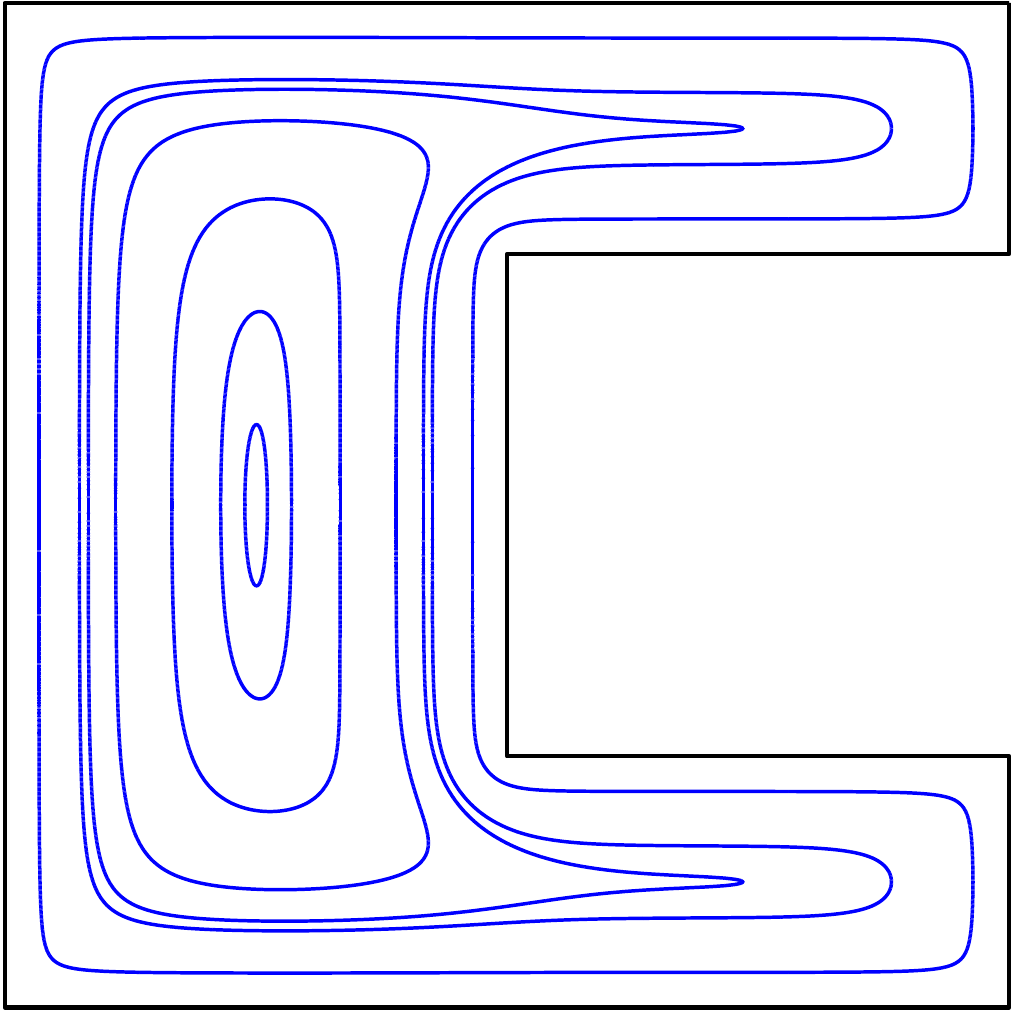}
		\caption{$a=0$}
		\label{fig:simC1}
	\end{subfigure}%
	~
	\begin{subfigure}[b]{0.32\textwidth}
		\centering
		\includegraphics[scale=0.4,trim=0 0 0 0,clip]{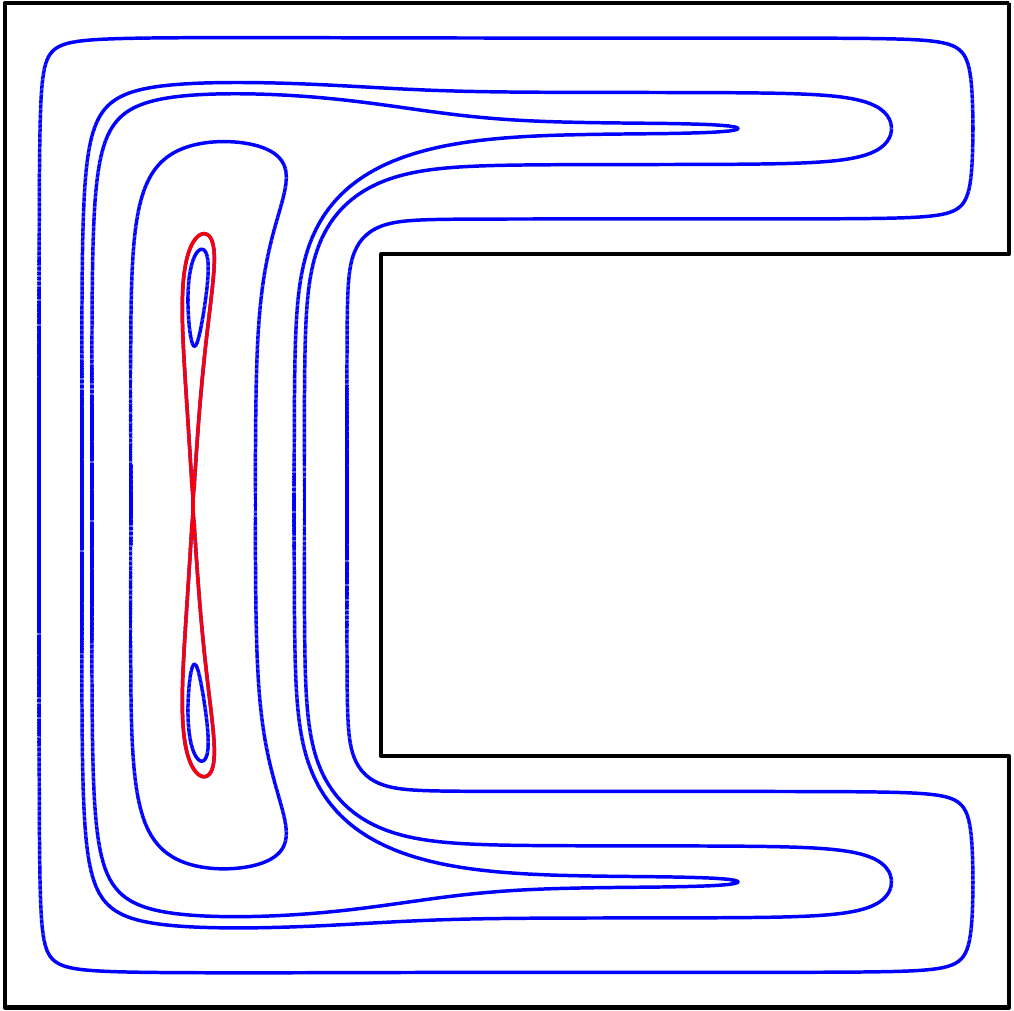}
		\caption{$a=-0.25$}
		\label{fig:simC2}
	\end{subfigure}%
	~
	\begin{subfigure}[b]{0.32\textwidth}
		\centering
		\includegraphics[scale=0.4,trim=0 0 0 0,clip]{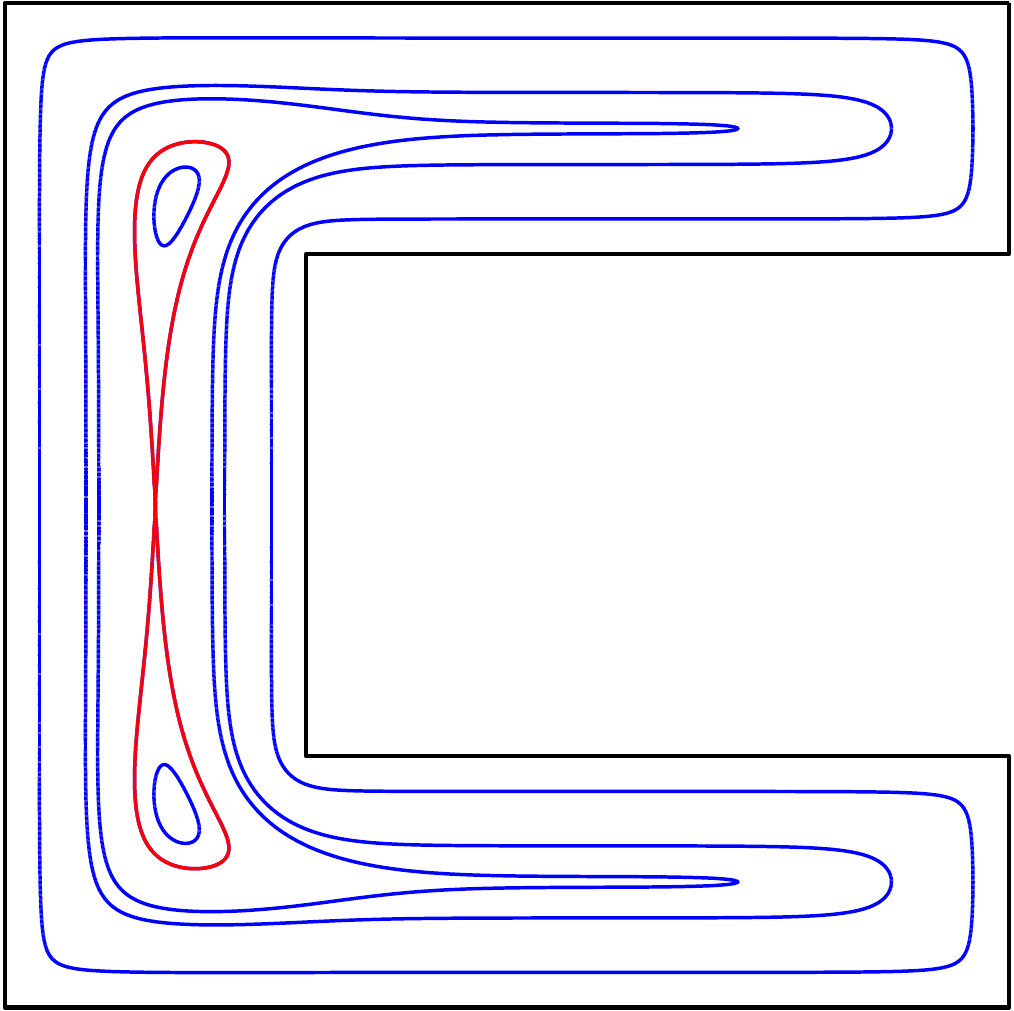}
		\caption{$a=-0.4$}
		\label{fig:simC3}
	\end{subfigure}%
	\caption{Vortex trajectories in a C-shaped polygon of vertices: $1+\i,-1+\i,-1-\i,1-\i,1-0.5\i,a-0.5\i,a+0.5\i,1+0.5\i$.}
	\label{fig:simC}
\end{figure}

\begin{figure}[!htb] %
	\centering
	\begin{subfigure}[b]{0.32\textwidth}
		\centering
		\includegraphics[scale=0.4,trim=0 0 0 0,clip]{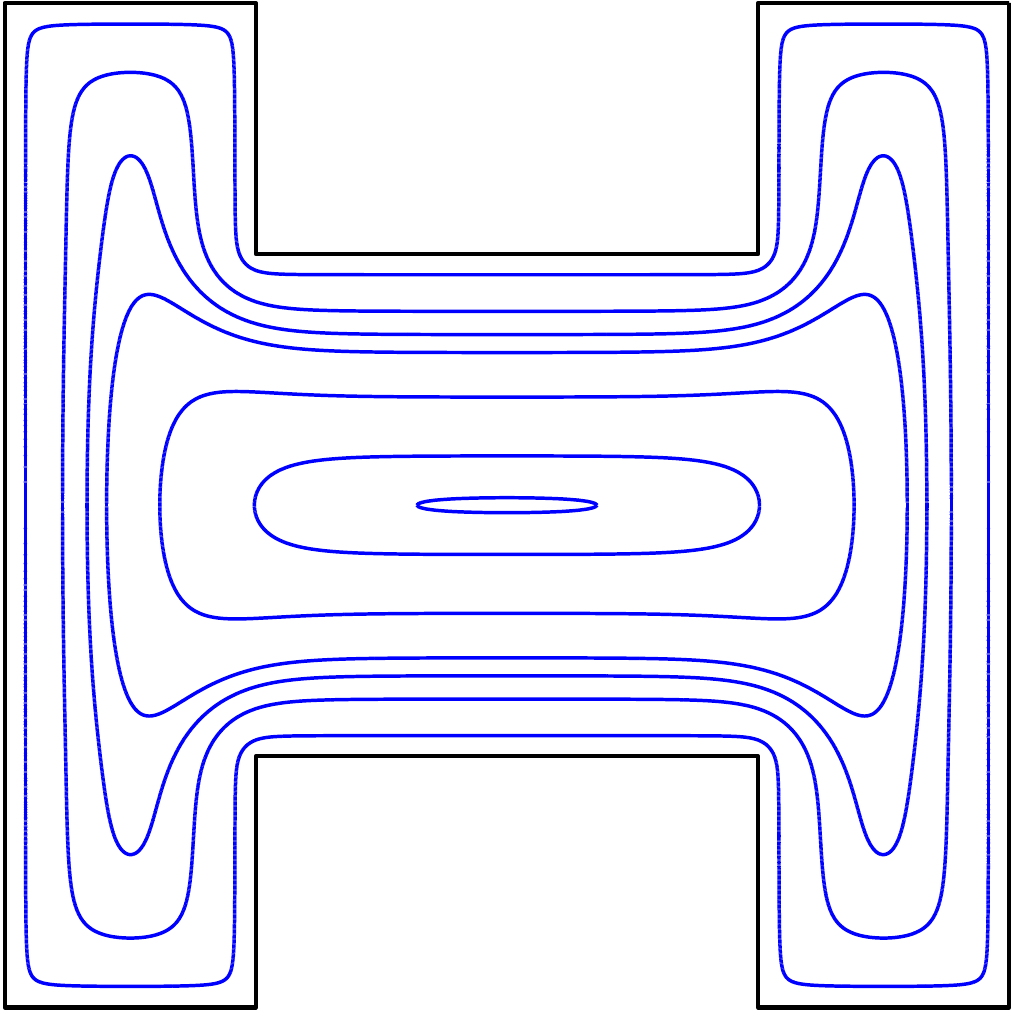}
		\caption{$a=0.5$}
		\label{fig:simH1}
	\end{subfigure}%
	~
	\begin{subfigure}[b]{0.32\textwidth}
		\centering
		\includegraphics[scale=0.4,trim=0 0 0 0,clip]{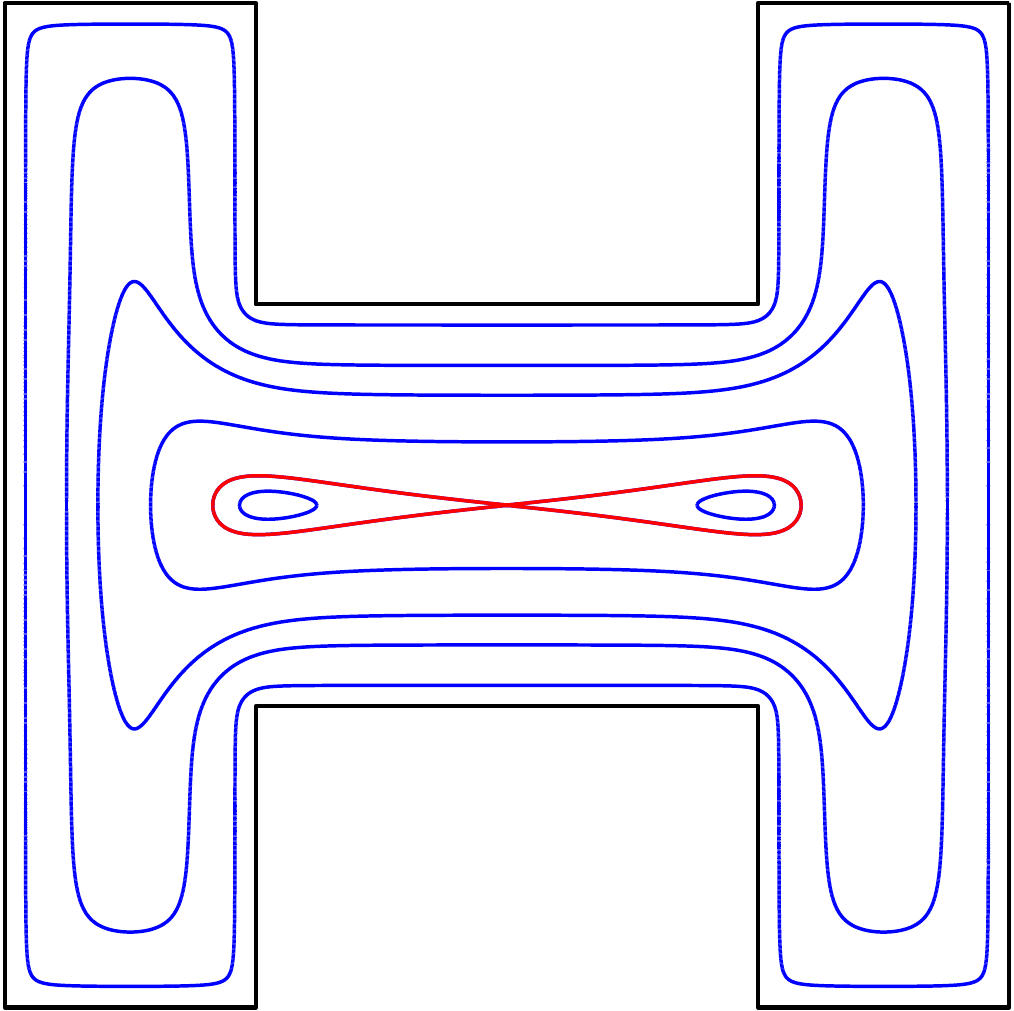}
		\caption{$a=0.4$}
		\label{fig:simH2}
	\end{subfigure}%
	~
	\begin{subfigure}[b]{0.32\textwidth}
		\centering
		\includegraphics[scale=0.4,trim=0 0 0 0,clip]{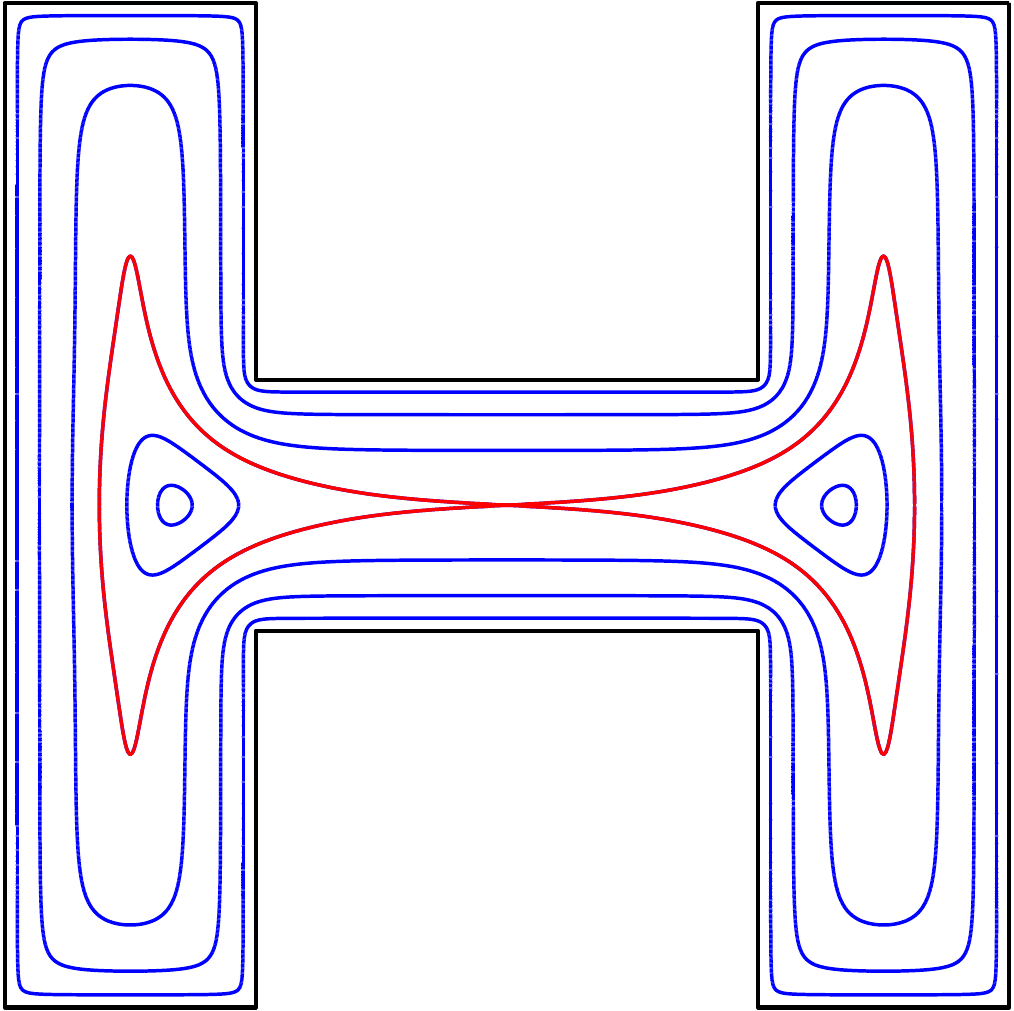}
		\caption{$a=0.25$}
		\label{fig:simH3}
	\end{subfigure}%
	\caption{Vortex trajectories in a H-shaped polygon of vertices: $1+\i,0.5+\i,0.5+a\i,-0.5+a\i,-0.5+\i,-1+\i,-1-\i,-0.5-\i,-0.5-a\i,0.5-a\i,0.5-\i,1-\i$.}
	\label{fig:simH}
\end{figure}

\begin{figure}[!htb] %
	\centering
	\begin{subfigure}[b]{0.32\textwidth}
		\centering
		\includegraphics[scale=0.4,trim=0 0 0 0,clip]{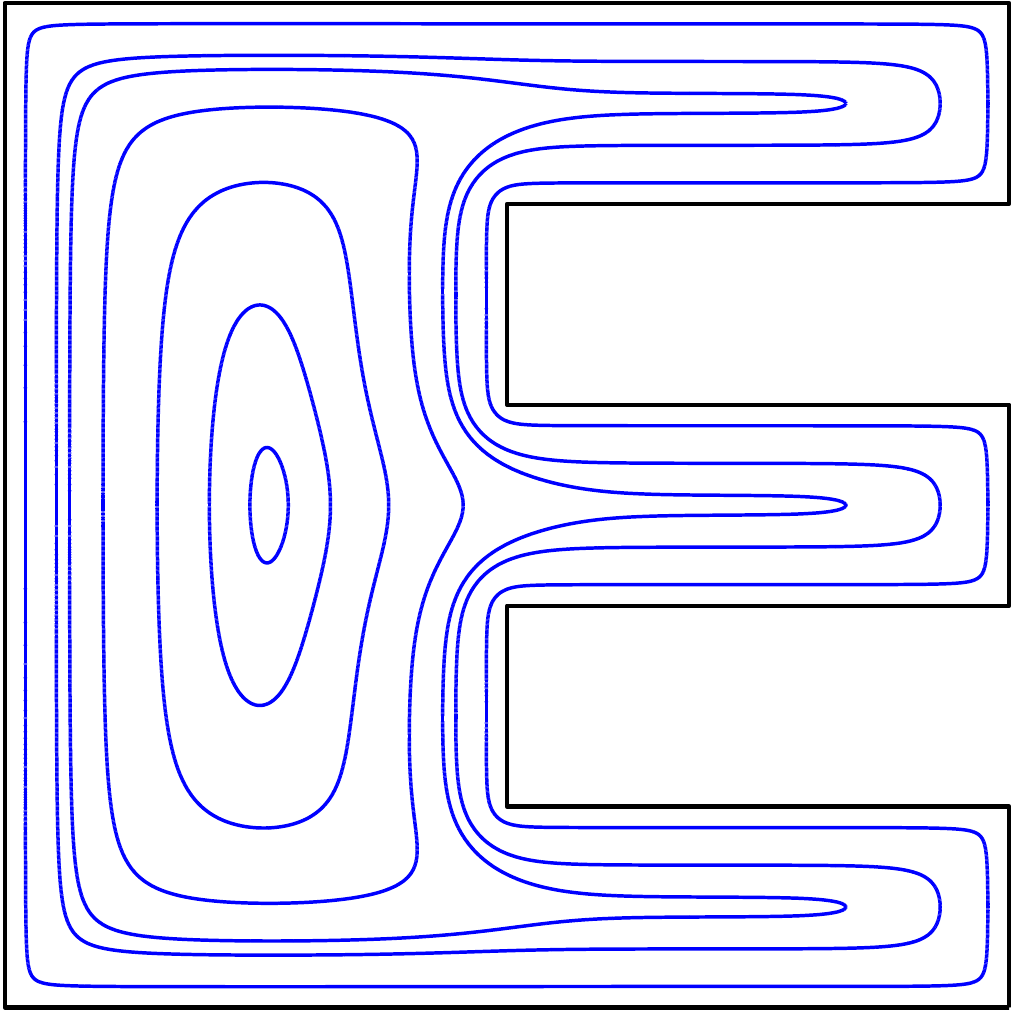}
		\caption{$a=1$}
		\label{fig:simE1}
	\end{subfigure}%
	~
	\begin{subfigure}[b]{0.32\textwidth}
		\centering
		\includegraphics[scale=0.4,trim=0 0 0 0,clip]{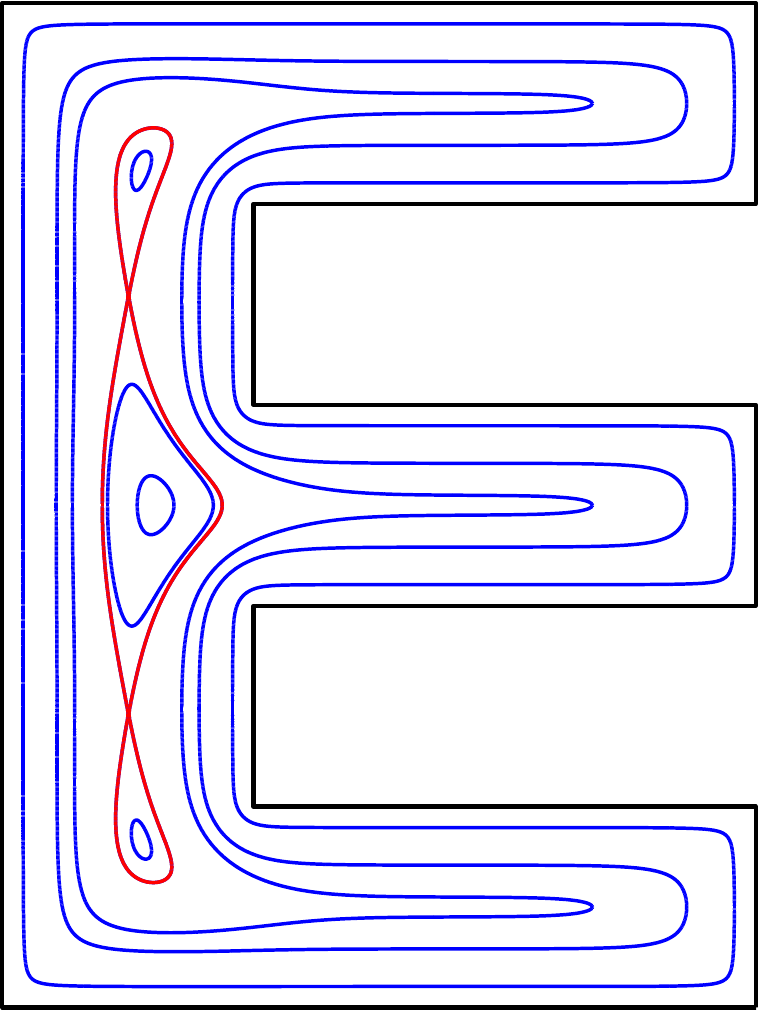}
		\caption{$a=0.5$}
		\label{fig:simE2}
	\end{subfigure}%
	~
	\begin{subfigure}[b]{0.32\textwidth}
		\centering
		\includegraphics[scale=0.4,trim=0 0 0 0,clip]{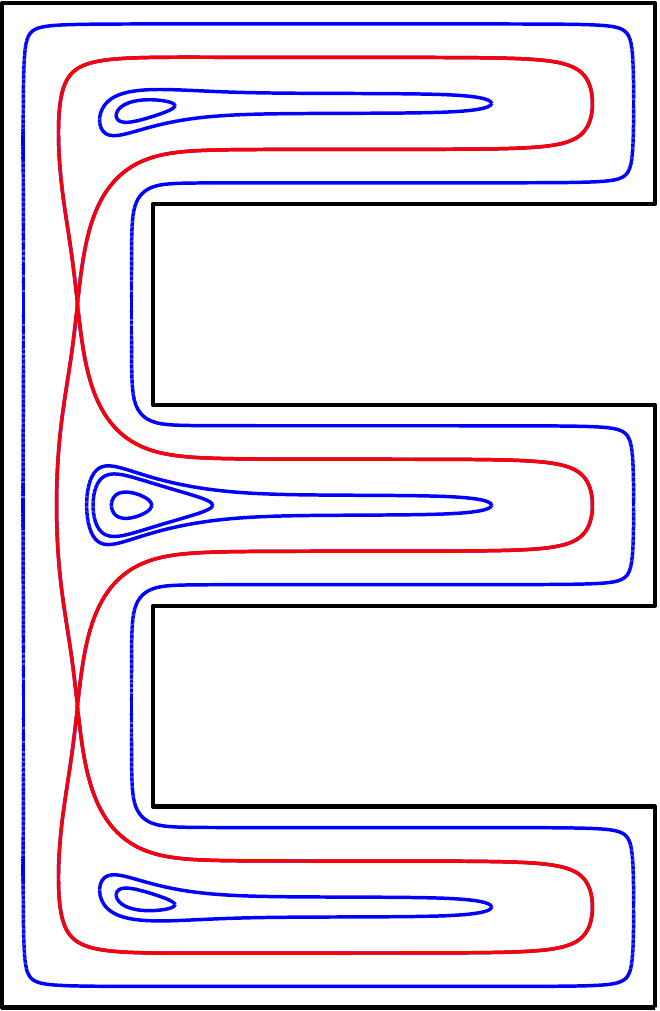}
		\caption{$a=0.3$}
		\label{fig:simE3}
	\end{subfigure}%
	\caption{Vortex trajectories in a E-shaped polygon of vertices: $1-\i,1-0.6\i,-0.6\i,-0.2\i,1-0.2\i,1+0.2\i,0.2\i,0.6\i,1+0.6\i,1+\i,-a+\i,-a-\i$.}
	\label{fig:simE}
\end{figure}

We consider in Figures~\ref{fig:simC}-\ref{fig:simE} three other examples of general bounded simply connected polygonal domains that are non convex. The C-shaped polygon in Figure~\ref{fig:simC} has eight vertices at $1+\i,-1+\i,-1-\i,1-\i,1-0.5\i,a-0.5\i,a+0.5\i,1+0.5\i$, the H-shaped polygon in Figure~\ref{fig:simH} has twelve vertices at  $1+\i,0.5+\i,0.5+a\i,-0.5+a\i,-0.5+\i,-1+\i,-1-\i,-0.5-\i,-0.5-a\i,0.5-a\i,0.5-\i,1-\i$, and the twelve vertices of the E-shaped polygon in Figure~\ref{fig:simE} are $1-\i,1-0.6\i,-0.6\i,-0.2\i,1-0.2\i,1+0.2\i,0.2\i,0.6\i,1+0.6\i,1+\i,-a+\i,-a-\i$.

The common dynamical feature between the three shapes is that the unique stable equilibrium point is always located at the center of the maximal rectangle that has antiknobs in the boundary. 
When this maximal rectangle gets narrower in either the C-shaped polygon or the H-shaped domain, the center point changes to a saddle connected by two homoclinic trajectories that surround two new vortex centers. Observe that this not the case in the E-shaped domain, as this center point still occurs but it is now surrounded by a heteroclinic orbit. In this example, two new saddle points and two new vortex centers are born making each group of elliptic and hyperbolic points vertically aligned.

\section{Doubly Connected Domains}
\label{sc:doub}
%-------------------------------------------------------------------

\begin{figure}[!htb]%{0.5\textwidth}
	\centering
	\begin{subfigure}[b]{0.25\textwidth}
		\centering
		\includegraphics[width=\linewidth,trim=0 0 0 0,clip]{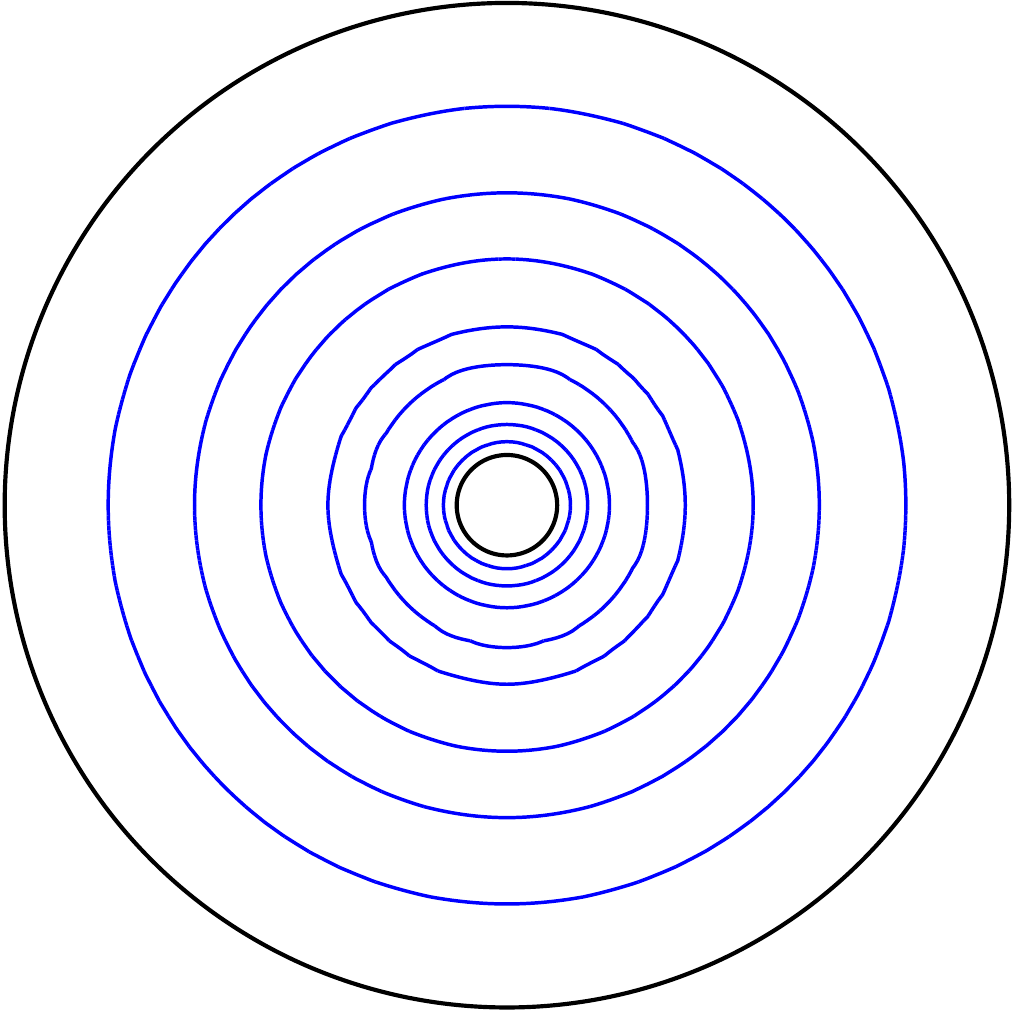}
		\caption{$p=0$}
		\label{fig:1c0}
	\end{subfigure}%
	\quad
	\begin{subfigure}[b]{0.25\textwidth}
		\centering
		\includegraphics[width=\linewidth,trim=0 0 0 0,clip]{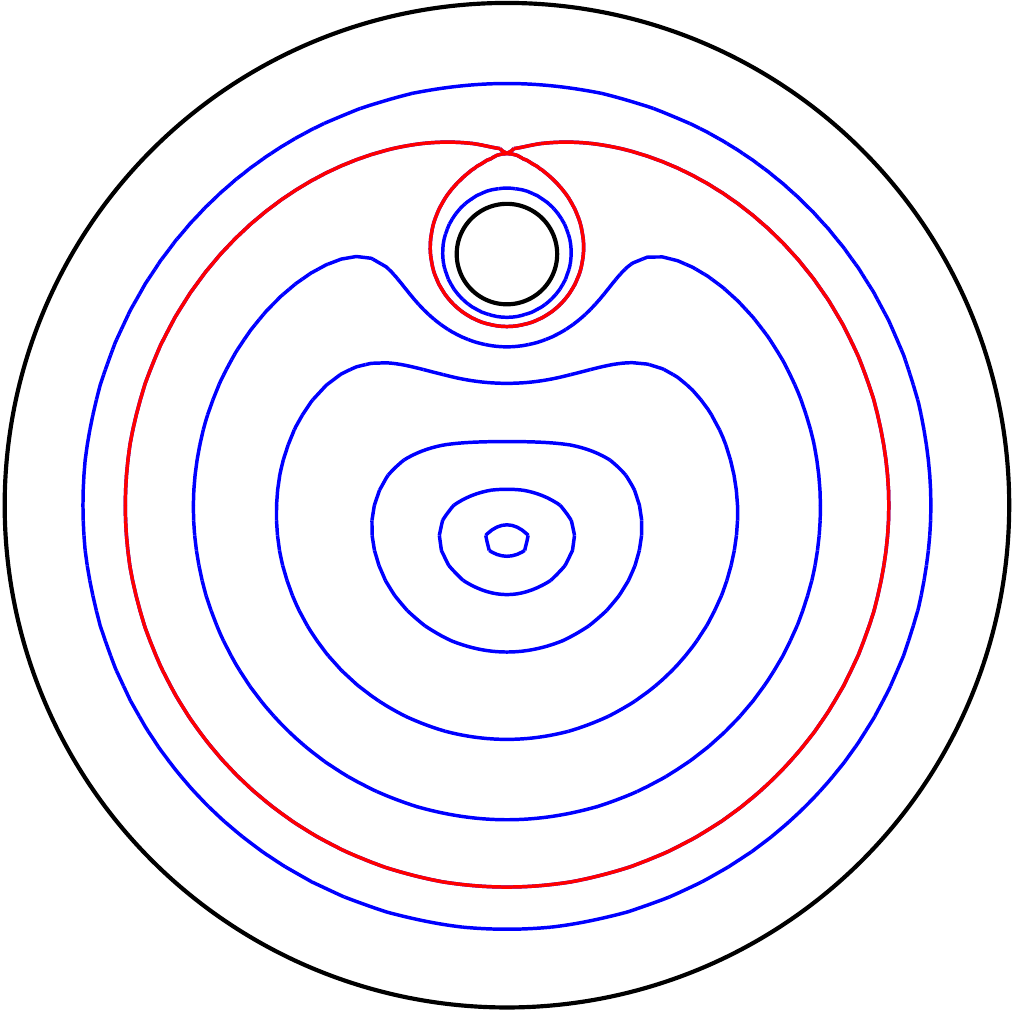}
		\caption{$p=0.5\i$}
		\label{fig:1c5}
	\end{subfigure}%
	\caption{Vortex trajectories in doubly connected circular domains. The outer circle is the unit circle $|\zeta|=1$ and the inner circle is $|\zeta - p|=0.1$.}
	\label{fig:1c}
\end{figure}

Sakajo~\cite{Tak-eq} studied the motion of a single point vortex in doubly connected circular domains. If the center of the inner circle is not the origin, then the Hamiltonian always has one elliptic center point and one saddle point connected to itself by two homoclinic vortex trajectories; see Figure~\ref{fig:1c5}. In this section, we discuss what happens to the dynamics of point vortex when we replace circular domains with polygonal domains. 

We assume the external boundary component of the domain $G$ is composed of the square with vertices $\pm 1 \pm\i$. For the inner obstacle shape, we consider four cases: a square, a rotated square, an equilateral triangle and a hexagon.
% all with the same radius of $0.1$.
In Figures~\ref{fig:1s}-\ref{fig:1h}, we show vortex motion in these doubly connected polygonal domains by plotting some contours of the Hamiltonian for different positions of the polygonal obstacle center. 
%The first dynamical feature is that the topological pattern is different from circular domain to polygonal domains and also between the polygonal domains.

\subsection{Concentric Domains}

\begin{figure}[!htb] %
	\centering
	\begin{subfigure}[b]{0.24\textwidth}
		\centering
		\includegraphics[width=\linewidth,trim=0 0 0 0,clip]{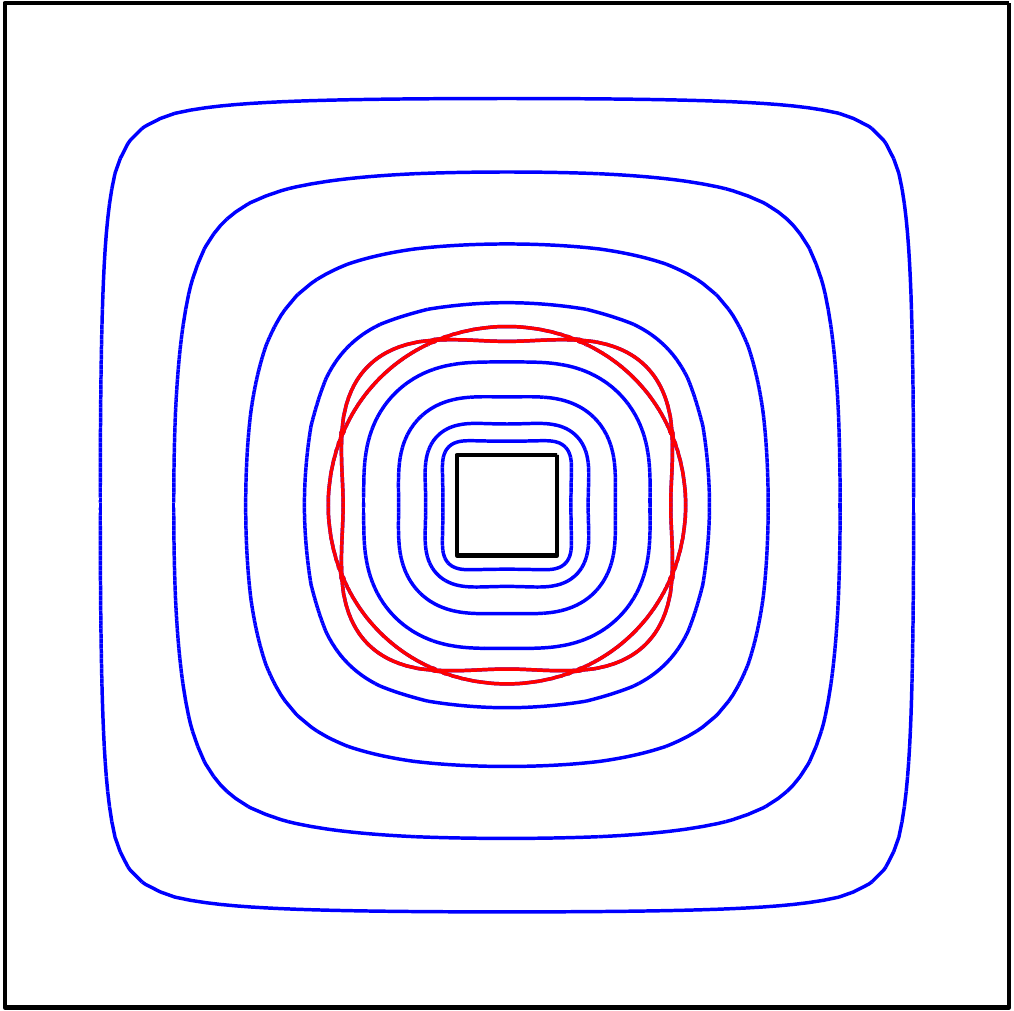}
		\caption{$p=0$}
		\label{fig:1s0}
	\end{subfigure}%
	~ 
	\begin{subfigure}[b]{0.24\textwidth}
		\centering
		\includegraphics[width=\linewidth,trim=0 0 0 0,clip]{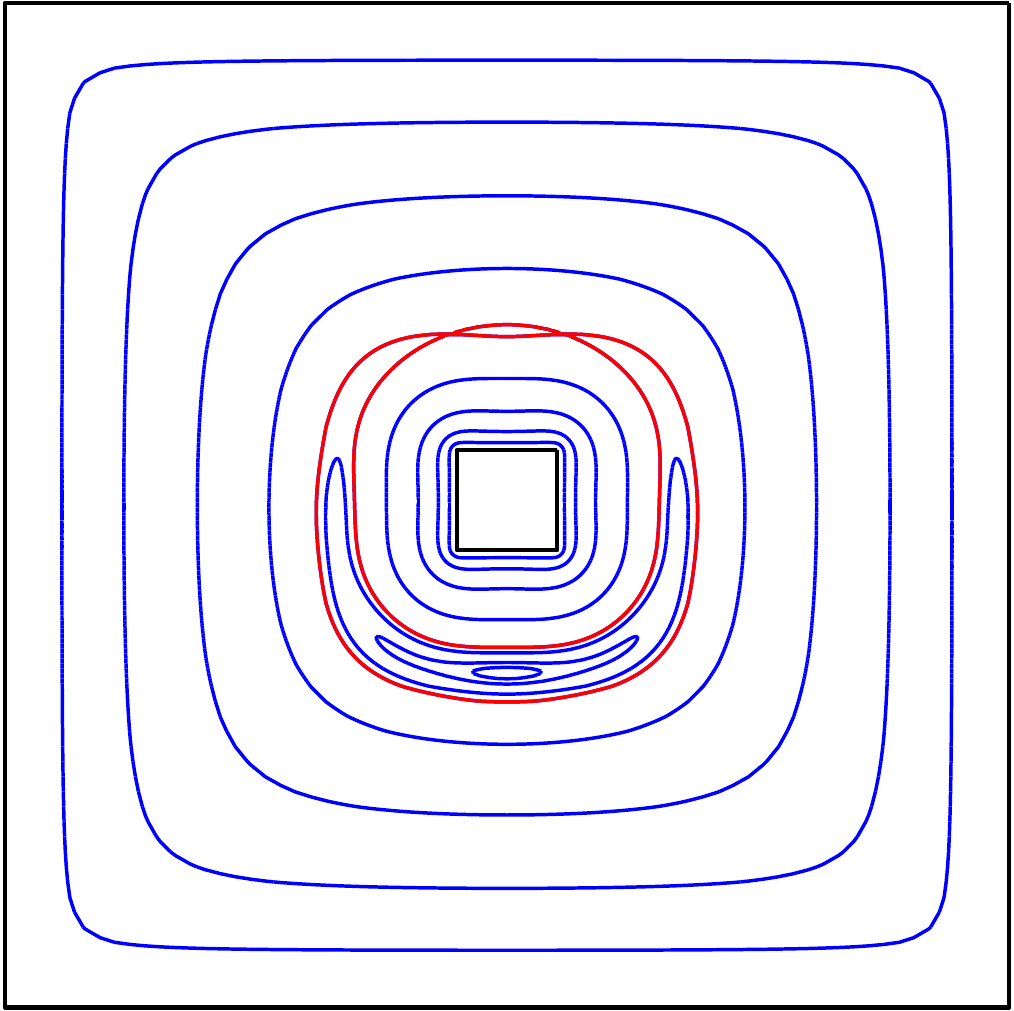}
		\caption{$p=0.01\i$}
		\label{fig:1s01}
	\end{subfigure}%
	~
	\begin{subfigure}[b]{0.24\textwidth}
		\centering
		\includegraphics[width=\linewidth,trim=0 0 0 0,clip]{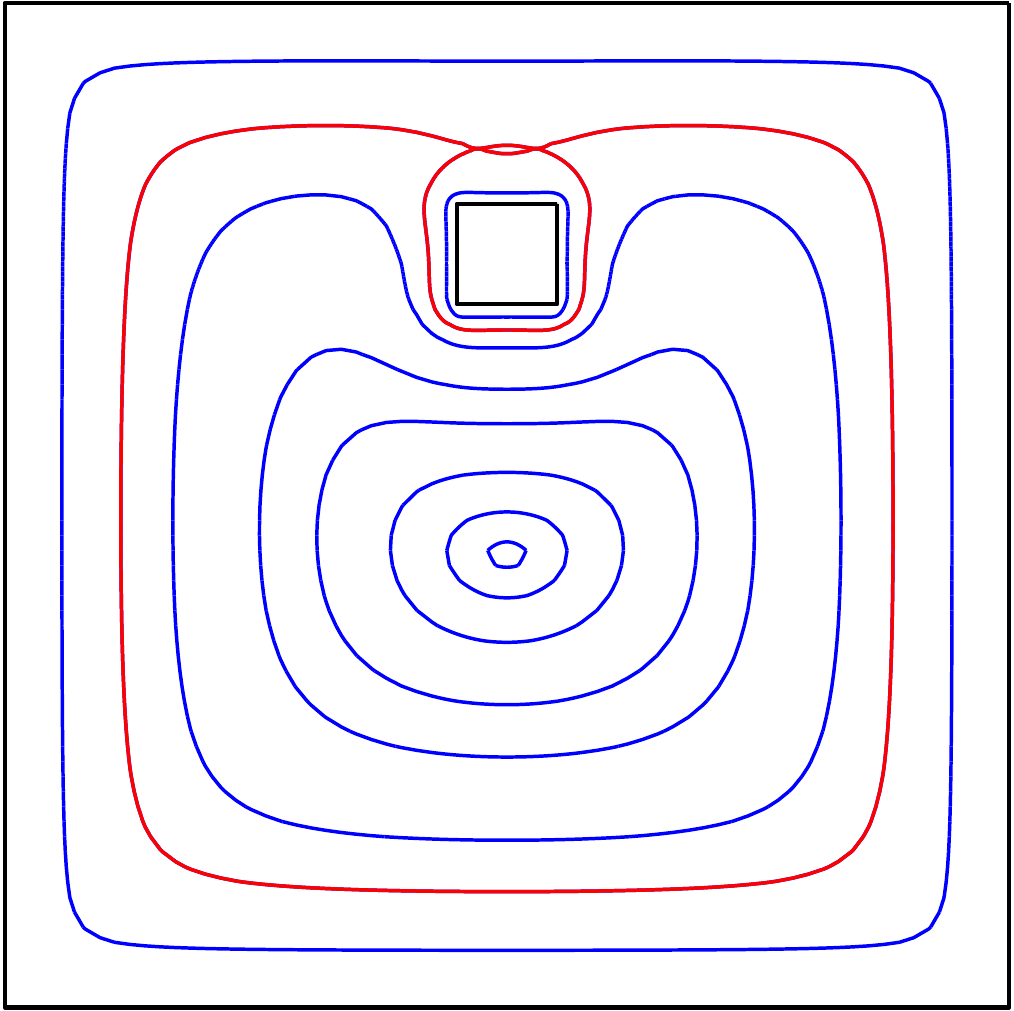}
		\caption{$p=0.5\i$}
		\label{fig:1s5}
	\end{subfigure}%
	~ 
	\begin{subfigure}[b]{0.24\textwidth}
		\centering
		\includegraphics[width=\linewidth,trim=0 0 0 0,clip]{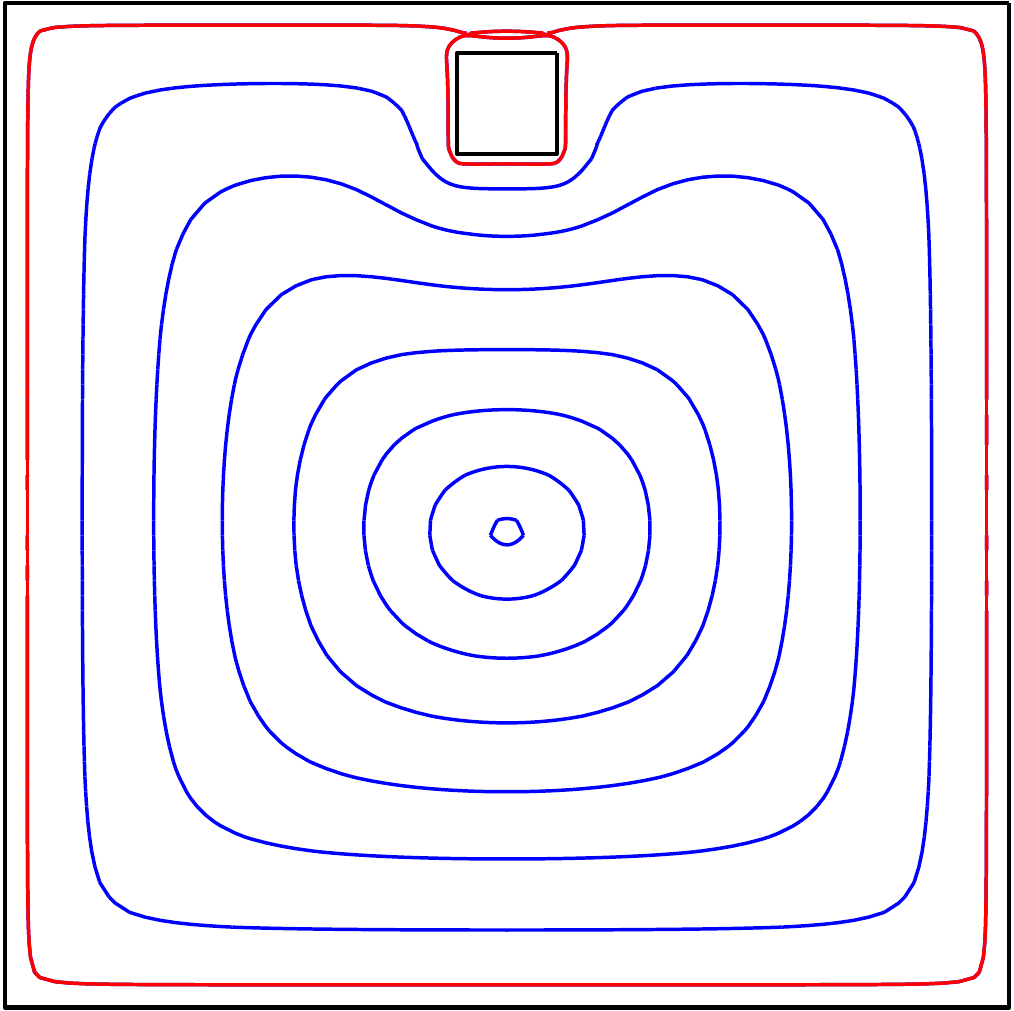}
		\caption{$p=0.8\i$}
		\label{fig:1s8}
	\end{subfigure}%
	\vskip\baselineskip
	\begin{subfigure}[b]{0.24\textwidth}
		\centering
		\includegraphics[width=\linewidth,trim=0 0 0 0,clip]{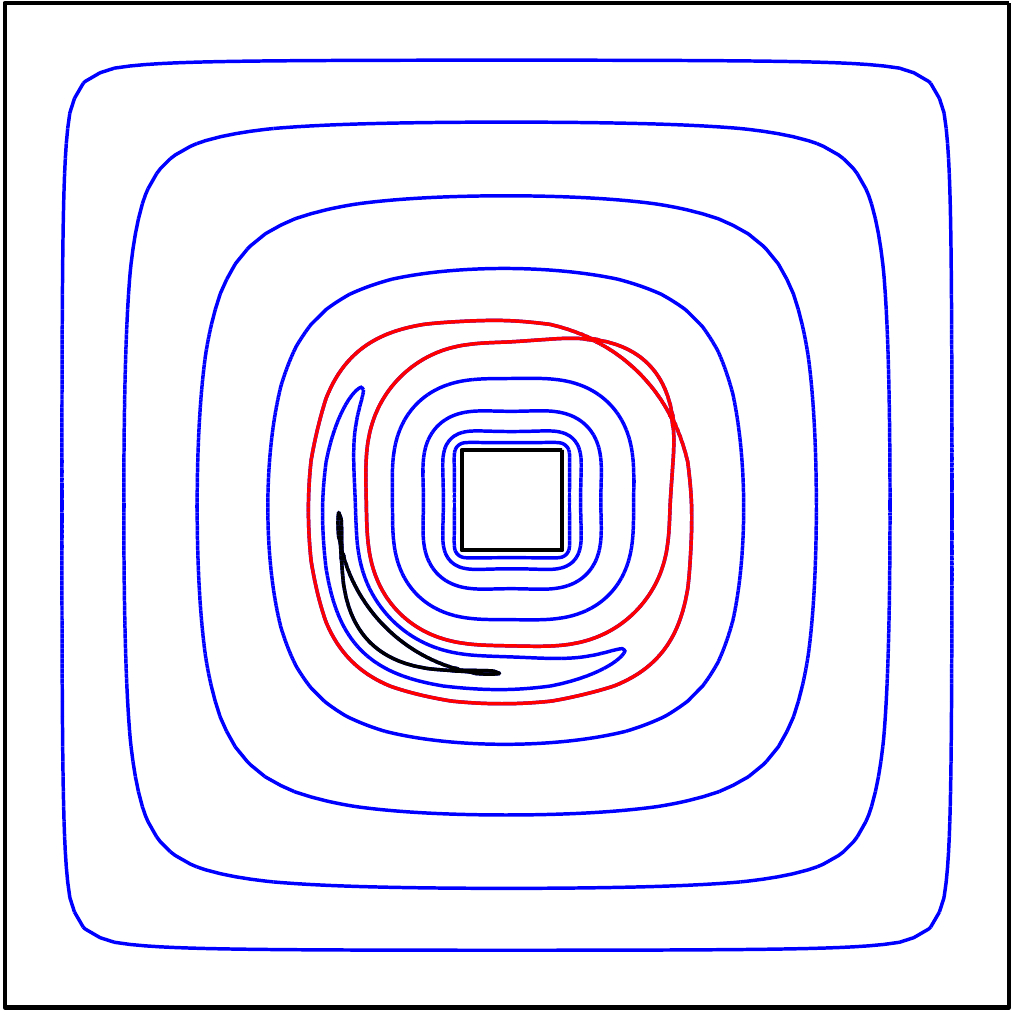}
		\caption{$p=0.01+0.01\i$}
		\label{fig:1s01d}
	\end{subfigure}%
	~ 
	\begin{subfigure}[b]{0.24\textwidth}
		\centering
		\includegraphics[width=\linewidth,trim=0 0 0 0,clip]{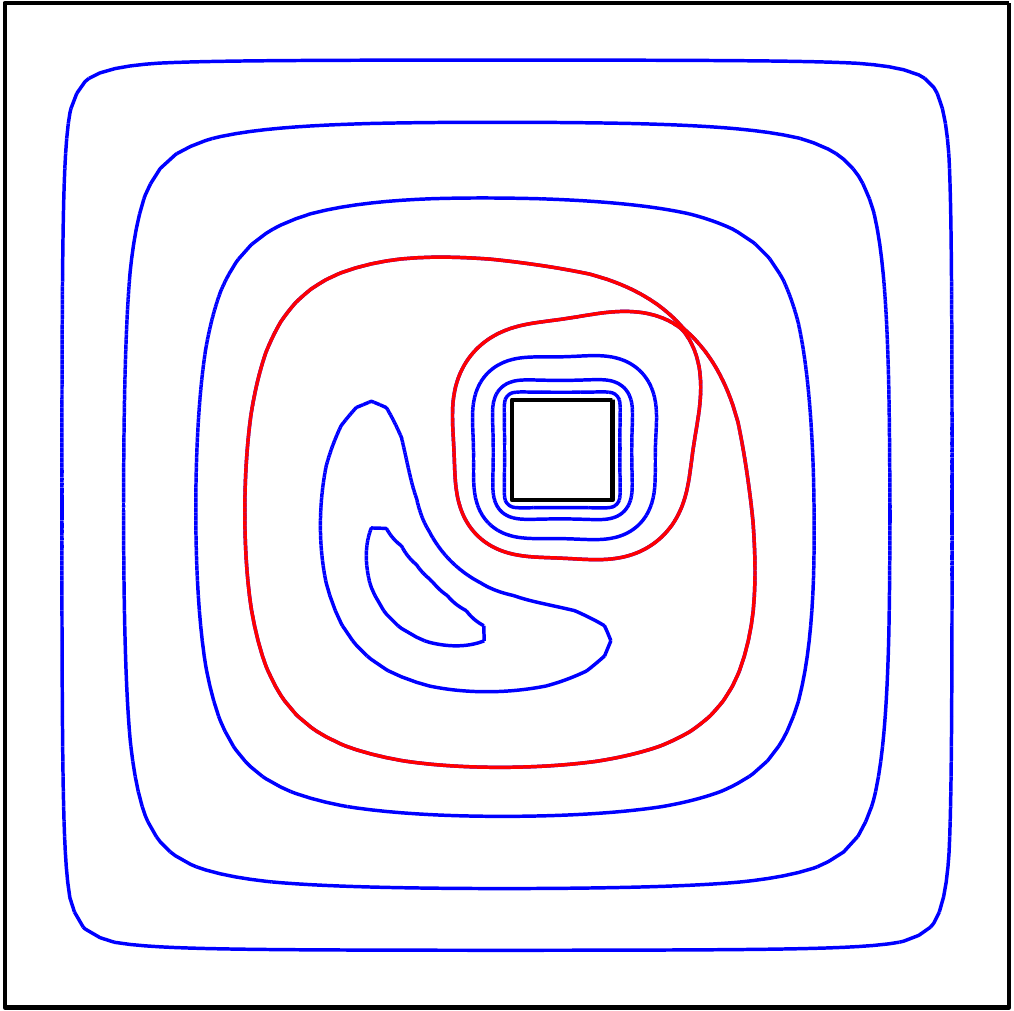}
		\caption{$p=0.11+0.11\i$}
		\label{fig:1s011d}
	\end{subfigure}%
	~
	\begin{subfigure}[b]{0.24\textwidth}
		\centering
		\includegraphics[width=\linewidth,trim=0 0 0 0,clip]{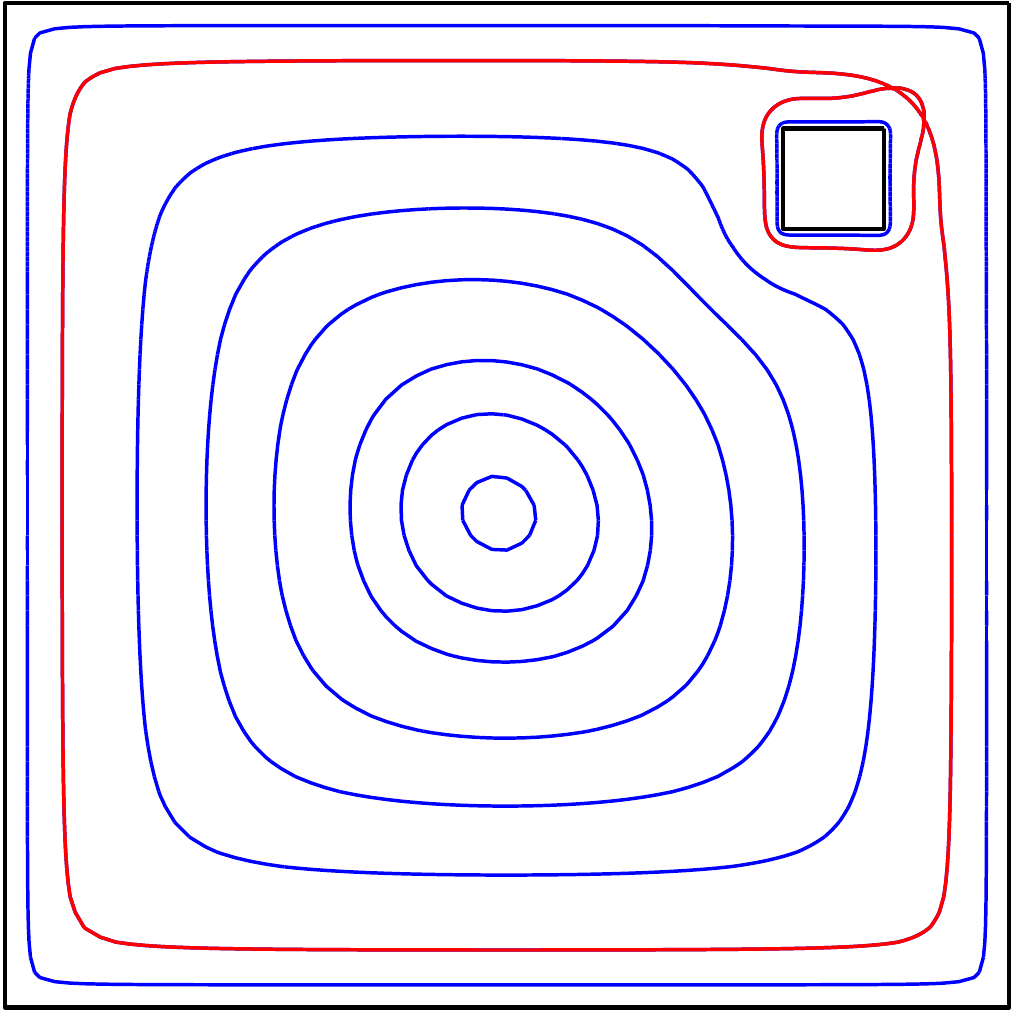}
		\caption{$p=0.65+0.65\i$}
		\label{fig:1s65d}
	\end{subfigure}%
	~ 
	\begin{subfigure}[b]{0.24\textwidth}
		\centering
		\includegraphics[width=\linewidth,trim=0 0 0 0,clip]{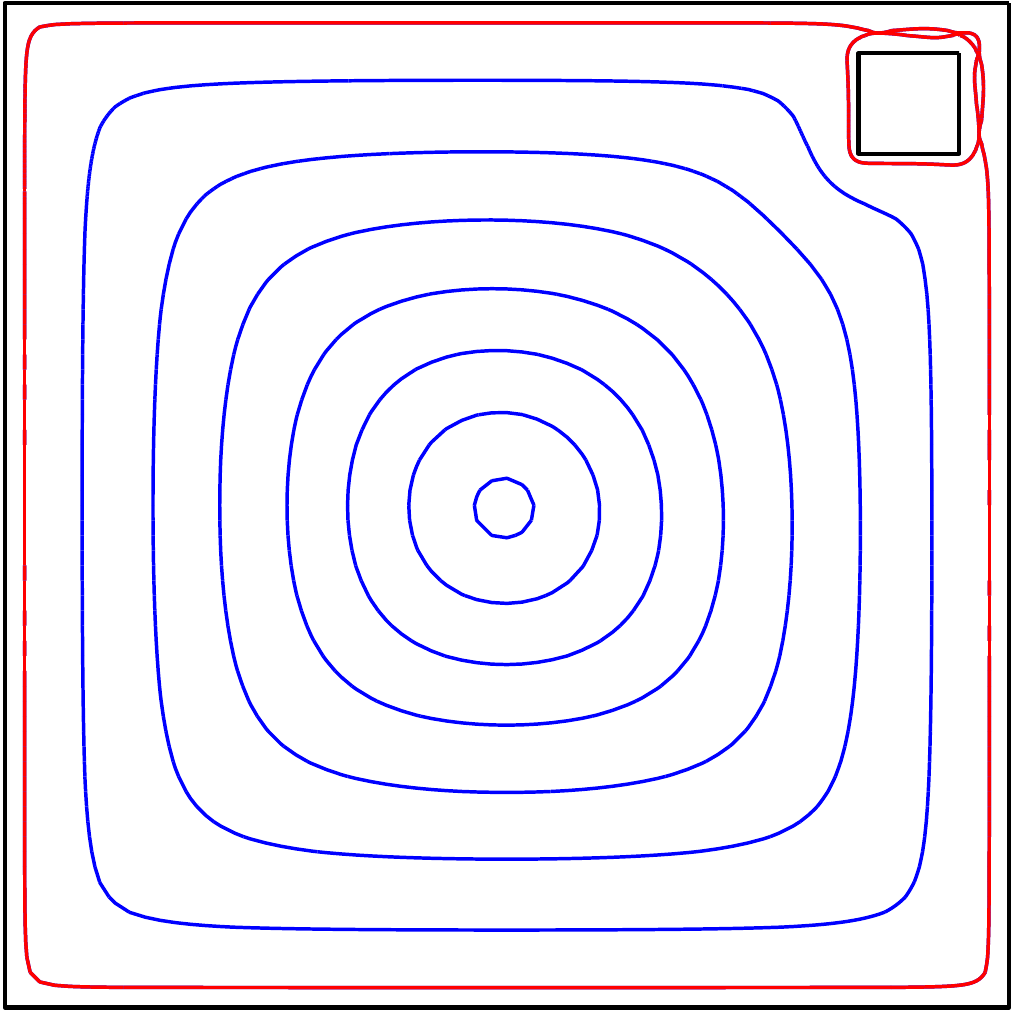}
		\caption{$p=0.8+0.8\i$}
		\label{fig:1s8d}
	\end{subfigure}%	
	\caption{Vortex trajectories in doubly connected square domains where the square obstacle has a center $p$ and a horizontal radius of $0.1$.}
	%the vertices of the square obstacle are at different positions $p+0.1+0.1\i,p+0.1-0.1\i,p-0.1-0.1\i,p-0.1+0.1\i$.}
	\label{fig:1s}
	%\scalebox{0.3}{\includegraphics[trim=0 0 0 0,clip]{Fig_1s_p11}}
	%\hfill	
\end{figure}

We first consider the case of concentric domains in Subfigures (a) of Figures~\ref{fig:1s}, \ref{fig:1d}, \ref{fig:1t} and \ref{fig:1h}.
Unlike the annulus domain (Figure~\ref{fig:1c0}) where there is no center or saddle, the polygonal structure gives rise to several equilibria of this type. The number of vortex centers of the Hamiltonian is the same as that of saddle points, which is again in accordance with Morse theory. The elliptic equilibria are surrounded by closed vortex trajectories and  the hyperbolic ones  are connected by either homoclinic or heteroclinic orbits. 

In  Table~\ref{tab:1a}, we show the number of saddle points and how they are linked by vortex trajectories in concentric polygonal doubly connected domains. By looking at these numbers there are aspects to be highlighted regarding the topology of vortex trajectories near the obstacle boundaries. For a square obstacle (standard and rotated), we observe a pure heteroclinic structure, which is not the case for triangle and hexagon obstacles. 
Near the square obstacle in Figure~\ref{fig:1s0}, there occur eight saddle points that are connected to each other by eight heteroclinic loops. At first this number seems to result from the total number of reflectional and rotational symmetries or the number of corner points in the domain. However, when we rotate the square obstacle by an angle of $\pi/4$ in Figure~\ref{fig:1d0} so that its sides become perpendicular to the diagonals of the outer square, this number gets reduced by half, although the total symmetries and corner points in the domain remain the same. 

\begin{table}[!htb]
	\centering
	\begin{tabular}{@{}cccc@{}}\toprule
		Obstacle shape	& Saddle  & Homoclinic  & Heteroclinic \\ \midrule
		Square    		& 8 & 0 & 8  \\
		Rotated square  & 4 & 0 & 4  \\
		Triangle   		& 3 & 4 & 1  \\
		Hexagon  		& 4 & 4 & 2   \\ \bottomrule
	\end{tabular}
	\caption{Number of saddle points and their connecting critical loops in doubly connected concentric polygonal domains with different obstacles shapes.}
	\label{tab:1a}%
\end{table}  

If we change the obstacle shape to an equilateral triangle as in Figure~\ref{fig:1t0}, there are only three saddle points that are connected by four homoclinic and one heteroclinic loops. For the case of a regular hexagon obstacle in Figure~\ref{fig:1h0}, the number of saddles increases again to four points connected by four homoclinic and two heteroclinic loops. This suggests that the number of equilibria and topology pattern  of vortex motion might be more related to the interaction between the vertices of the polygon obstacle and the outer square boundaries. 

In the next subsections, we are interested in the manner the vortex motion changes due to obstacle movement.

\subsection{A Square Obstacle}

After breaking the symmetry with respect to the outer square in Figure~\ref{fig:1s01} by a perturbation of the center $p$ of the square obstacle with a translation of $0.01$ unit along the imaginary axis, the topology pattern of vortex motion changes dramatically. Only two saddle points which are situated on the side of the obstacle movement still occur. The saddle connection is made by two heteroclinic loops encircling two center points, which are symmetrically located on the imaginary axis with one above the obstacle and the other below it. The same configuration appears if we continue moving the square obstacle northward as shown in Figures~\ref{fig:1s5} and \ref{fig:1s8}. 

On the other hand, if we choose to break the symmetry by shifting the obstacle center to the upper right corner of the outer square as in Figure~\ref{fig:1s01d}, then a small perturbation changes again the topology pattern, but in a slightly different way. Besides the heteroclinic cycle containing two saddles in the direction of the obstacle movement as in the previous case, 
%two center points occur on the diagonal line $y=x$ with the one located  near the upper right corner of the obstacle is surrounded by a heteroclinic loop that connects two saddle points. 
the other elliptic point which is situated on the left lower side of the diagonal line $y=x$ is now the center of a new heteroclinic loop itself joining two new saddles.   Each of the latter hyperbolic points belongs to one homoclinic orbit surrounding a vortex center, which gives a total of eight critical points. 

When displacing the center of the obstacle to $p=0.11+0.11\i$ as shown in Figure~\ref{fig:1s011d}, the vortex motion loses all its rest points except one saddle and one center, which are respectively located on the upper right and lower left sides of the diagonal line $y=x$. 
Observe also that a topological configuration similar to the vortex motion in Figures~\ref{fig:1s5} and \ref{fig:1s8} appears but on the diagonal instead of the imaginary axis when $p=0.65+0.65\i$; see Figure~\ref{fig:1s65d}. 
After the separation of the opposite corners in the outer and inner squares gets smaller as shown in Figure~\ref{fig:1s8d}, the motion dynamics consists of four saddles and three center points in the separation region plus a fourth elliptic point situated near the center of the outer square. It is worth emphasizing that homoclinic orbits are only present in the phase portraits in Figures~\ref{fig:1s01d} and~\ref{fig:1s011d} while all other cases contain pure heteroclinic cycles.

\subsection{A Rotated Square Obstacle}

\begin{figure}[!htb] %
	\centering
	\begin{subfigure}[b]{0.32\textwidth}
		\centering
		\includegraphics[width=\linewidth,trim=0 0 0 0,clip]{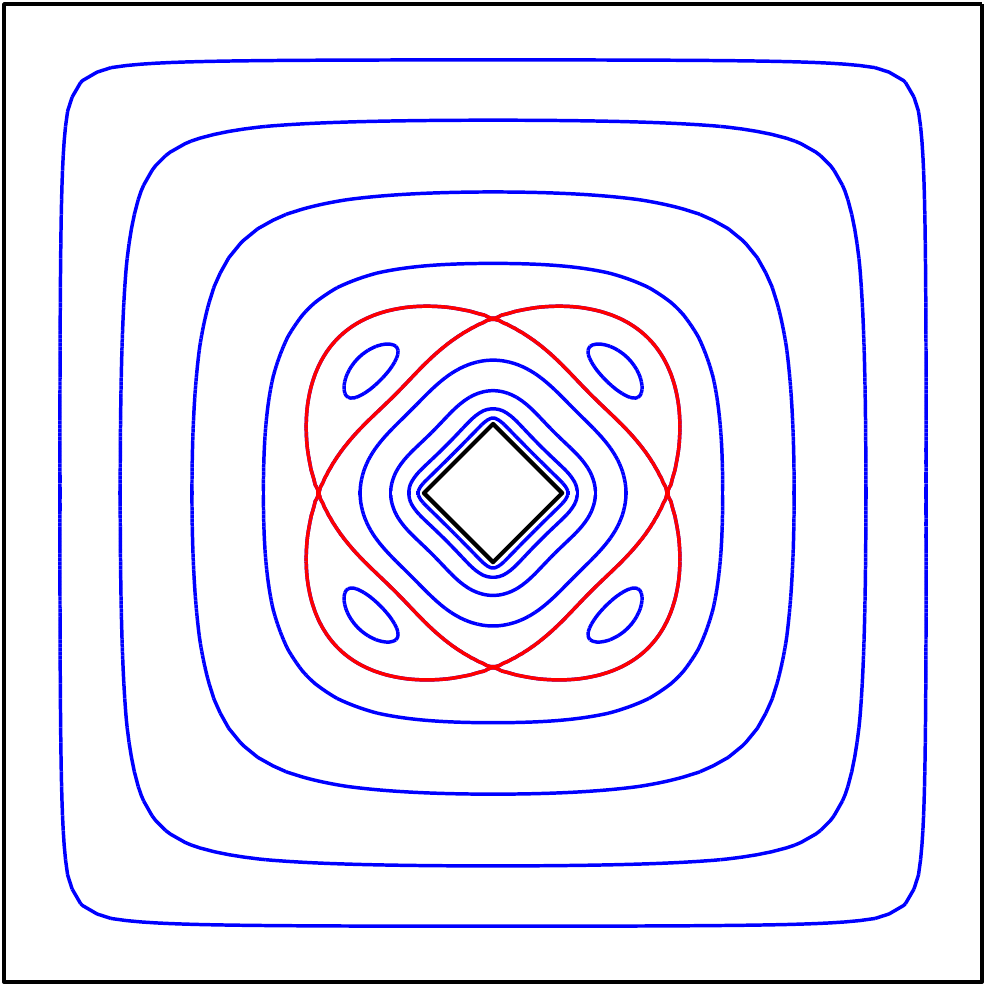}
		\caption{$p=0$}
		\label{fig:1d0}
	\end{subfigure}%
	~ 
	\begin{subfigure}[b]{0.32\textwidth}
		\centering
		\includegraphics[width=\linewidth,trim=0 0 0 0,clip]{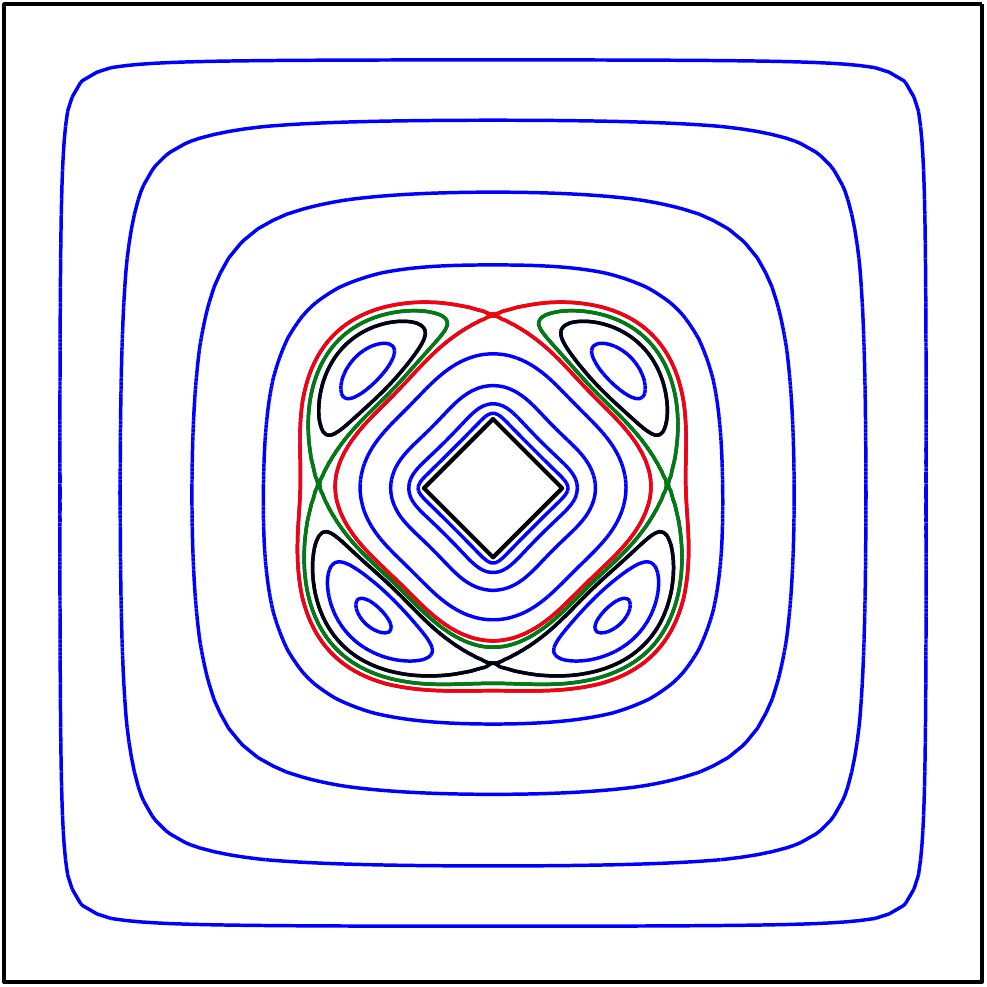}
		\caption{$p=0.01\i$}
		\label{fig:1d01}
	\end{subfigure}%
	~
	\begin{subfigure}[b]{0.32\textwidth}
		\centering
		\includegraphics[width=\linewidth,trim=0 0 0 0,clip]{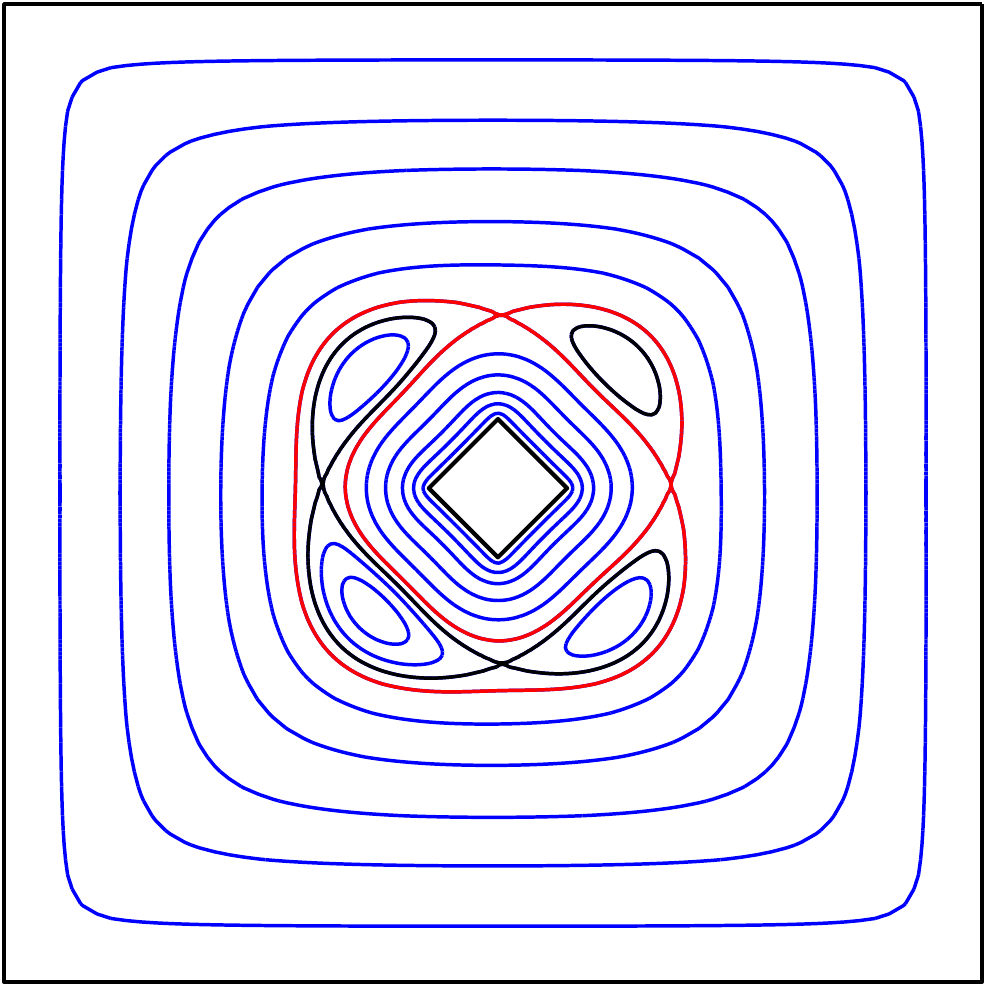}
		\caption{$p=0.01+0.01\i$}
		\label{fig:1d01d}
	\end{subfigure}%
	\vskip\baselineskip
	\begin{subfigure}[b]{0.32\textwidth}
		\centering
		\includegraphics[width=\linewidth,trim=0 0 0 0,clip]{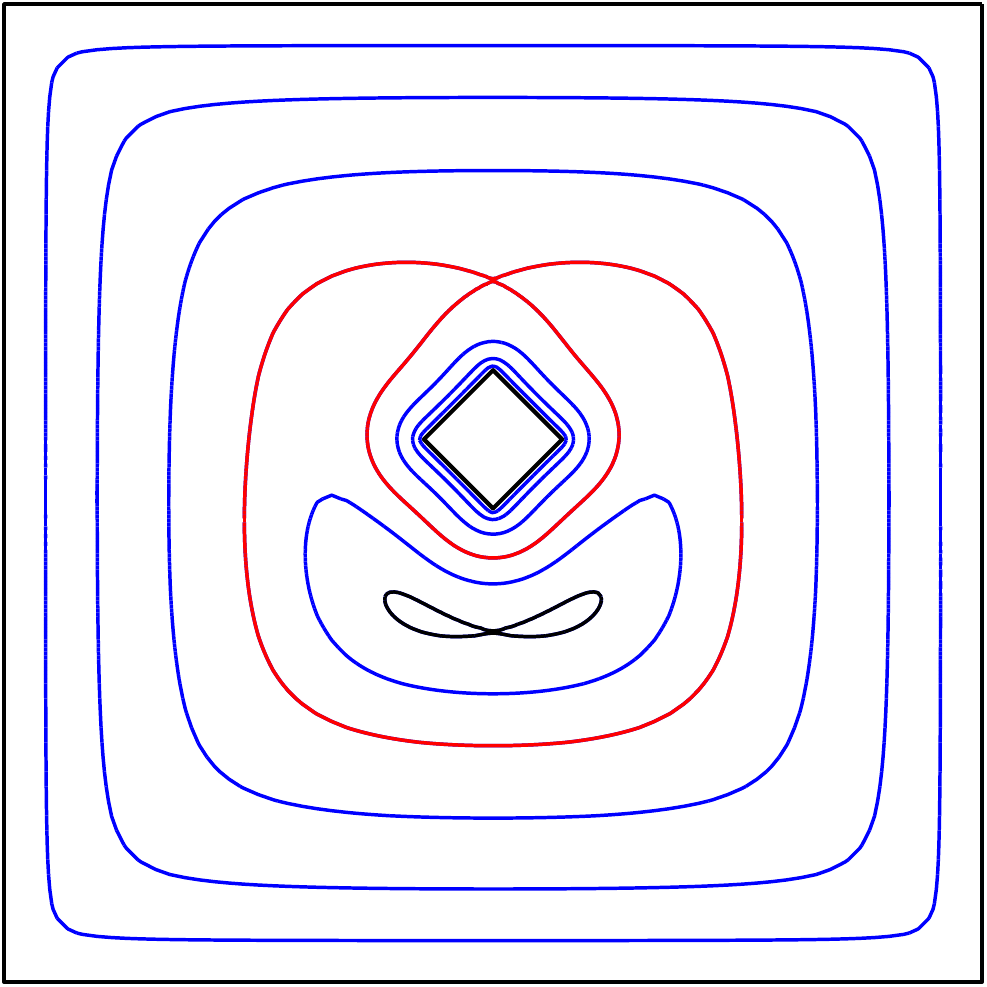}
		\caption{$p=0.11\i$}
		\label{fig:1d11}
	\end{subfigure}%
	~ 
	\begin{subfigure}[b]{0.32\textwidth}
		\centering
		\includegraphics[width=\linewidth,trim=0 0 0 0,clip]{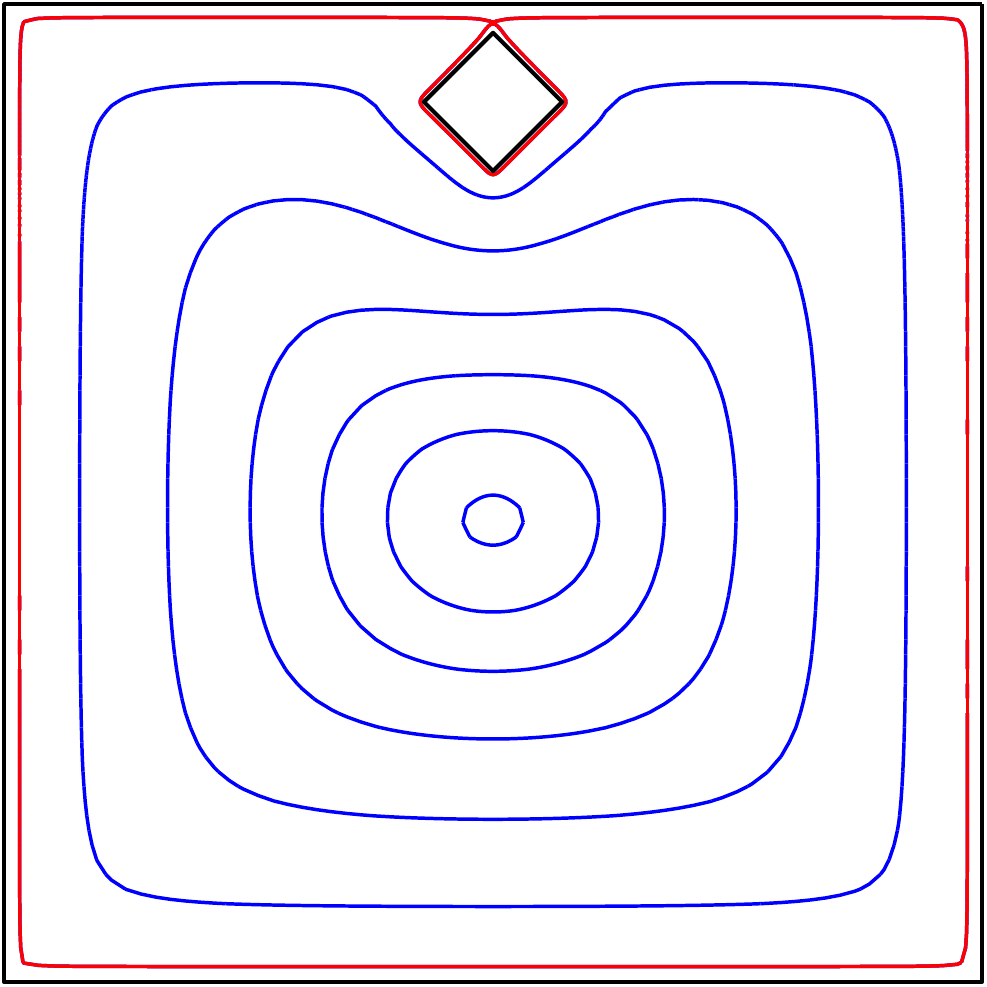}
		\caption{$p=0.8\i$}
		\label{fig:1d8}
	\end{subfigure}%	
	~ 
	\begin{subfigure}[b]{0.32\textwidth}
		\centering
		\includegraphics[width=\linewidth,trim=0 0 0 0,clip]{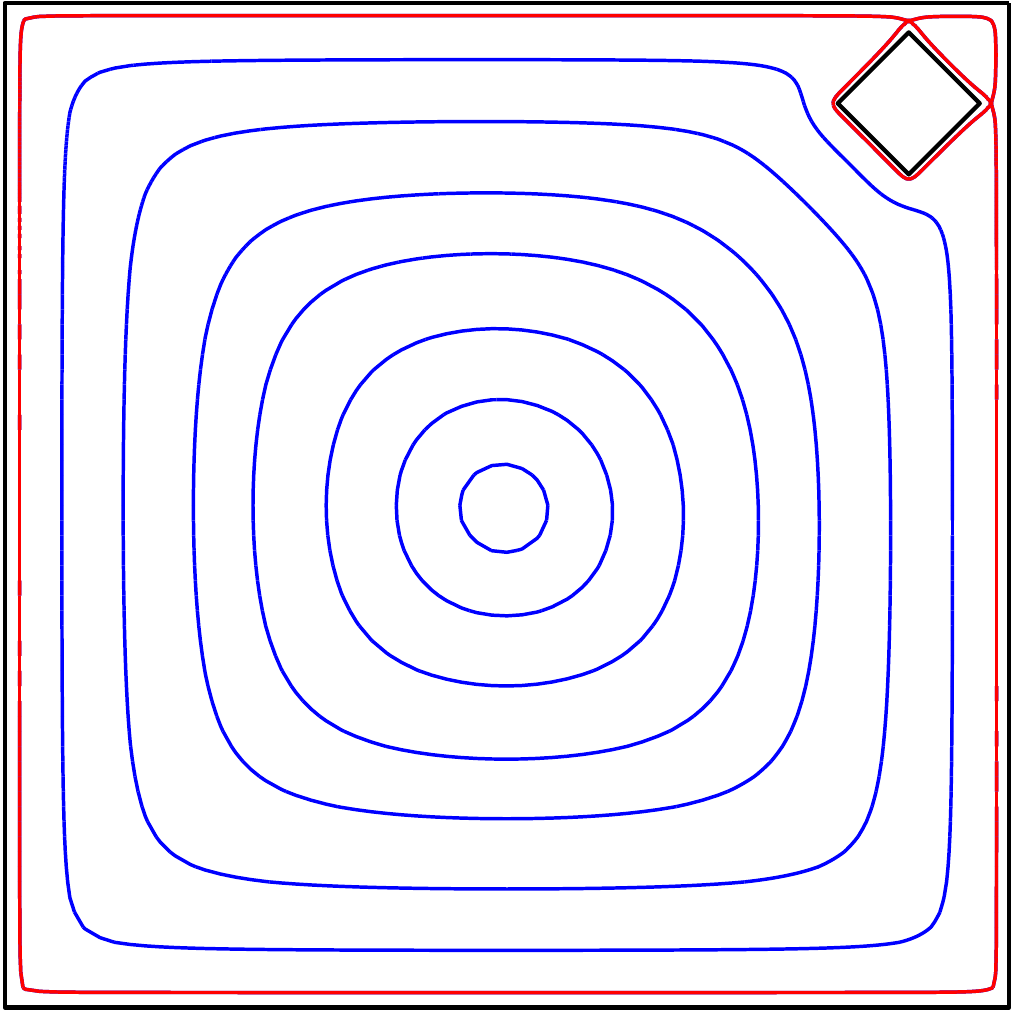}
		\caption{$p=0.8+0.8\i$}
		\label{fig:1d8d}
	\end{subfigure}%	
	\caption{Vortex trajectories in doubly connected polygonal domains where the vertices of the rotated square obstacle are on the circle of center $p$ and radius $0.1$.}
	\label{fig:1d}
\end{figure}

We now examine the effect of symmetry breaking when the obstacle is the rotated square in Figure~\ref{fig:1d}. The small displacement of the center along the imaginary axis in Figure~\ref{fig:1d01} does not reduce the number of equilibria as in the square obstacle. However, this breakdown destroys the heteroclinic structure seen in Figure~\ref{fig:1d0}, which confirms the structurally instability of heteroclinic cycles. Indeed, none of the four saddles is anymore the intersection of two heteroclinic loops. However, each saddle gives rise to the bifurcation of two homoclinic orbits  that form a separatrix. 

On the other hand, a perturbation of $\sqrt{2}\times 0.01$ toward the top right corner of the outer square, as done in Figure~\ref{fig:1d01d}, seems to damage less the heteroclinic structure since we still have two heteroclinic loops surrounding the two centers on the main diagonal line. The two other heteroclinic loops change to homoclinic orbits as they get separated from the two saddles near the diagonal line $y=x$.

Figure~\ref{fig:1d11} reveals that when we move the obstacle center to $p=0.11\i$, the vortex motion only involves homoclinic trajectories that intersect at two hyperbolic fixed points situated on the imaginary axis. Observe also that the two bottom homoclinic orbits encircle two elliptic points. 

We next consider the point vortex motion when the rotated square obstacle interacts more closely with the outer square boundaries. As the obstacle gets closer to the upper outer boundary in Figure~\ref{fig:1d8}, the lower saddle point changes to a center yielding a configuration similar to the circular domain in Figure~\ref{fig:1c5}, while the homoclinic pattern is maintained. This seems related to the interaction between the obstacle top vertex and the nearest outer square side. Figure~\ref{fig:1d8d} confirms this observation as two saddle points occur when two obstacle vertices become closer to the outer boundary. Note also that the saddle connections are of a heteroclinic nature in this case.

\subsection{An Equilateral Triangle Obstacle}

\begin{figure}[!htb] %
	\centering
	\begin{subfigure}[b]{0.32\textwidth}
		\centering
		\includegraphics[width=\linewidth,trim=0 0 0 0,clip]{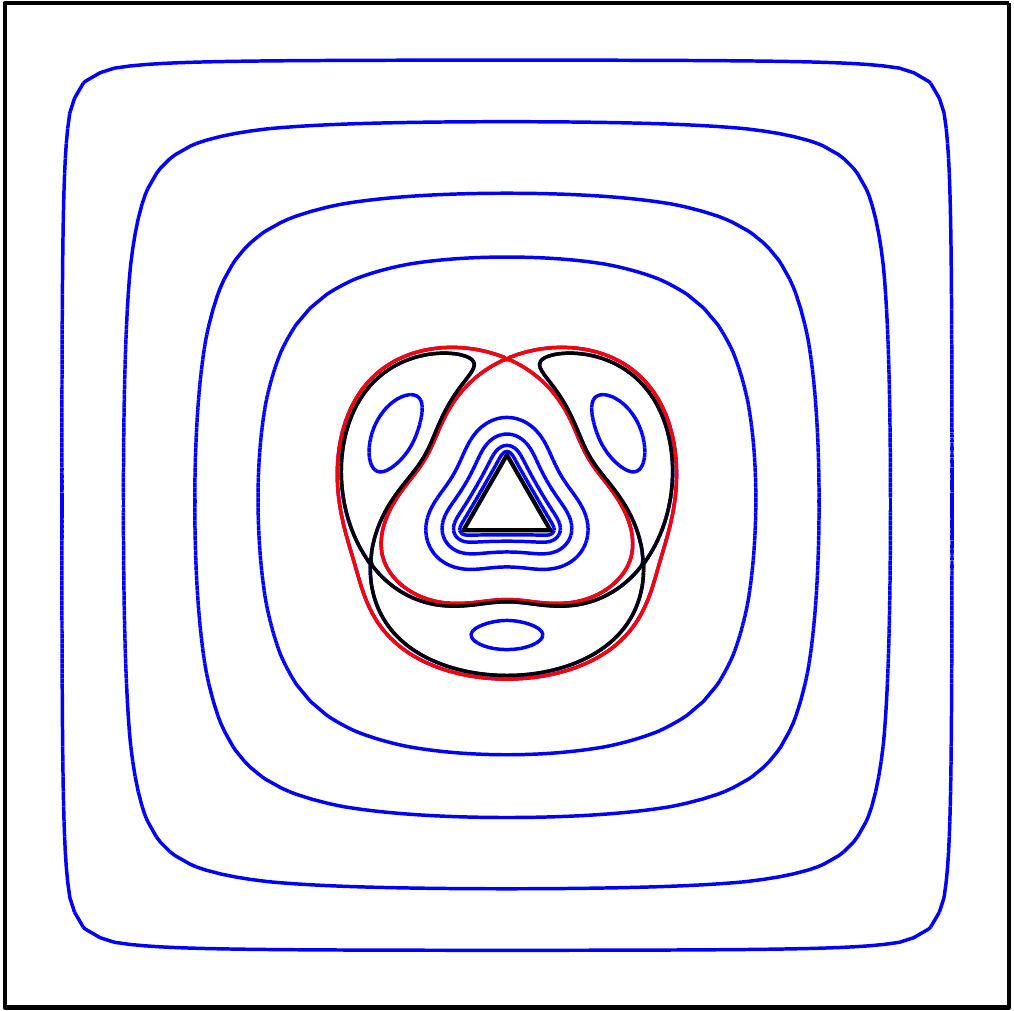}
		\caption{$p=0$}
		\label{fig:1t0}
	\end{subfigure}%
	~
	\begin{subfigure}[b]{0.32\textwidth}
		\centering
		\includegraphics[width=\linewidth,trim=0 0 0 0,clip]{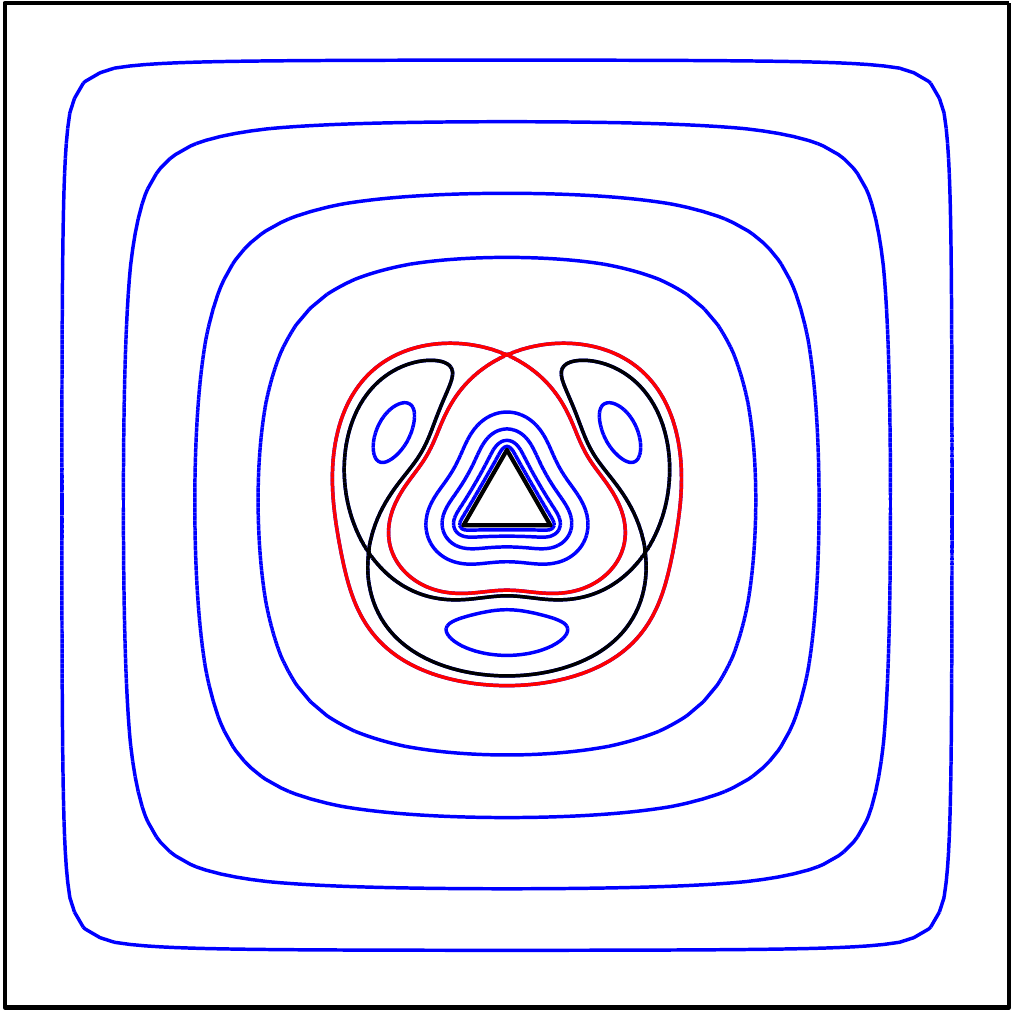}
		\caption{$p=0.01\i$}
		\label{fig:1t01u}
	\end{subfigure}%
	~ 
	\begin{subfigure}[b]{0.32\textwidth}
		\centering
		\includegraphics[width=\linewidth,trim=0 0 0 0,clip]{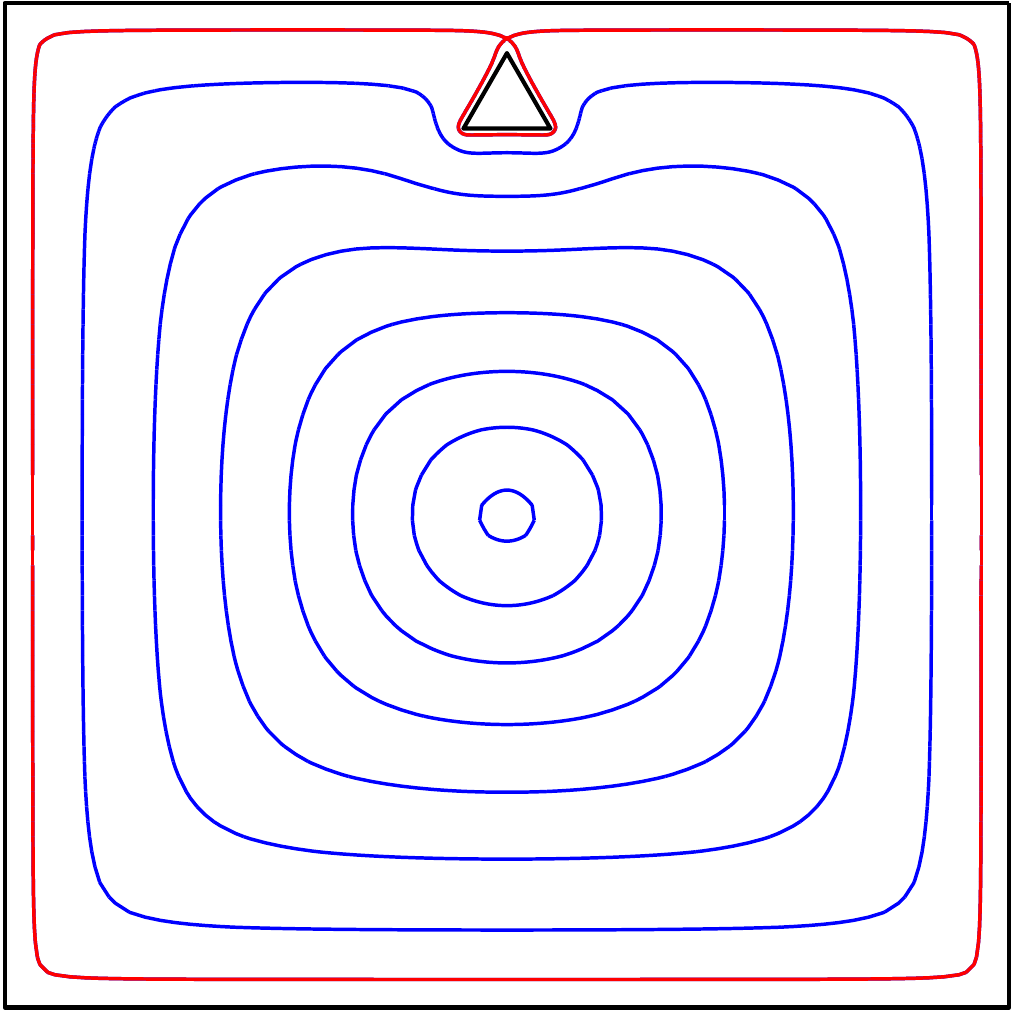}
		\caption{$p=0.8\i$}
		\label{fig:1t8u}
	\end{subfigure}%
	\vskip\baselineskip
	\begin{subfigure}[b]{0.32\textwidth}
		\centering
		\includegraphics[width=\linewidth,trim=0 0 0 0,clip]{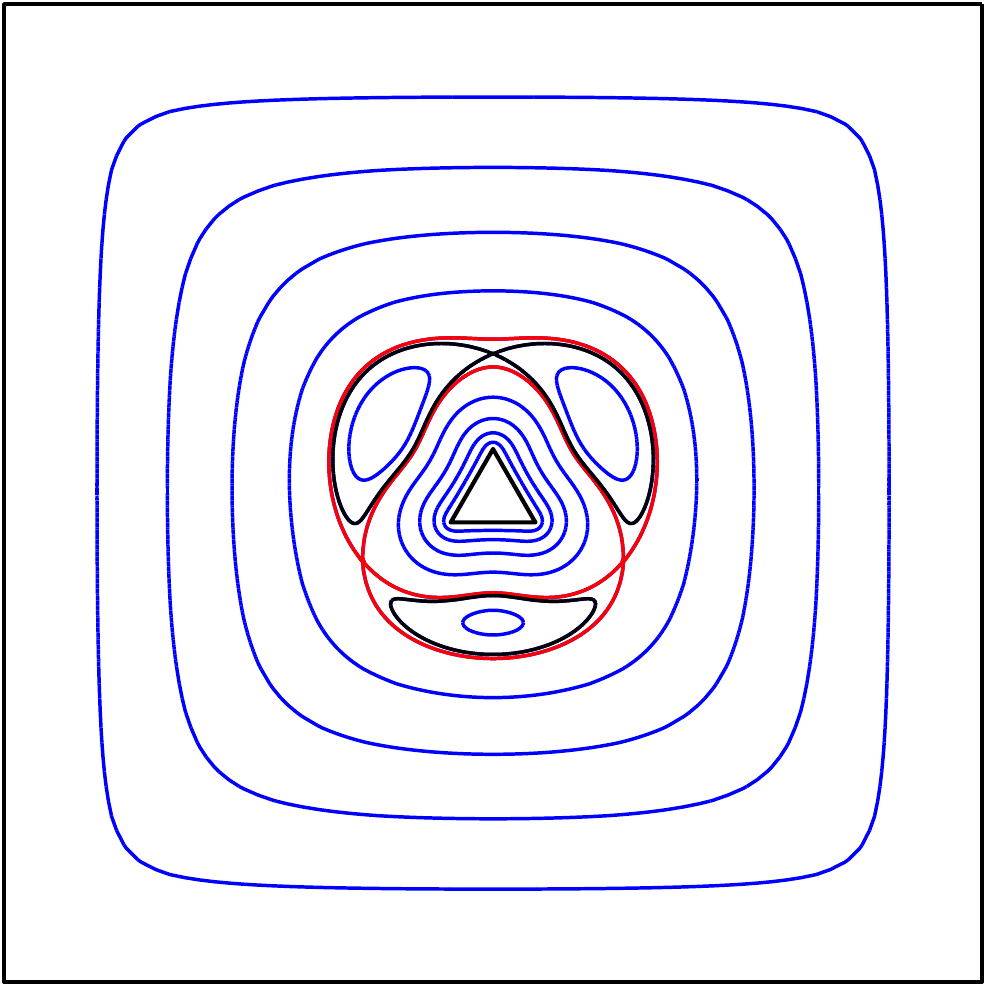}
		\caption{$p=-0.01\i$}
		\label{fig:1t01l}
	\end{subfigure}%
	~ 
	\begin{subfigure}[b]{0.32\textwidth}
		\centering
		\includegraphics[width=\linewidth,trim=0 0 0 0,clip]{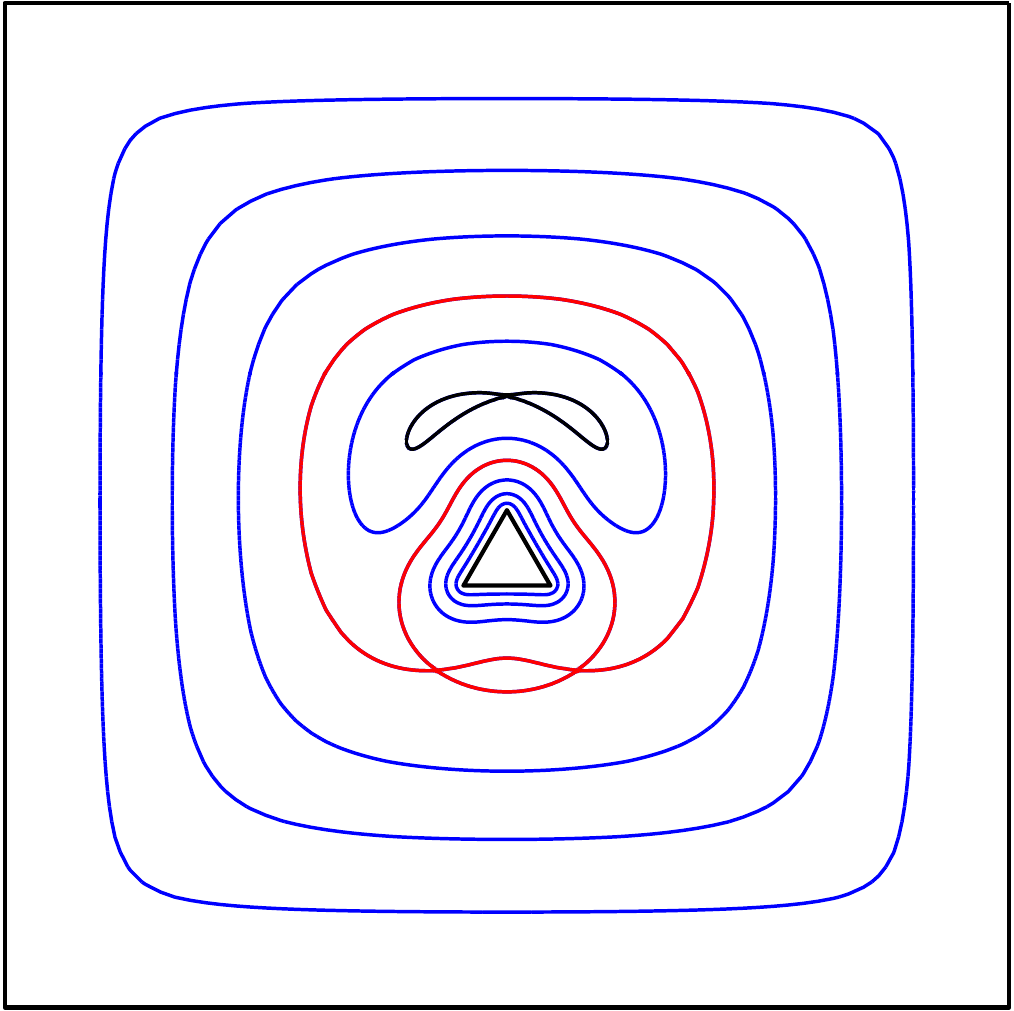}
		\caption{$p=-0.11\i$}
		\label{fig:1t11l}
	\end{subfigure}%
	~
	\begin{subfigure}[b]{0.32\textwidth}
		\centering
		\includegraphics[width=\linewidth,trim=0 0 0 0,clip]{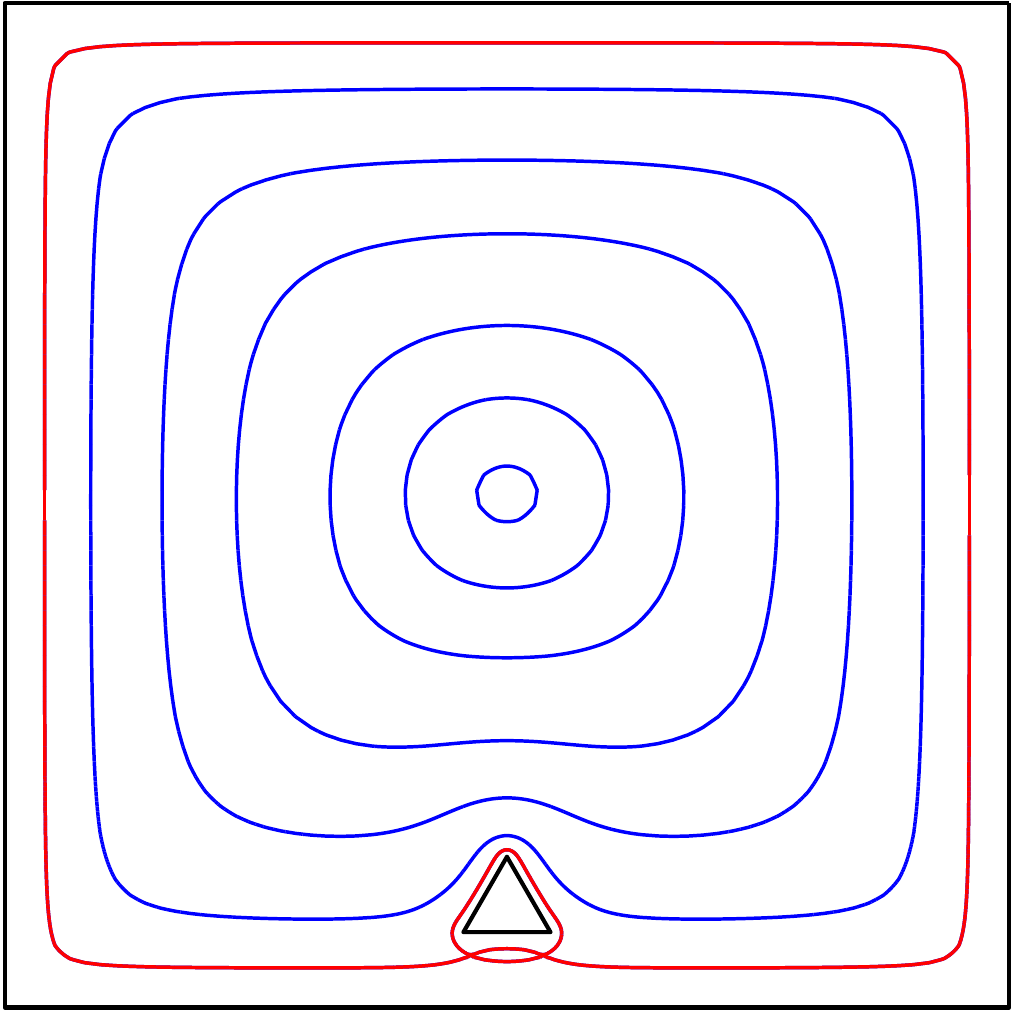}
		\caption{$p=-0.8\i$}
		\label{fig:1t8l}
	\end{subfigure}%	
	\vskip\baselineskip
	\begin{subfigure}[b]{0.32\textwidth}
		\centering
		\includegraphics[width=\linewidth,trim=0 0 0 0,clip]{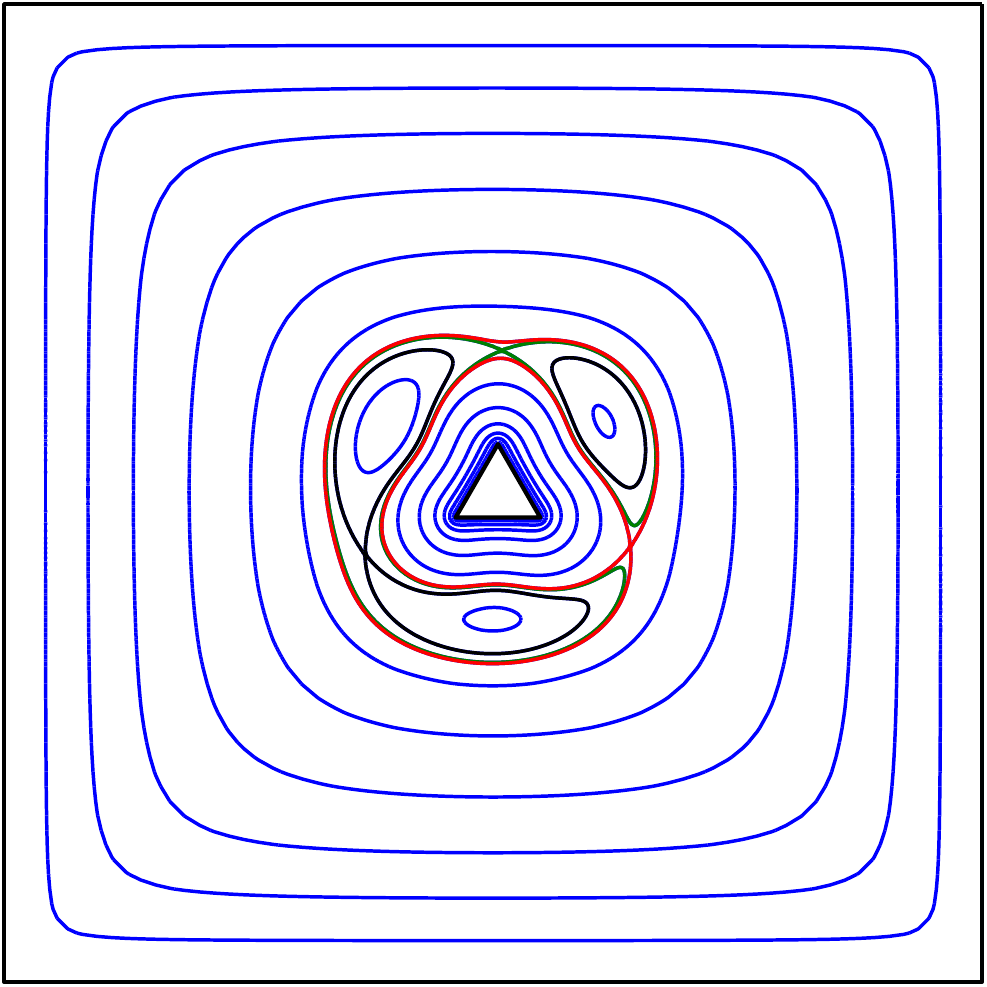}
		\caption{$p=0.01$}
		\label{fig:1t01r}
	\end{subfigure}%
	~ 
	\begin{subfigure}[b]{0.32\textwidth}
		\centering
		\includegraphics[width=\linewidth,trim=0 0 0 0,clip]{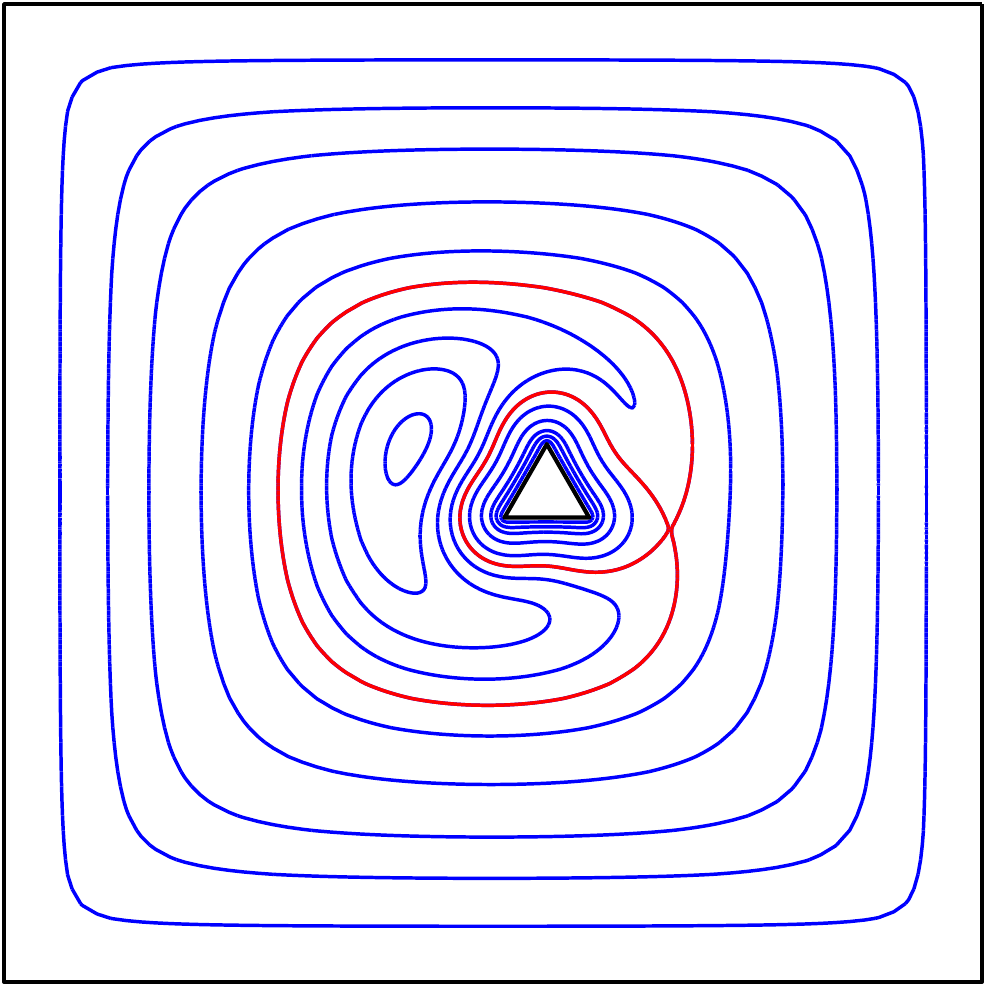}
		\caption{$p=0.11$}
		\label{fig:1t11r}
	\end{subfigure}%
	~
	\begin{subfigure}[b]{0.32\textwidth}
		\centering
		\includegraphics[width=\linewidth,trim=0 0 0 0,clip]{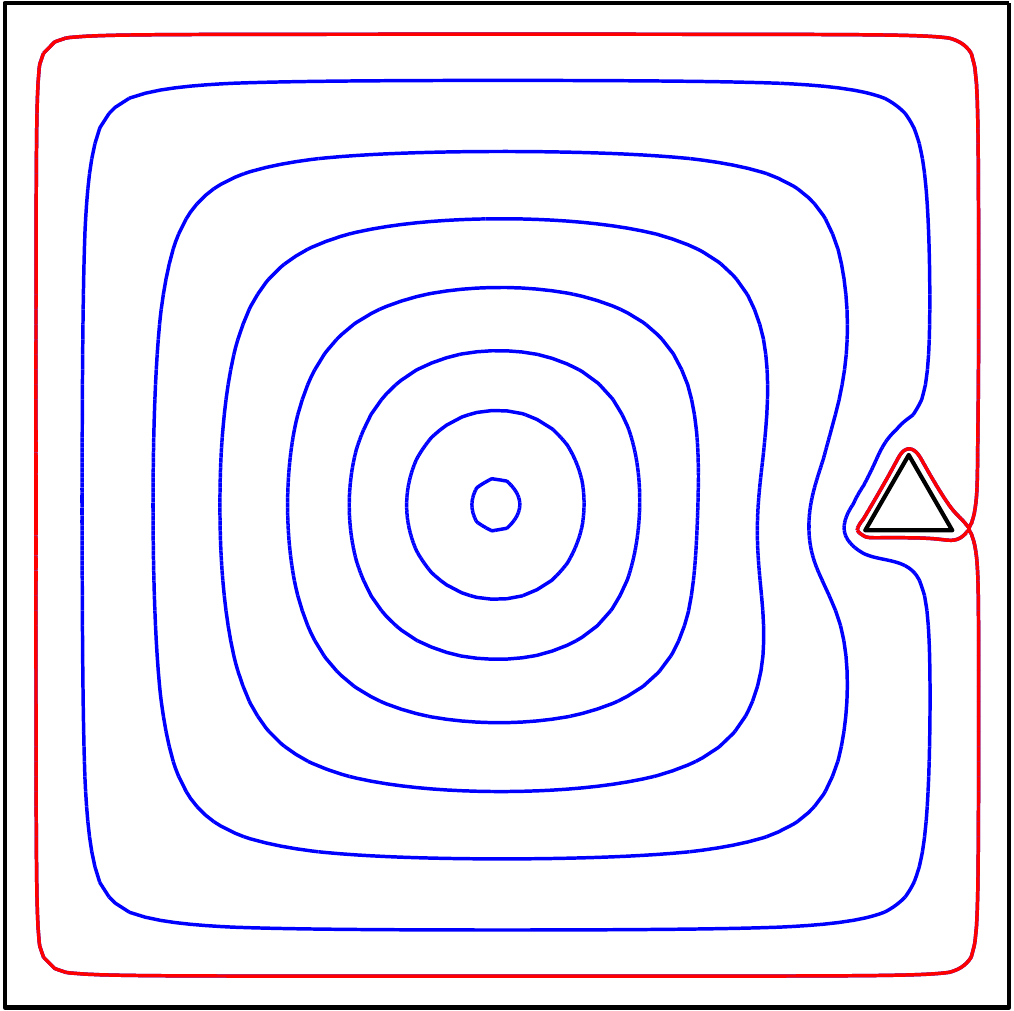}
		\caption{$p=0.8$}
		\label{fig:1t8r}
	\end{subfigure}%	
	\vskip\baselineskip
	\begin{subfigure}[b]{0.32\textwidth}
		\centering
		\includegraphics[width=\linewidth,trim=0 0 0 0,clip]{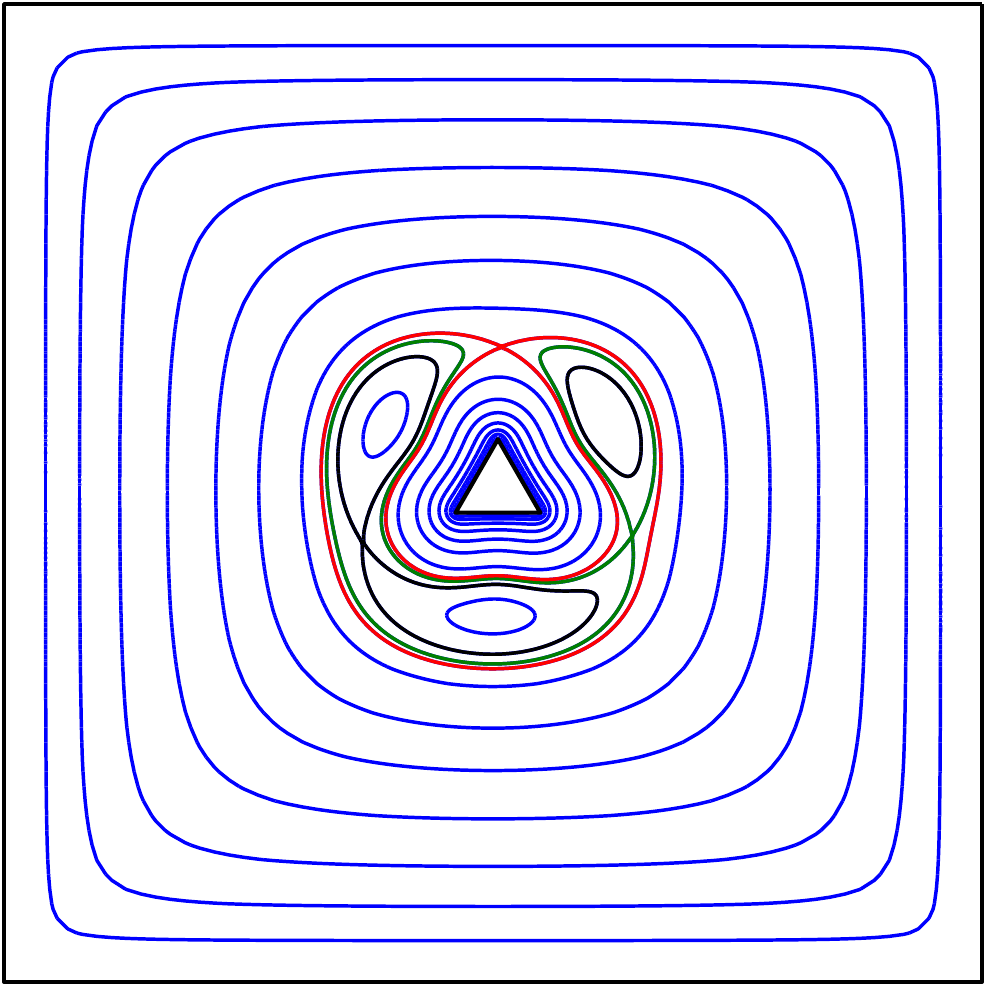}
		\caption{$p=0.01+0.01\i$}
		\label{fig:1t01ur}
	\end{subfigure}%
	~
	\begin{subfigure}[b]{0.32\textwidth}
		\centering
		\includegraphics[width=\linewidth,trim=0 0 0 0,clip]{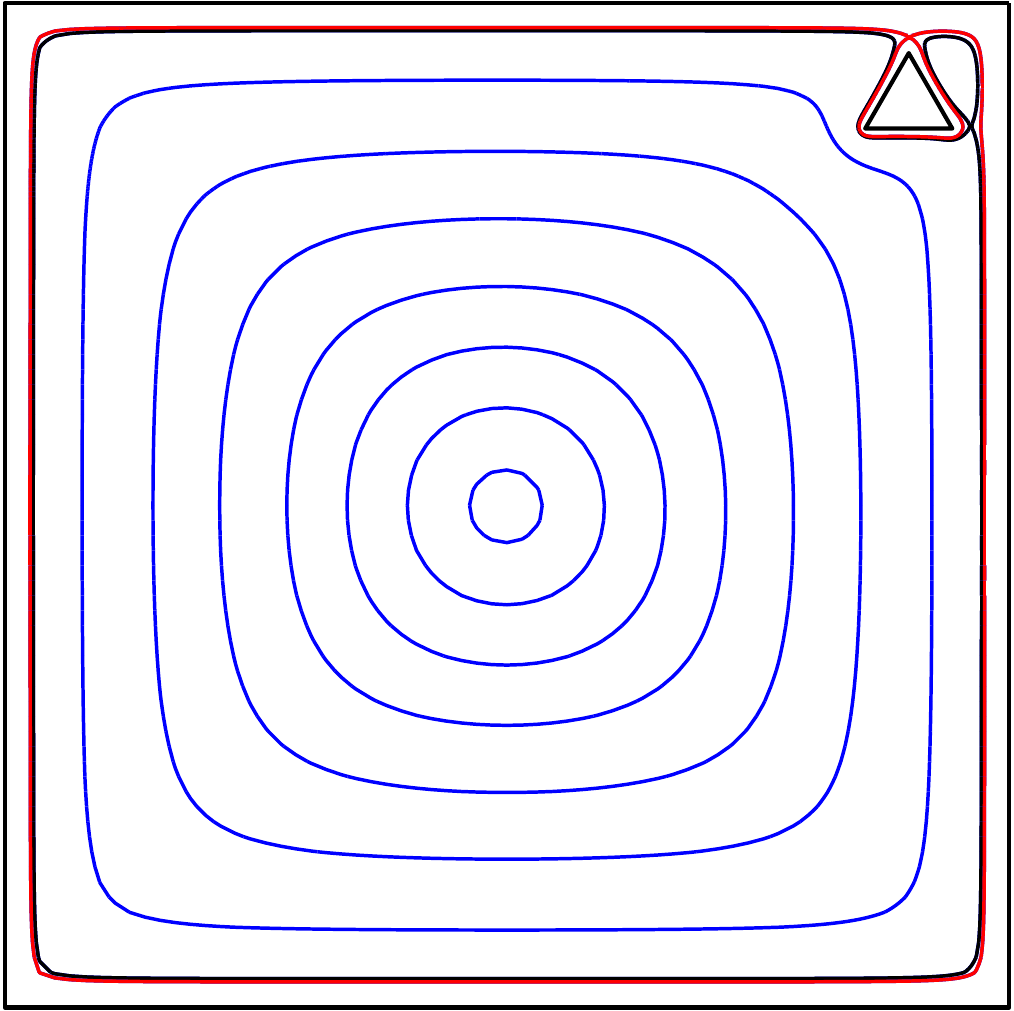}
		\caption{$p=0.8+0.8\i$}
		\label{fig:1t8ur}
	\end{subfigure}%
	~ 
	\begin{subfigure}[b]{0.32\textwidth}
		\centering
		\includegraphics[width=\linewidth,trim=0 0 0 0,clip]{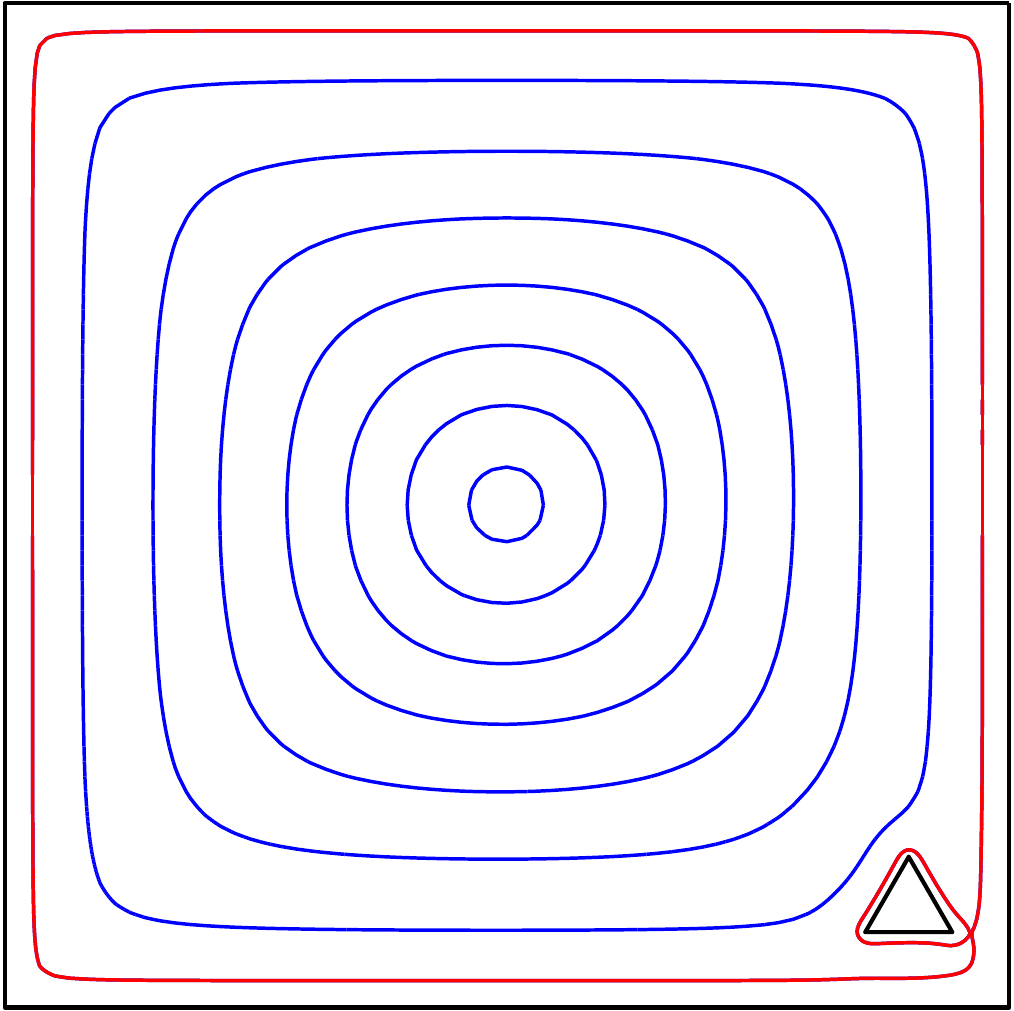}
		\caption{$p=0.8-0.8\i$}
		\label{fig:1t8lr}
	\end{subfigure}%	
	\caption{Vortex trajectories in doubly connected square domains where the vertices of the triangle obstacle are $p+0.1e^{\pi\i/2},p+0.1e^{-\pi\i/6},p+0.1e^{-5\pi\i/6}$.}
	\label{fig:1t}
	%\scalebox{0.3}{\includegraphics[trim=0 0 0 0,clip]{Fig_1s_p11}}
	%\hfill	
\end{figure} 

In Figure~\ref{fig:1t}, we show the effect of obstacle displacement on vortex motion for the case of the equilateral triangle. 
We note first that the imaginary axis is the unique line of reflectional symmetry in the starting configuration of this example; that is when the obstacle center is at $p=0$; see Figure~\ref{fig:1t0}.
This configuration seems structurally more stable to obstacle movement to the upper side external boundary. Indeed, a small perturbation by a $0.01$ unit does not change at all the topology pattern of vortex trajectories. 

\begin{figure}[!htb]  %
	\centering
	\begin{subfigure}[b]{0.16\textwidth}
		\centering
		\includegraphics[width=\linewidth,trim=0 0 0 0,clip]{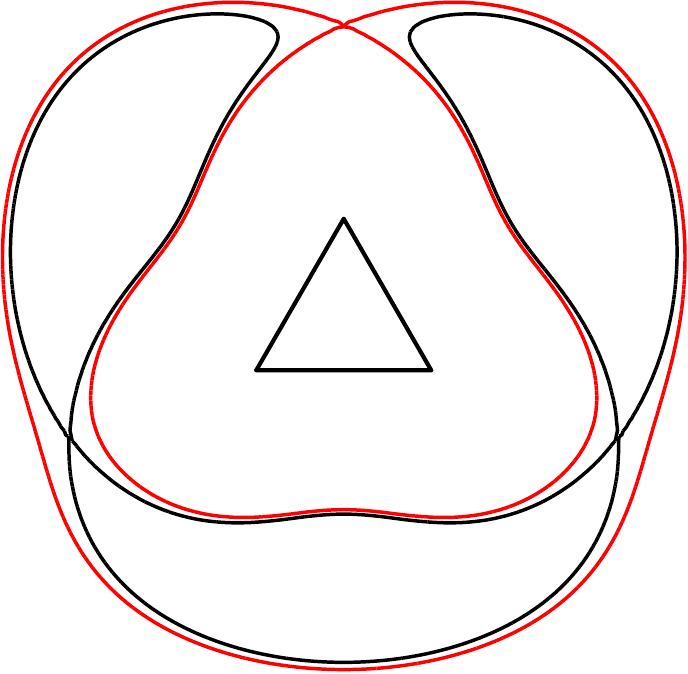}
		\caption{$p=0$}
		\label{fig:c1t0}
	\end{subfigure}%
	~
	\begin{subfigure}[b]{0.16\textwidth}
		\centering
		\includegraphics[width=\linewidth,trim=0 0 0 0,clip]{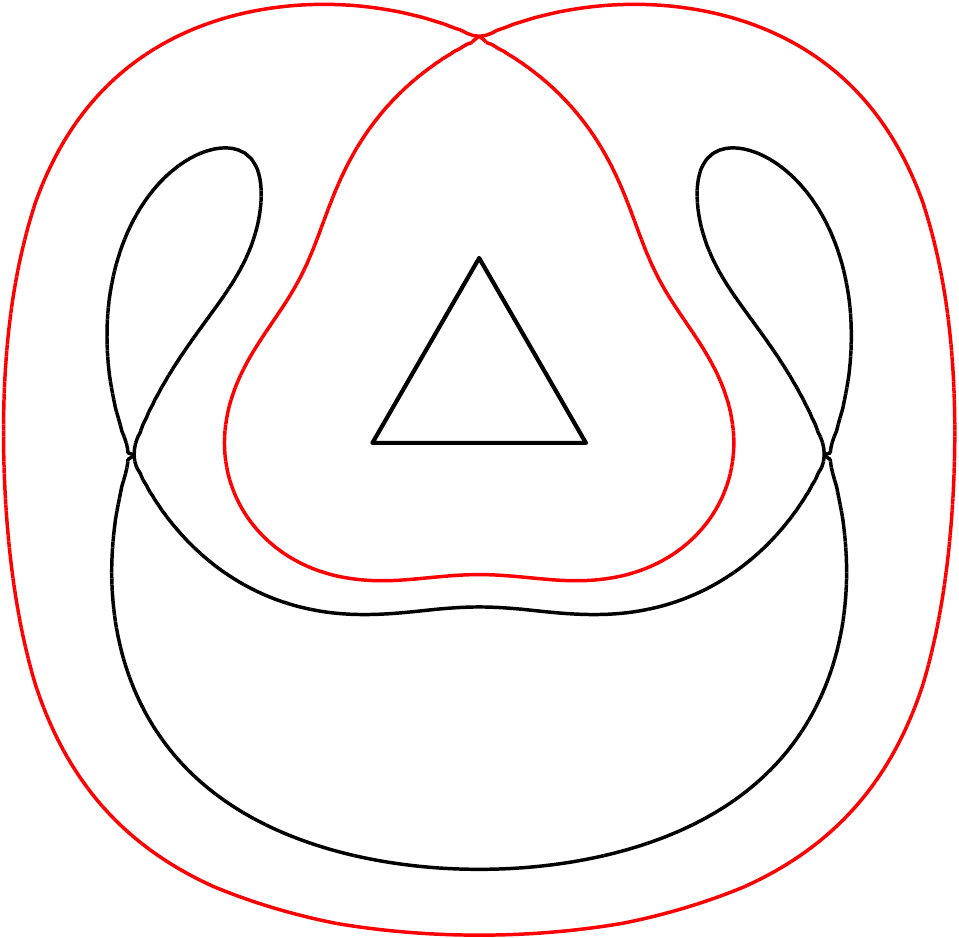}
		\caption{$p=\beta\i$}
		\label{fig:1t05u}
	\end{subfigure}%
	~ 
	\begin{subfigure}[b]{0.16\textwidth}
		\centering
		\includegraphics[width=\linewidth,trim=0 0 0 0,clip]{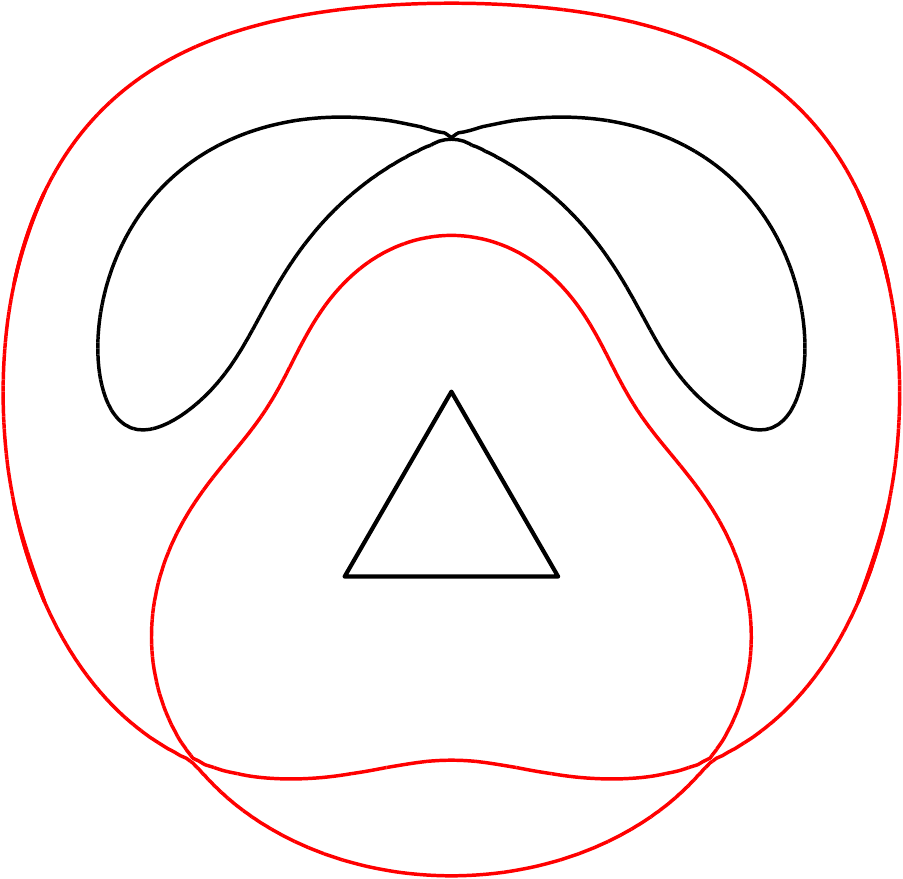}
		\caption{$p=-\beta\i$}
		\label{fig:1t05l}
	\end{subfigure}%
	~
	\begin{subfigure}[b]{0.16\textwidth}
		\centering
		\includegraphics[width=\linewidth,trim=0 0 0 0,clip]{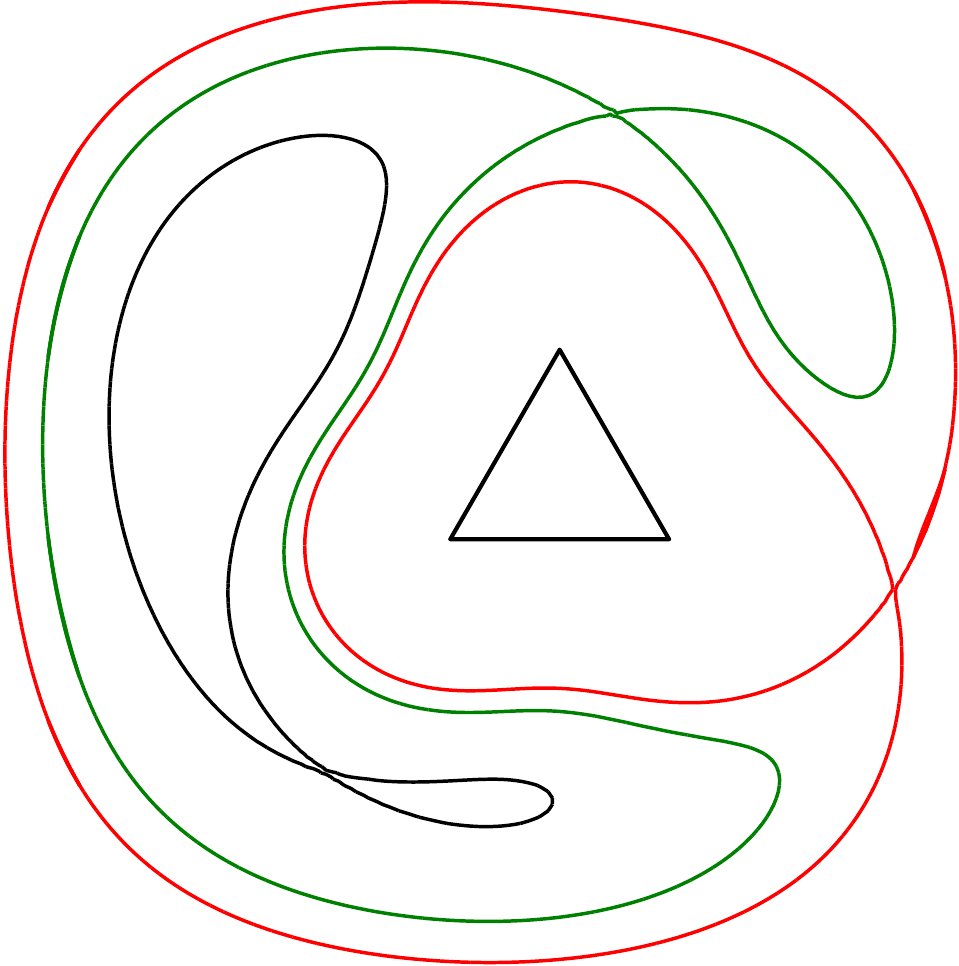}
		\caption{$p=\beta$}
		\label{fig:1t05r}
	\end{subfigure}%
	~
	\begin{subfigure}[b]{0.16\textwidth}
		\centering
		\includegraphics[width=\linewidth,trim=0 0 0 0,clip]{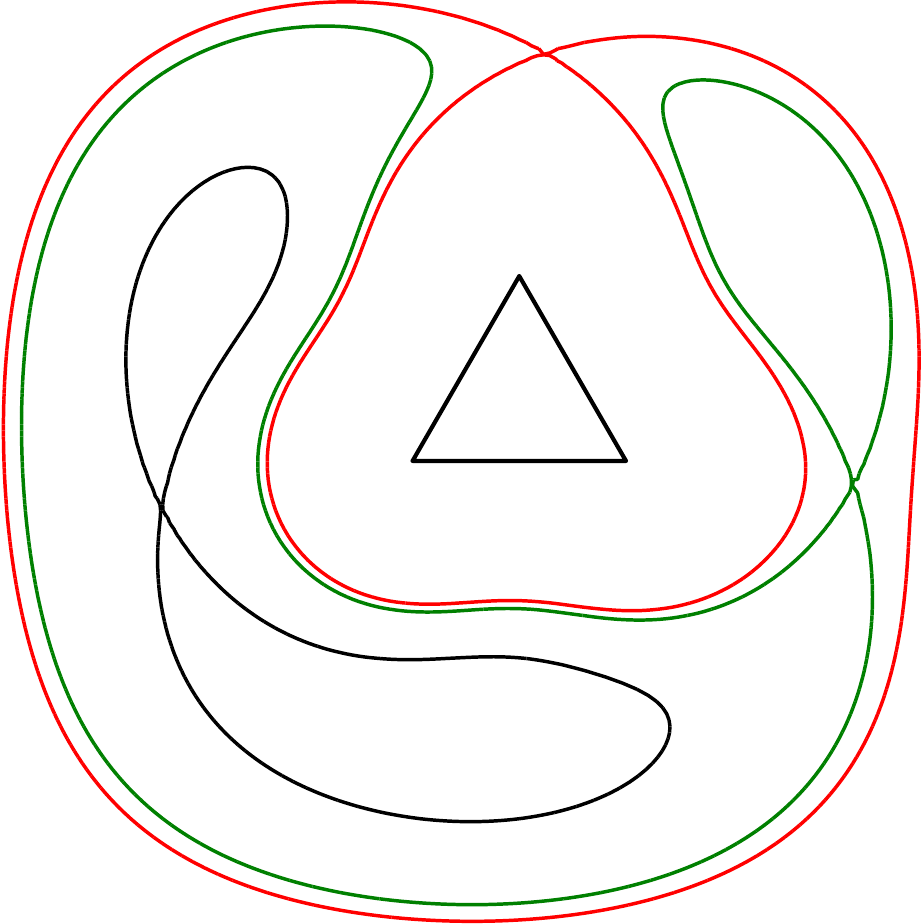}
		\caption{$p=\beta e^\frac{\pi\i}{4}$}
		\label{fig:1t05ur}
	\end{subfigure}%
	~ 
	\begin{subfigure}[b]{0.16\textwidth}
		\centering
		\includegraphics[width=\linewidth,trim=0 0 0 0,clip]{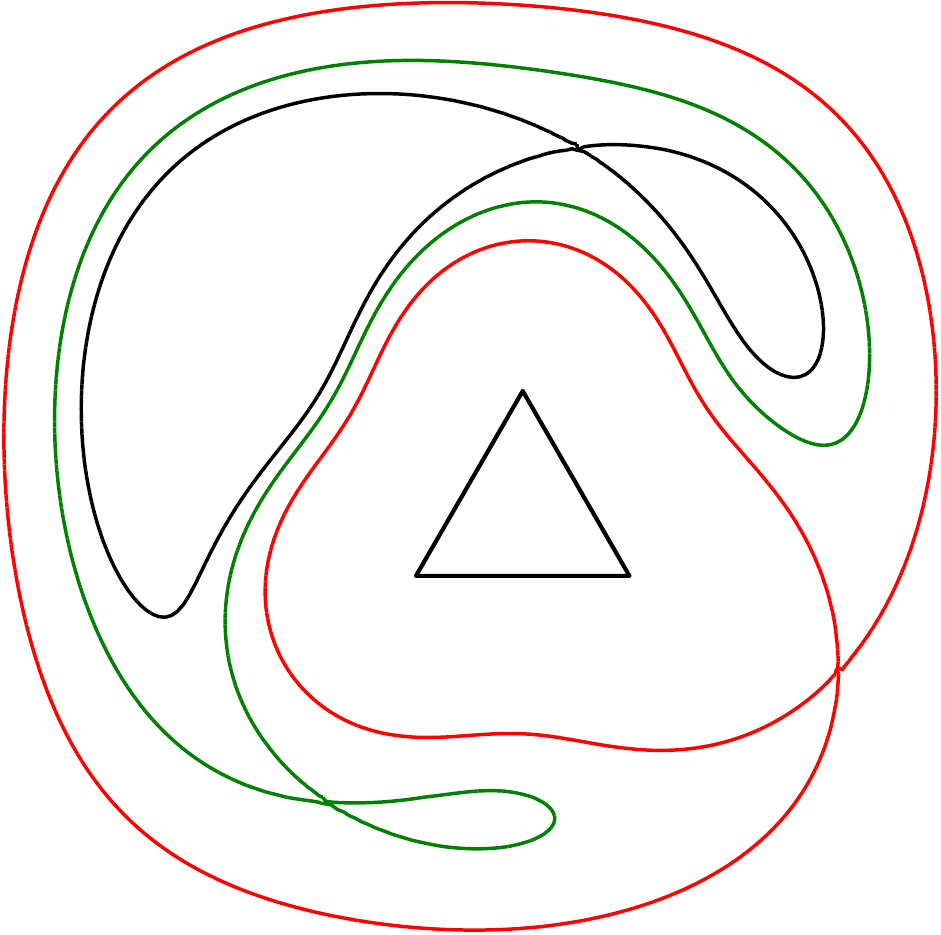}
		\caption{$p=\beta e^{-\frac{\pi\i}{4}}$}
		\label{fig:1t05lr}
	\end{subfigure}%	
	\caption{Critical vortex trajectories after a small displacement ($\beta = 0.05$) of the triangle obstacle in different directions. The vertices of the triangle are $p+0.1e^{\pi\i/2},p+0.1e^{-\pi\i/6},p+0.1e^{-5\pi\i/6}$.}
	\label{fig:1t-small}
\end{figure}

In Figure~\ref{fig:1t-small}, we illustrate how a small perturbation of the triangle obstacle location in different directions impacts the topology pattern of vortex trajectories. We can see that the configuration does not change for a vertical translation of $0.05$ while it does with the same translation on the opposite direction. On the other hand, we observe a change to the same configuration when a small perturbation is applied along the horizontal and diagonal directions; see Figure~\ref{fig:1t05r}-\ref{fig:1t05lr}. 

\begin{figure}[!htb] %
	\begin{subfigure}[b]{0.195\textwidth}
		\centering
		\includegraphics[width=\linewidth,trim=0 0 0 0,clip]{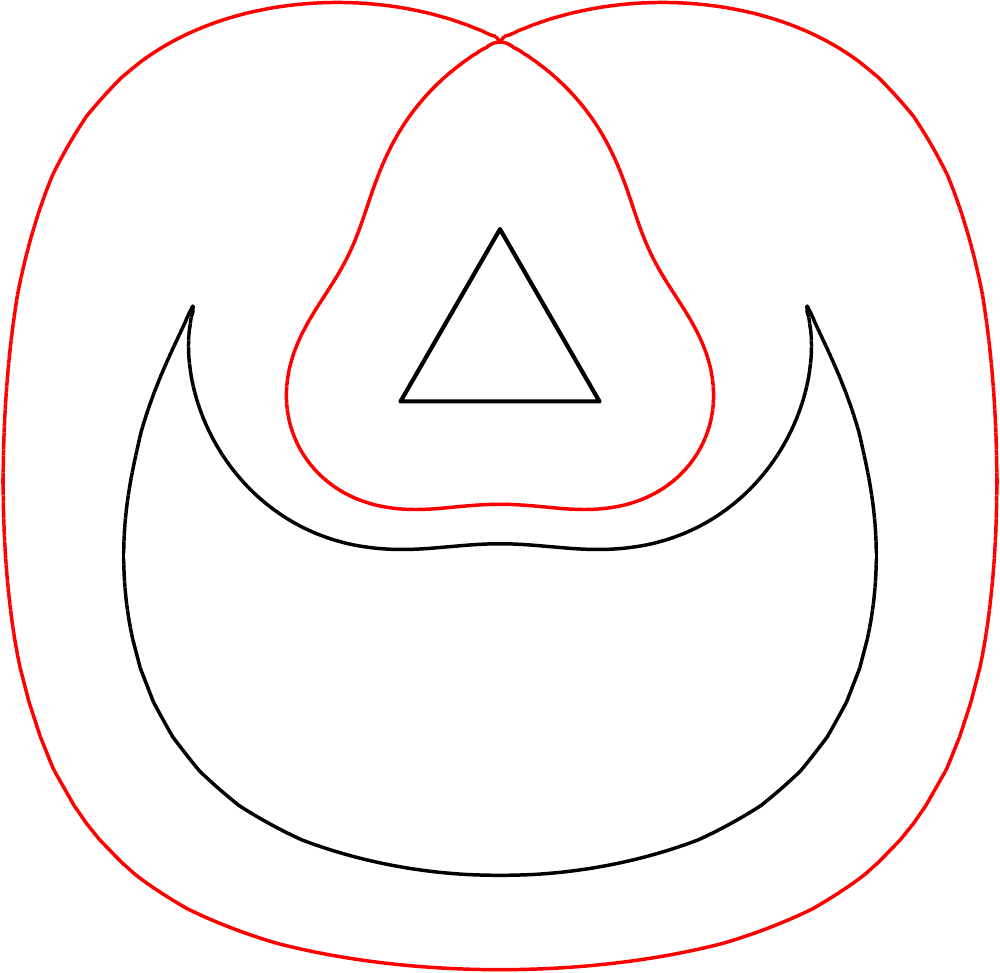}
		\caption{$p=0.102\i$}
		\label{fig:1t102u}
	\end{subfigure}%
	~
	\begin{subfigure}[b]{0.195\textwidth}
		\centering
		\includegraphics[width=\linewidth,trim=0 0 0 0,clip]{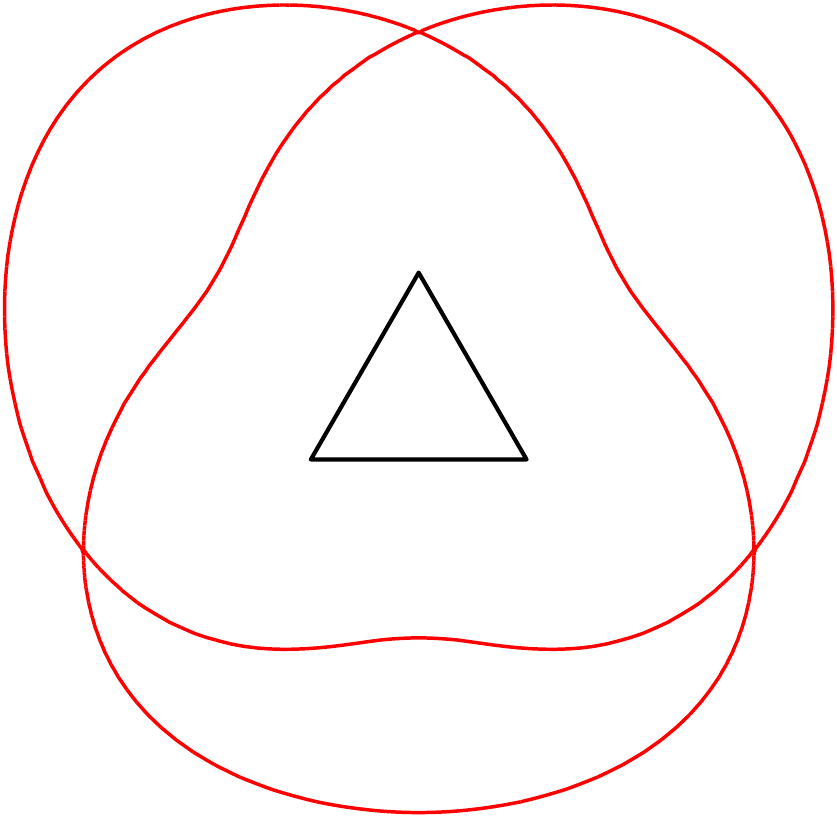}
		\caption{\scalebox{0.9}{$p=-0.004985\i$}}
		\label{fig:1t4985u}
	\end{subfigure}%
	~
	\begin{subfigure}[b]{0.195\textwidth}
		\centering
		\includegraphics[width=\linewidth,trim=0 0 0 0,clip]{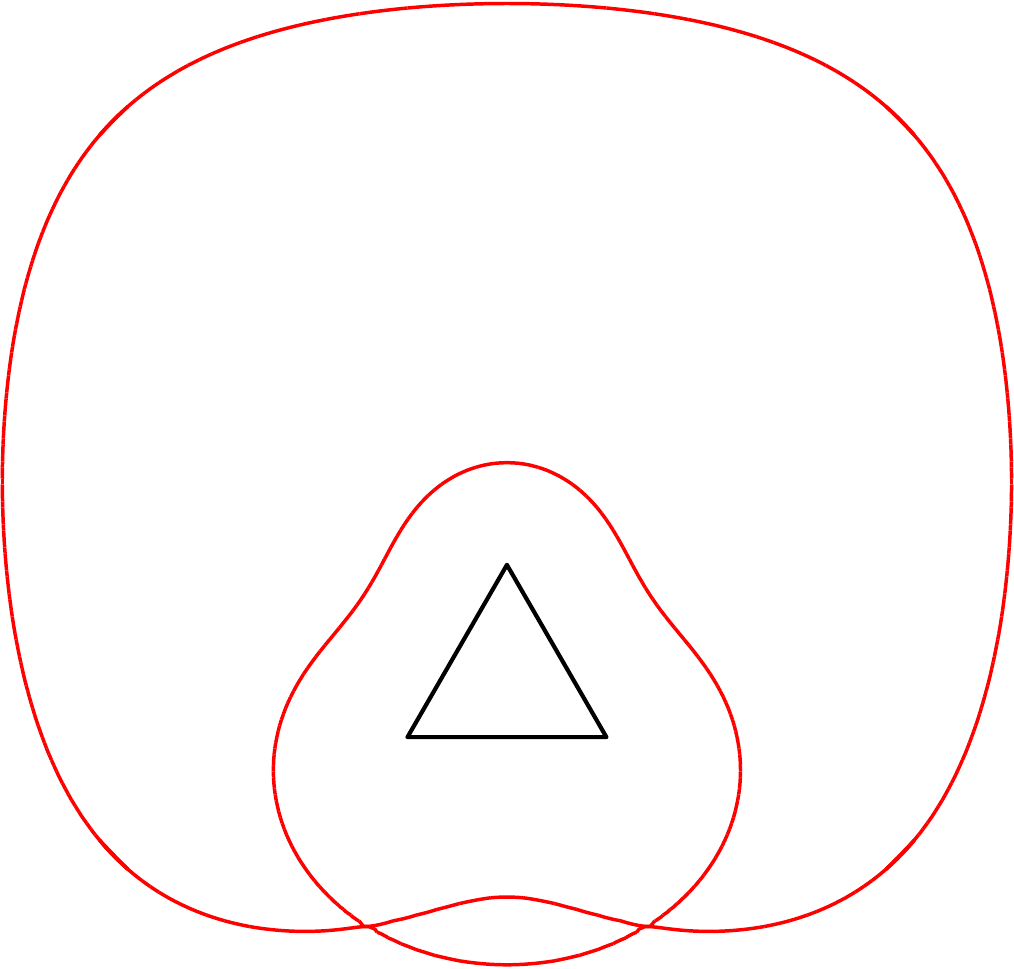}
		\caption{$p=-0.145\i$}
		\label{fig:1t145l}
	\end{subfigure}%
	~
	\begin{subfigure}[b]{0.195\textwidth}
		\centering
		\includegraphics[width=\linewidth,trim=0 0 0 0,clip]{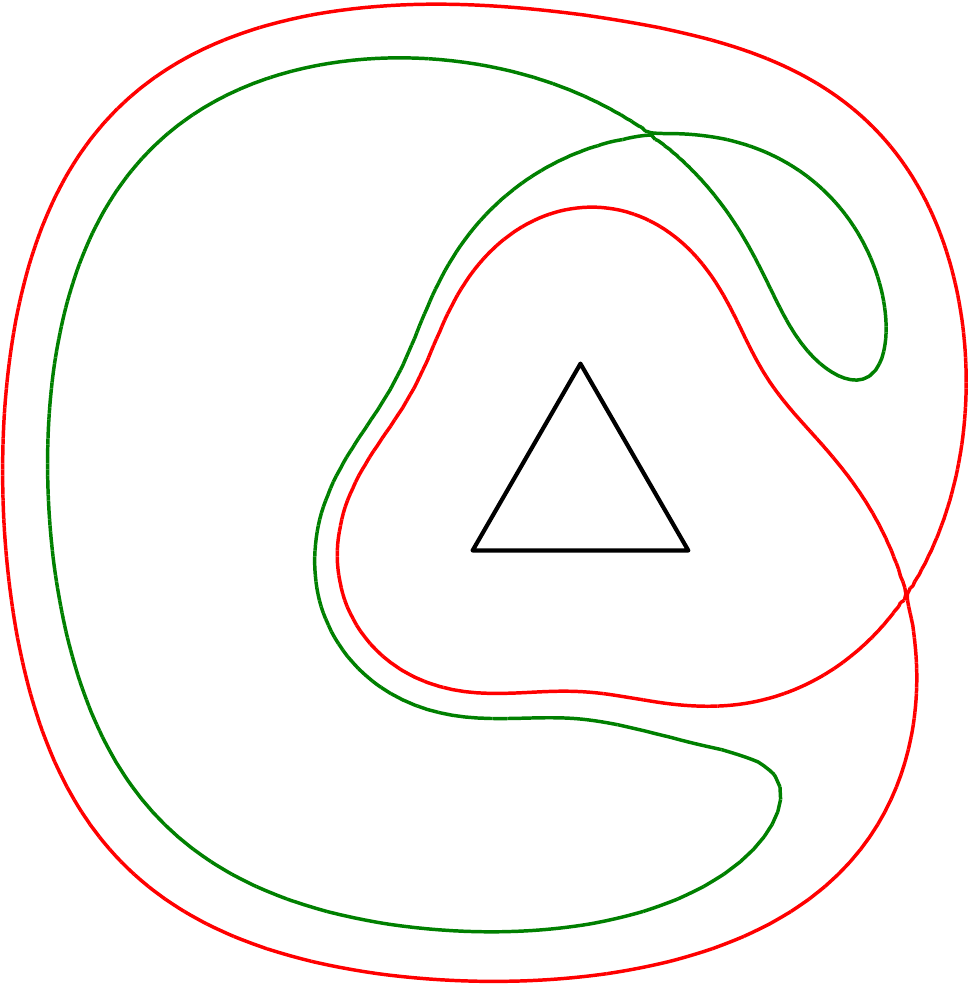}
		\caption{$p=0.063$}
		\label{fig:1t063r}
	\end{subfigure}%	
	~
	\begin{subfigure}[b]{0.195\textwidth}
		\centering
		\includegraphics[width=\linewidth,trim=0 0 0 0,clip]{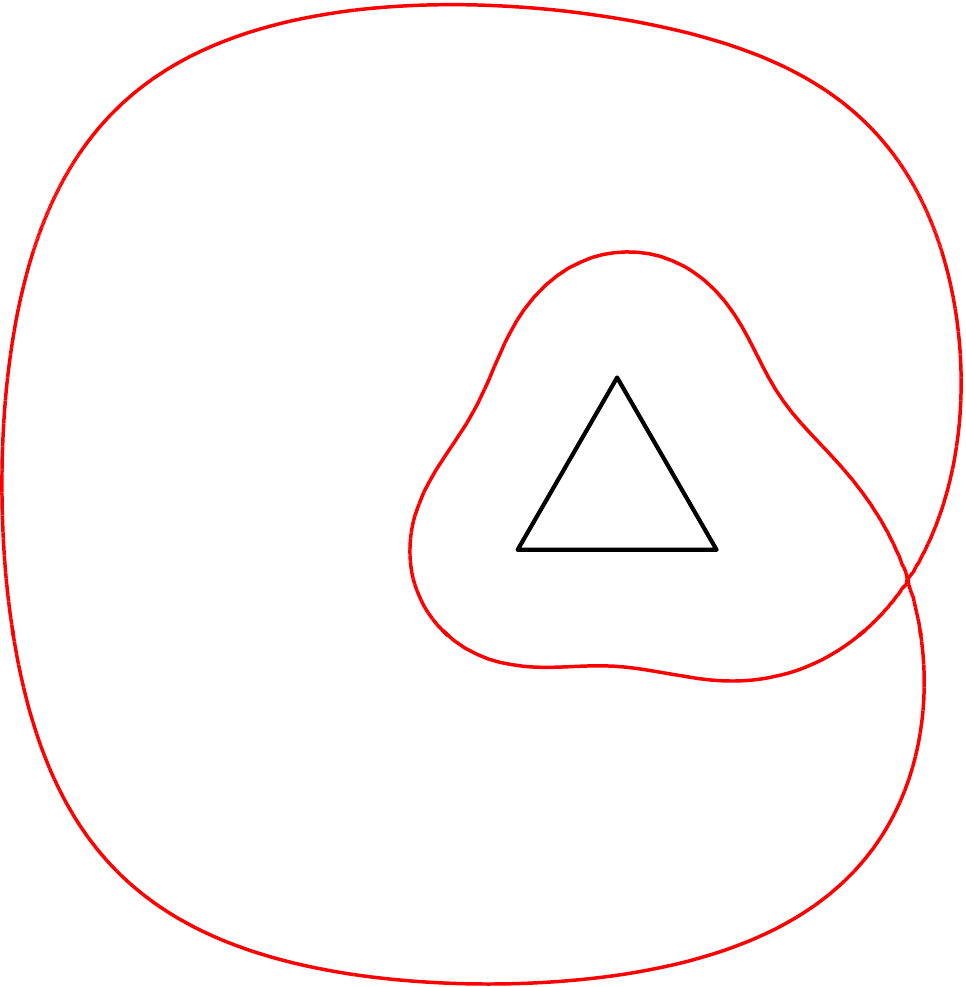}
		\caption{$p=0.102$}
		\label{fig:1t102r}
	\end{subfigure}%		
	\caption{Critical vortex trajectories for bifurcation locations of the triangle obstacle. The vertices of the triangle are $p+0.1e^{\pi\i/2},p+0.1e^{-\pi\i/6},p+0.1e^{-5\pi\i/6}$.}
	\label{fig:1t-small2}
\end{figure}

Next, we explore closely how the topological pattern of vortex motion changes after displacement of the obstacle triangle. First, for the vertical translation to the upper side boundary, the bifurcation from the two lower saddles and three centers to a unique center point does not happen until an approximate displacement of $0.102$; see Figure~\ref{fig:1t102u}. Afterwards, the same configuration continues to hold when moving the obstacle close to the boundary as shown in Figure~\ref{fig:1t8u}. 

An interesting dynamical feature in this example arises if we move the obstacle on the other side of the imaginary axis. A merging of two pairs of homoclinic orbits and one heteroclinic loop to a new structure composed of three heteroclinic connections and one triheteroclinic loop occurs at a very small bifurcation location of the triangle center equal to $p=-0.004985$; see Figure~\ref{fig:1t4985u}. This configuration changes immediately after this value to the topology pattern shown in Figure~\ref{fig:1t05l}. Indeed, the upper saddle point gets separated from the two lower saddles and becomes an intersection of a pair of homoclinic orbits. The same topological pattern appears also for a displacement of $0.11$ unit in Figure~\ref{fig:1t11l}. The upper saddle changes to a center point for an approximate bifurcation displacement of $0.145$ yielding the pattern given in Figure \ref{fig:1t145l}. This configuration stays the same all the way as the obstacle approaches the lower boundary; see Figure~\ref{fig:1t8l}.  

Now, when we break the line symmetry with respect to the imaginary axis by moving the obstacle triangle with $0.01$ unit eastward in Figure~\ref{fig:1t01r}, the heteroclinic loop disappears as each of the three saddles becomes an intersection point of homoclinic orbits. The inner saddle point disappears first at an approximate bifurcation displacement of $0.063$ (Figure~\ref{fig:1t063r}), and then the other inner saddle changes to a center point at almost $p=0.102$ (Figure~\ref{fig:1t102r}). Afterwards, the vortex motion maintains a unique unstable rest point until the obstacle comes close the right boundary; see Figure~\ref{fig:1t11r}-\ref{fig:1t8r}. 

Note also that as we said before, a small perturbation along the line $y=x$ yields a topologically equivalent pattern to the horizontal translation case. However, when the obstacle triangle approaches the upper right corner of the outer square, we get a different configuration to the horizontal movement case as the vortex motion keeps two unstable equilibria; see Figures~\ref{fig:1t01ur}-\ref{fig:1t8ur}. It is worth mentioning that this does not happen when the obstacle comes close to the lower right corner. The configuration in this case is topologically equivalent to the horizontal translation as shown in Figure~\ref{fig:1t8lr}.

\subsection{A Hexagon Obstacle}

\begin{figure}[!htb] %
	\centering
	\begin{subfigure}[b]{0.32\textwidth}
		\centering
		\includegraphics[width=\linewidth,trim=0 0 0 0,clip]{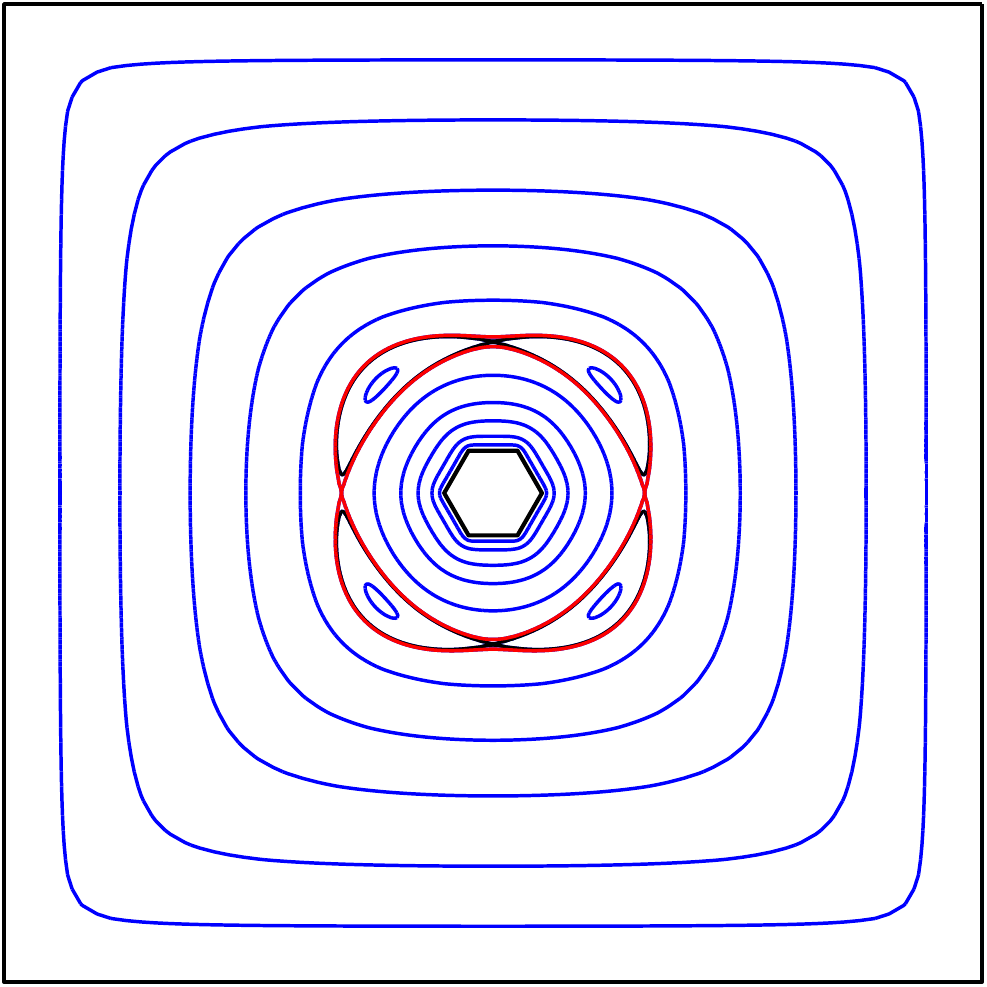}
		\caption{$p=0$}
		\label{fig:1h0}
	\end{subfigure}%
	\vskip\baselineskip
	\begin{subfigure}[b]{0.32\textwidth}
		\centering
		\includegraphics[width=\linewidth,trim=0 0 0 0,clip]{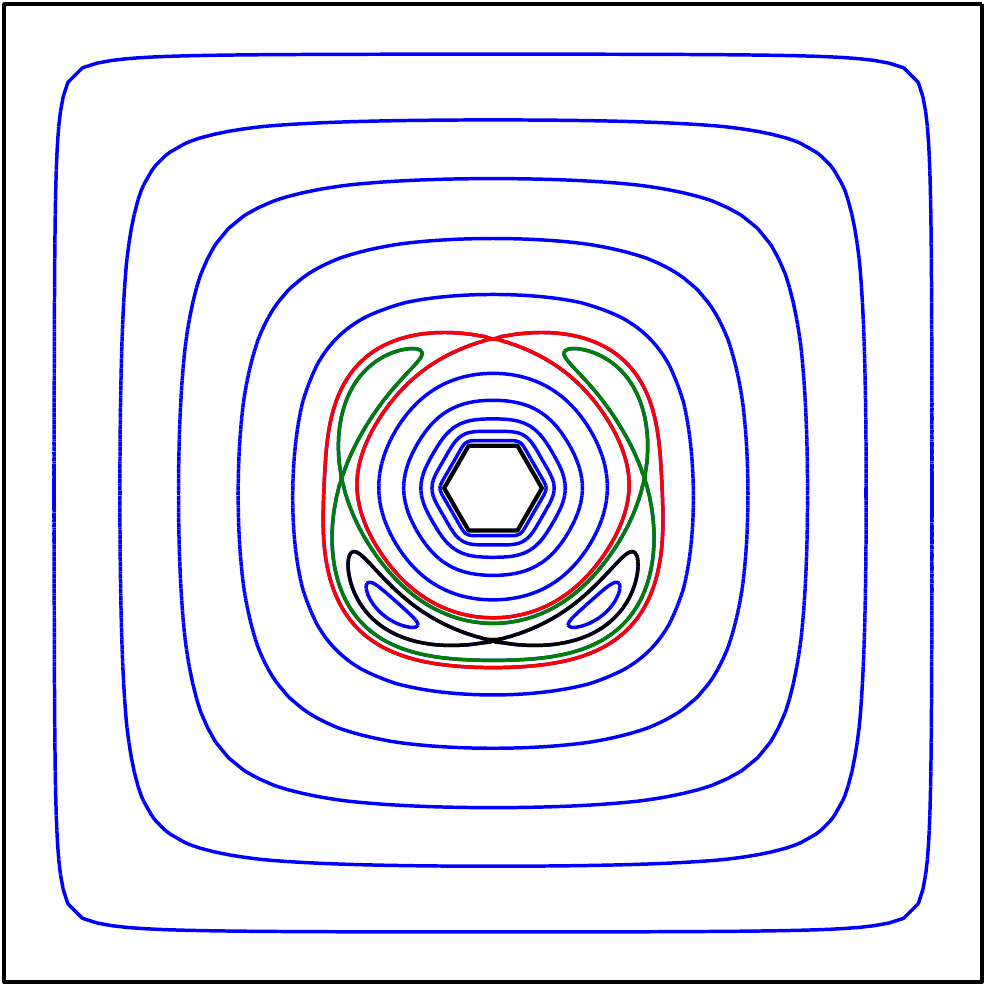}
		\caption{$p=0.01\i$}
		\label{fig:1h01}
	\end{subfigure}%
	~ 
	\begin{subfigure}[b]{0.32\textwidth}
		\centering
		\includegraphics[width=\linewidth,trim=0 0 0 0,clip]{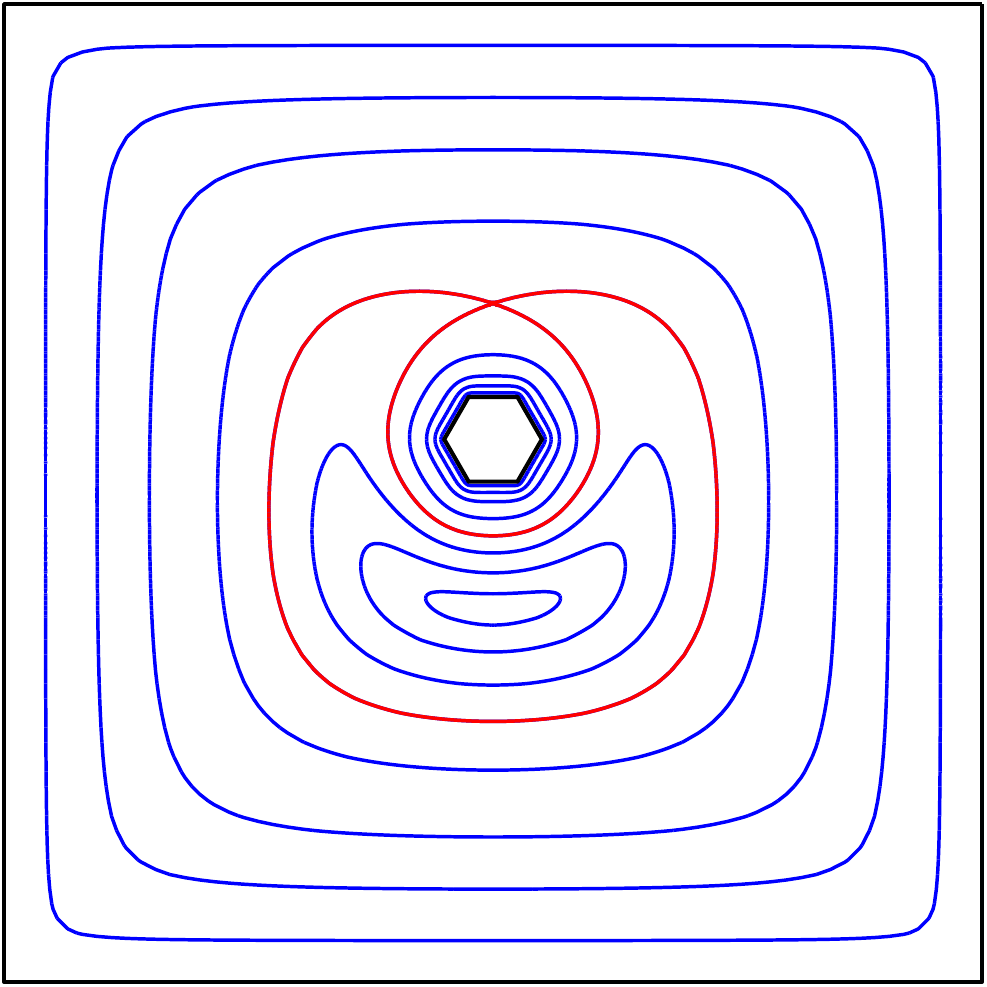}
		\caption{$p=0.11\i$}
		\label{fig:1h11}
	\end{subfigure}%
	~
	\begin{subfigure}[b]{0.32\textwidth}
		\centering
		\includegraphics[width=\linewidth,trim=0 0 0 0,clip]{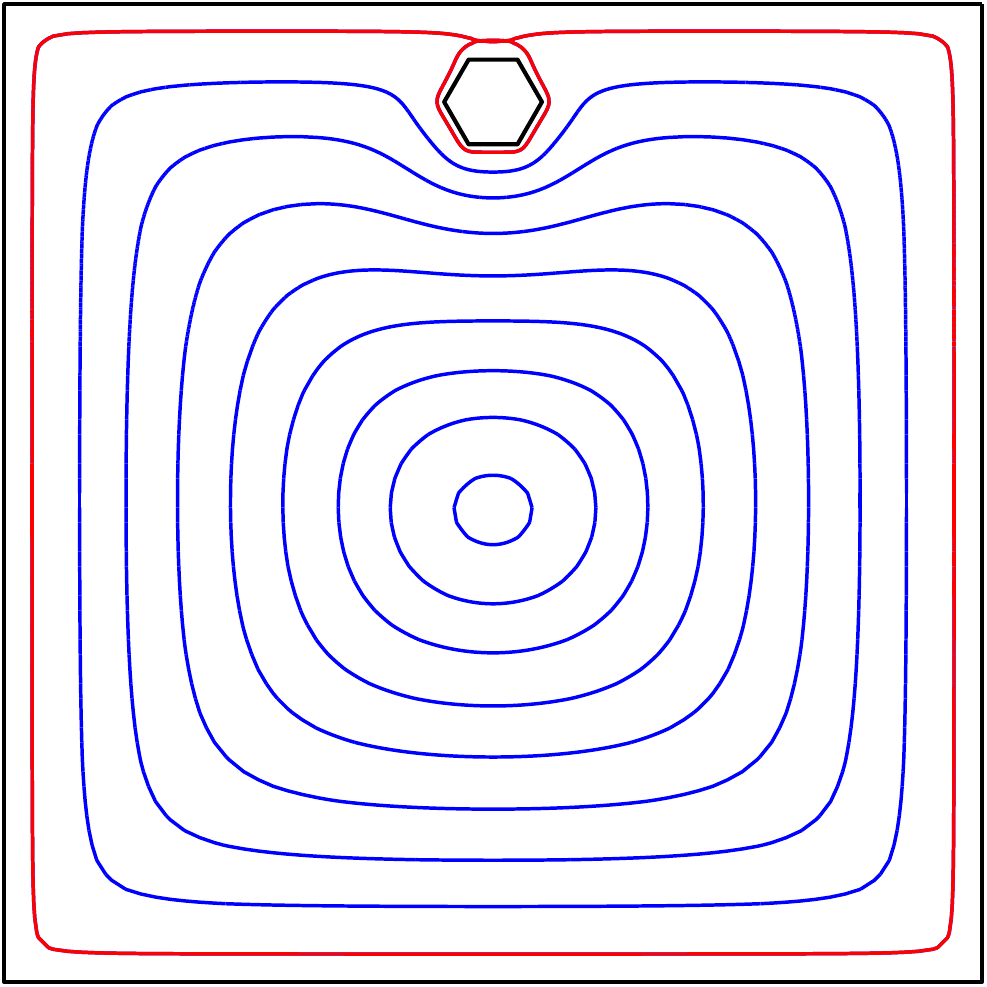}
		\caption{$p=0.8\i$}
		\label{fig:1h8}
	\end{subfigure}%
	\vskip\baselineskip
	\begin{subfigure}[b]{0.32\textwidth}
		\centering
		\includegraphics[width=\linewidth,trim=0 0 0 0,clip]{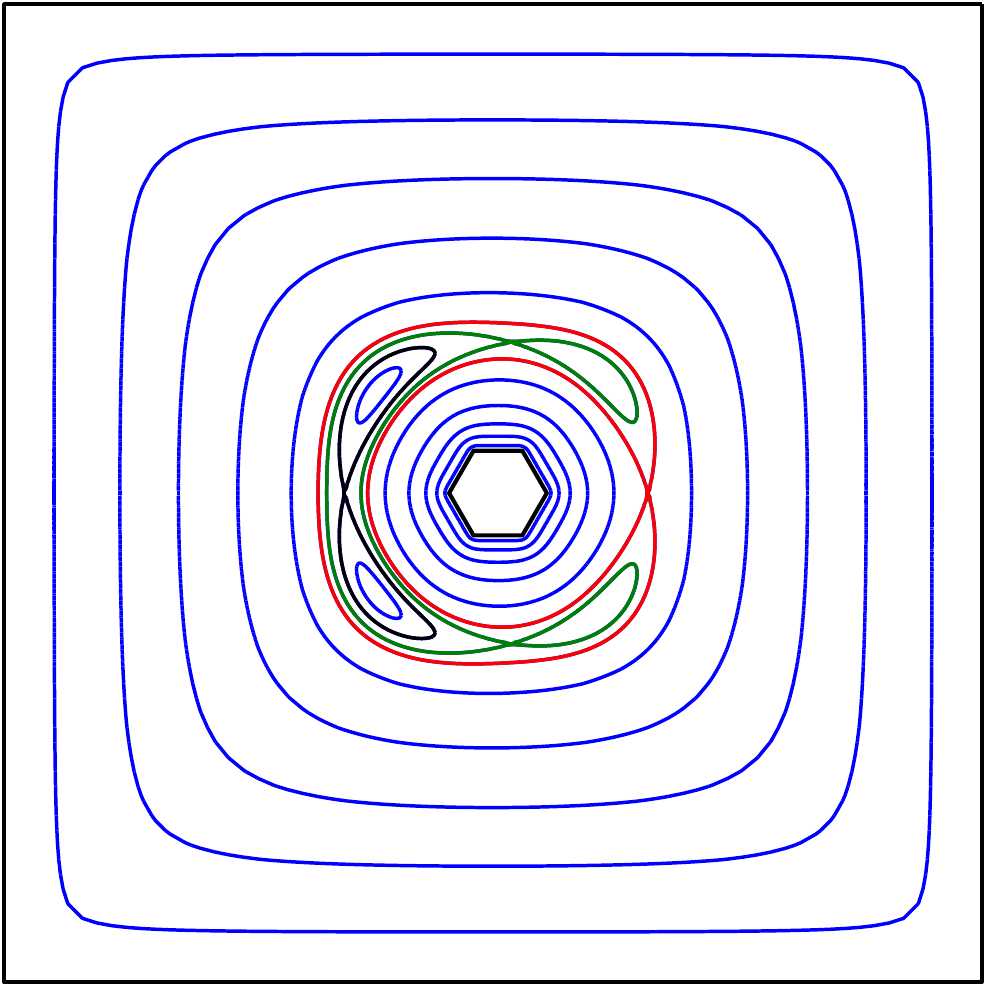}
		\caption{$p=0.01$}
		\label{fig:1h01r}
	\end{subfigure}%
	~ 
	\begin{subfigure}[b]{0.32\textwidth}
		\centering
		\includegraphics[width=\linewidth,trim=0 0 0 0,clip]{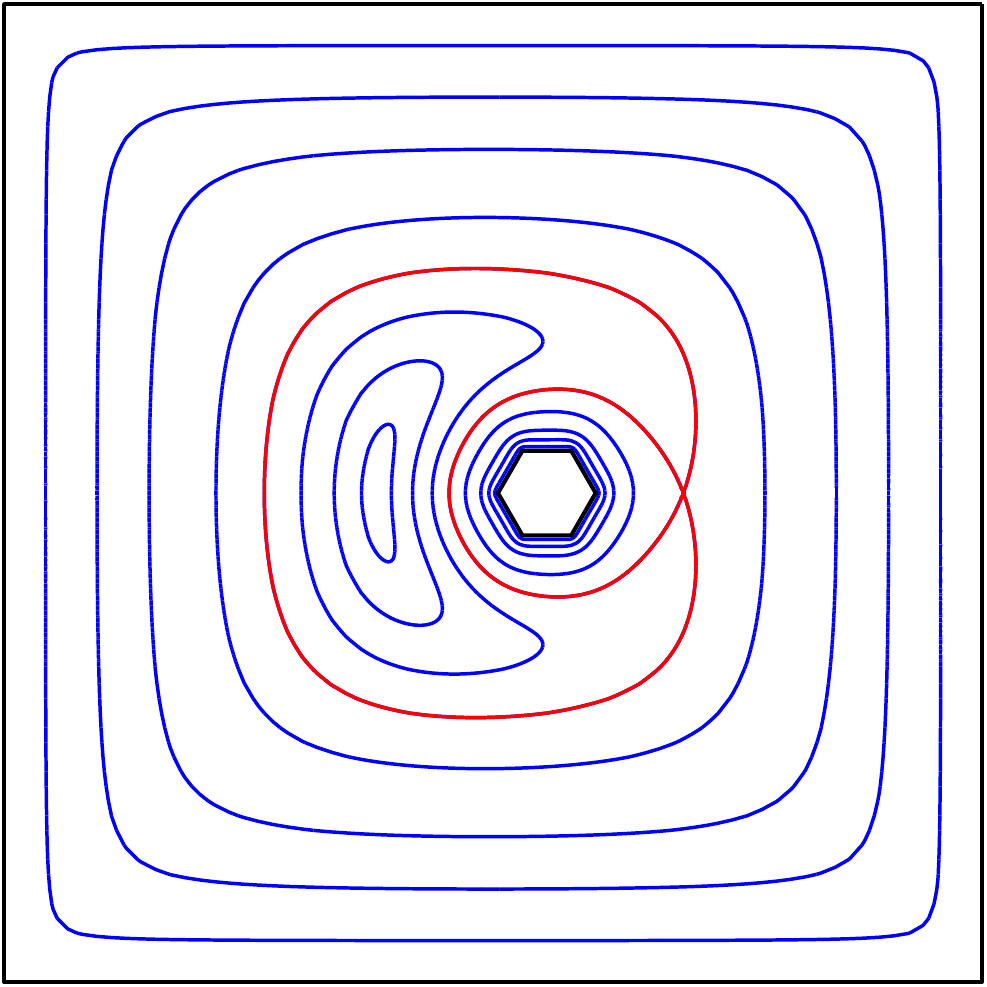}
		\caption{$p=0.11$}
		\label{fig:1h11r}
	\end{subfigure}%
	~ 
	\begin{subfigure}[b]{0.32\textwidth}
		\centering
		\includegraphics[width=\linewidth,trim=0 0 0 0,clip]{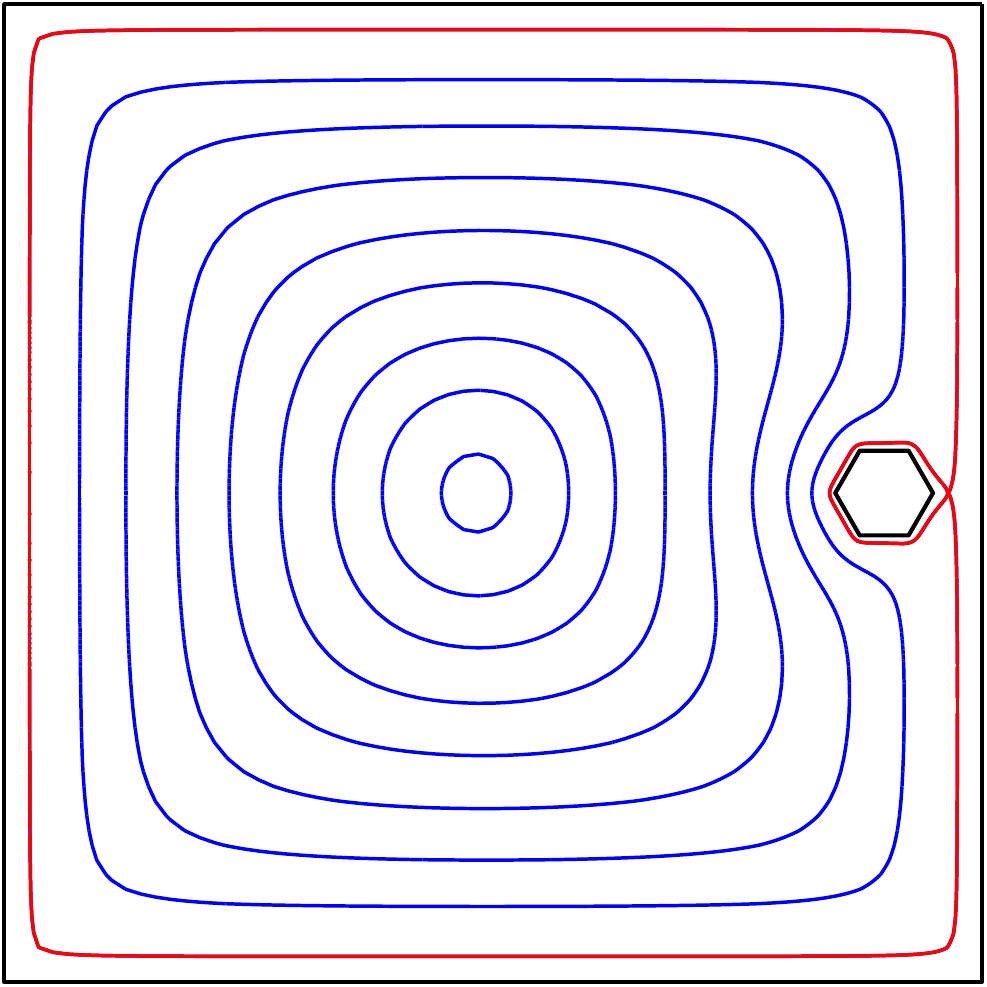}
		\caption{$p=0.8$}
		\label{fig:1h8r}
	\end{subfigure}%	
	\vskip\baselineskip
	\begin{subfigure}[b]{0.32\textwidth}
		\centering
		\includegraphics[width=\linewidth,trim=0 0 0 0,clip]{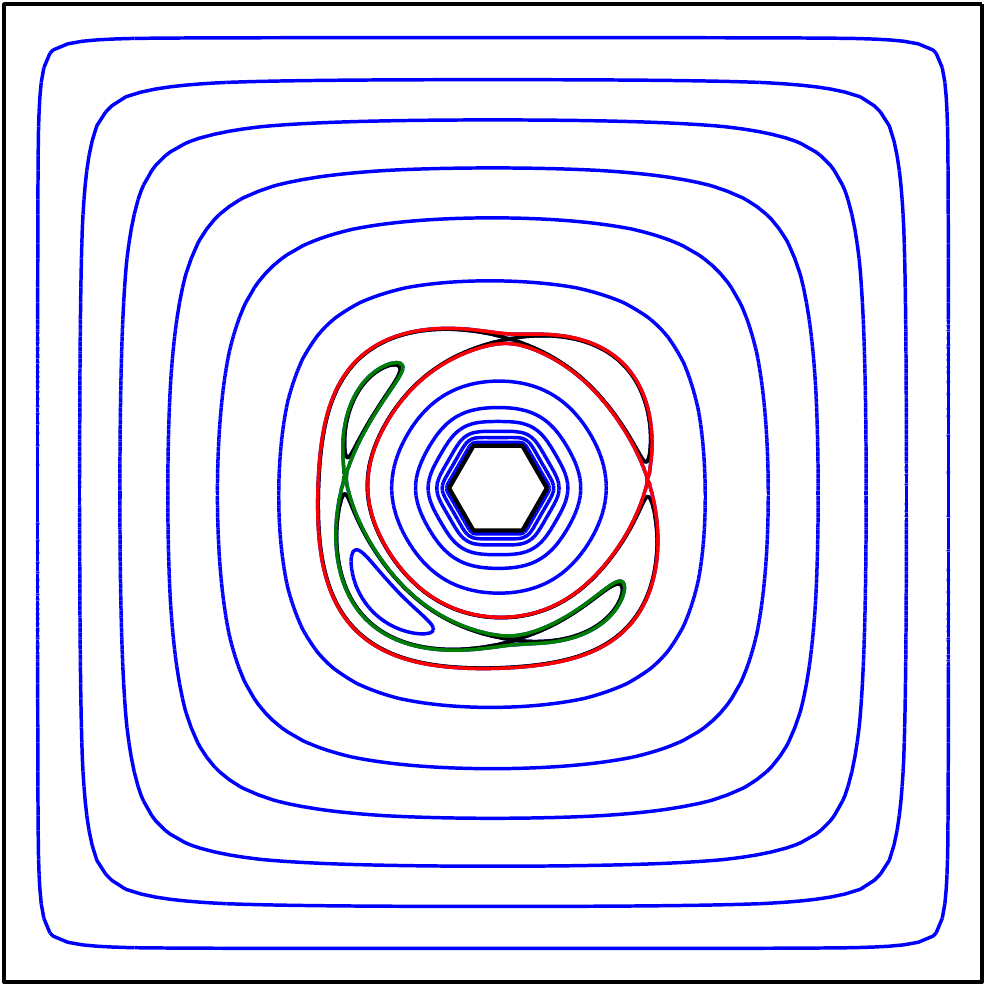}
		\caption{$p=0.01+0.01\i$}
		\label{fig:1h01ur}
	\end{subfigure}%
	~ 
	\begin{subfigure}[b]{0.32\textwidth}
		\centering
		\includegraphics[width=\linewidth,trim=0 0 0 0,clip]{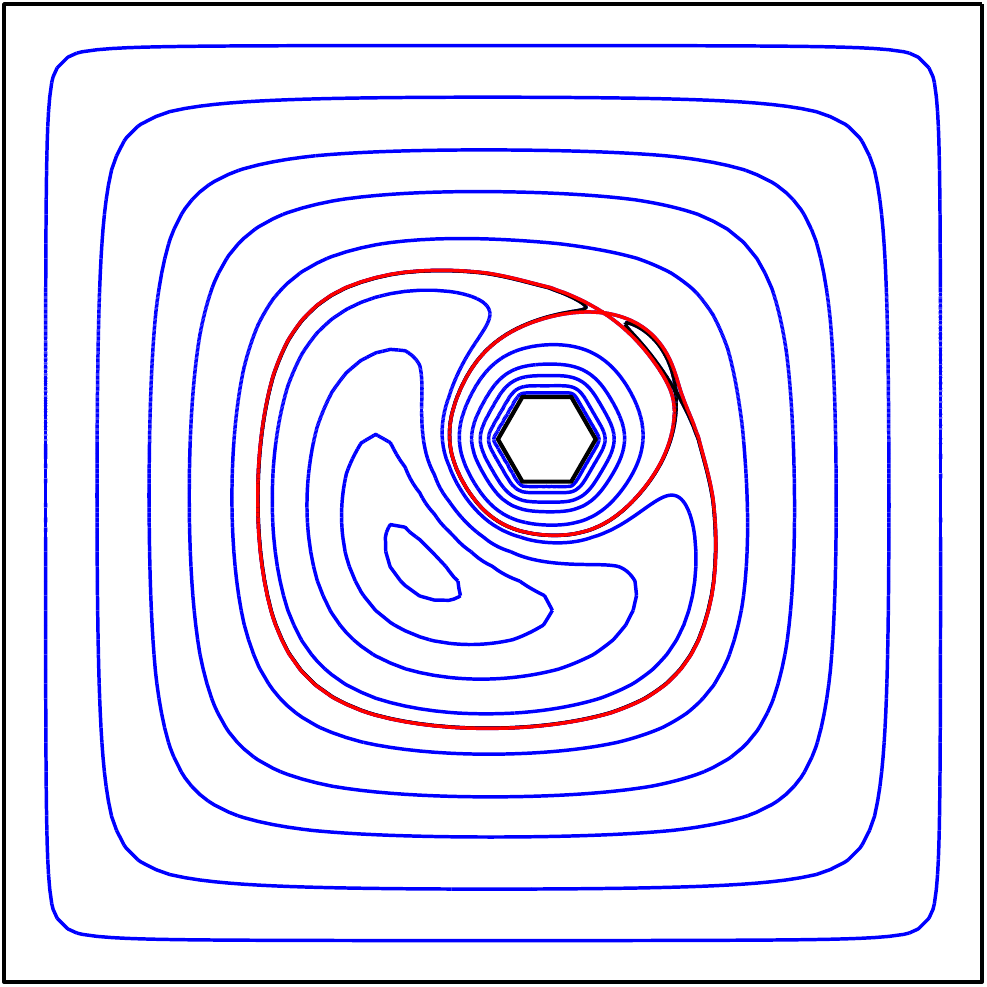}
		\caption{$p=0.11+0.11\i$}
		\label{fig:1h11ur}
	\end{subfigure}%
	~
	\begin{subfigure}[b]{0.32\textwidth}
		\centering
		\includegraphics[width=\linewidth,trim=0 0 0 0,clip]{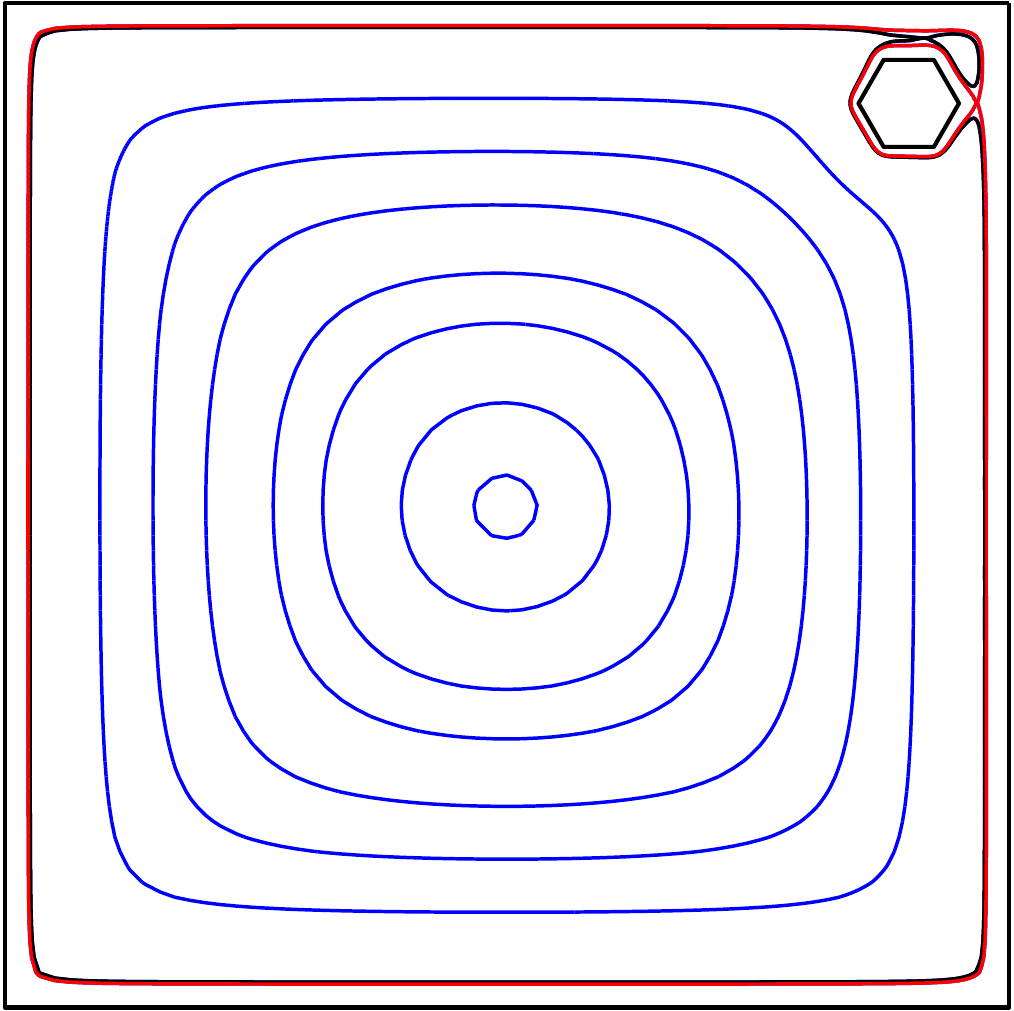}
		\caption{$p=0.8+0.8\i$}
		\label{fig:1h8ur}
	\end{subfigure}%	
	\caption{Vortex trajectories in doubly connected polygonal domains where the vertices of the hexagonal obstacle are $p+0.1,p+0.1e^{-\pi\i/3},p+0.1e^{-2\pi\i/3},p-0.1,p+0.1e^{-4\pi\i/3},p+0.1e^{-5\pi\i/3}$.}
	\label{fig:1h}
	%\scalebox{0.3}{\includegraphics[trim=0 0 0 0,clip]{Fig_1s_p11}}
	%\hfill	
\end{figure}   

In this example we consider how the displacement of the hexagon obstacle in Figure~\ref{fig:1h} impacts the point vortex motion. 

First, the upper heteroclinic loop connecting the two saddles on the real axis gets splitted into two homoclinic orbits after a small perturbation of the  center $p$ of the hexagon by $0.01$ unit along the imaginary axis; see Figure~\ref{fig:1h01}.  The upper saddle on the imaginary axis becomes unique after the three other points disappear at a vertical displacement of $0.11$ unit. When approaching the upper boundary side, this saddle changes to a center point of a heteroclinic loop linking two new saddles.  

When we displace the center of the triangle by $0.01$ unit eastward in Figure~\ref{fig:1h01r}, a topologically equivalent pattern to the $(6,1)$ homoclinic-heteroclinic structure we got after a small vertical perturbation of $p$ appears again. The only difference in the horizontal displacement is that the pattern appearing at $p=0.11$ stays the same even after approaching the right side boundary. 

On the other hand, a diagonal displacement of the hexagon obstacle changes the dynamics of the point vortex in a different manner. A small perturbation as in Figure~\ref{fig:1h01ur} destroys the heteroclinic structure  completely, and all four saddles become intersection points of homoclinic orbits. At $p=0.11+0.11\i$, the pair of saddle points on the side of the obstacle movement are maintained with the same homoclinic structure, but the other pair of saddles and pair of centers disappear except one vortex center on the main diagonal line. This pattern remains topologically invariant as the obstacle approaches the upper right corner.

\section{Multiply Connected Domains}
\label{sc:mul}
%-------------------------------------------------------------------

\begin{figure}[!htb] %
	\centering
	\begin{subfigure}[b]{0.24\textwidth}
		\centering
		\includegraphics[width=\linewidth,trim=0 0 0 0,clip]{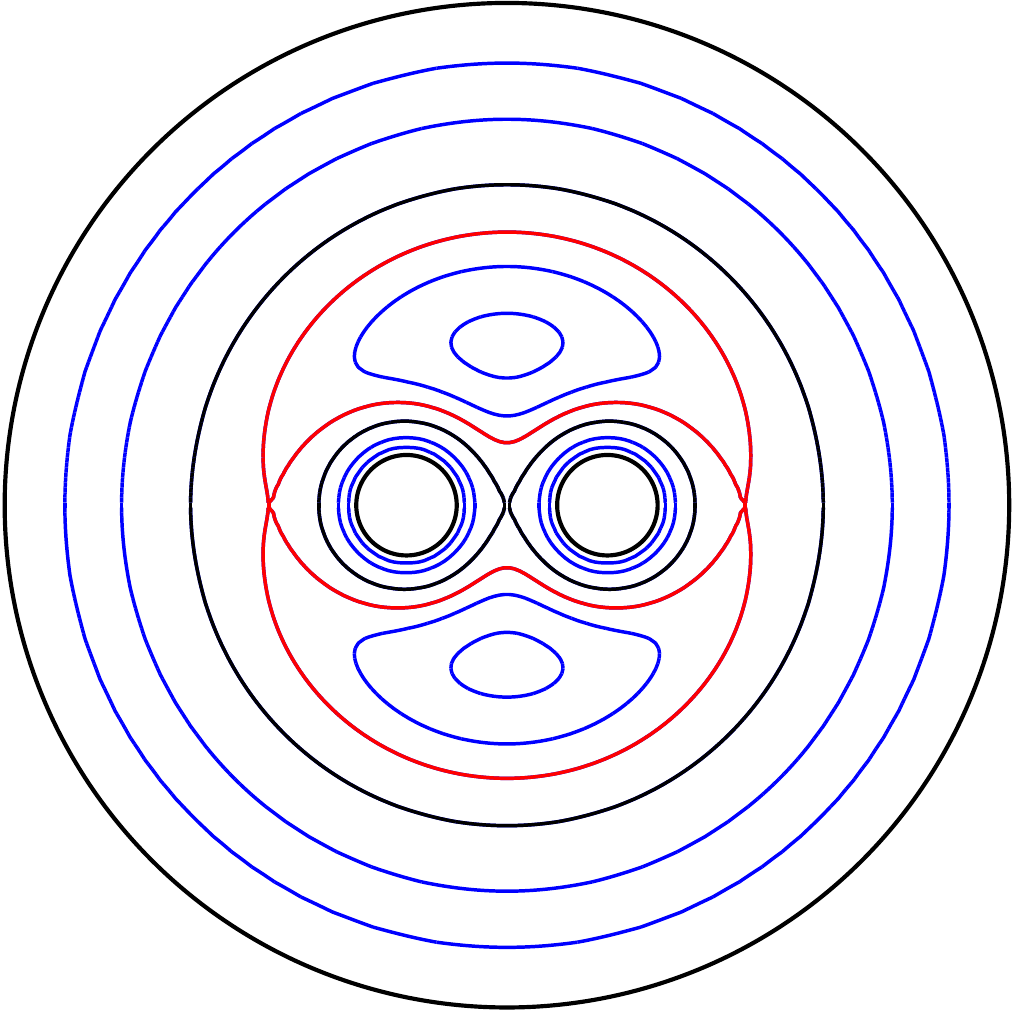}
		\caption{$p=0.2$}
		\label{fig:2c2}
	\end{subfigure}%
	~ 
	\begin{subfigure}[b]{0.24\textwidth}
		\centering
		\includegraphics[width=\linewidth,trim=0 0 0 0,clip]{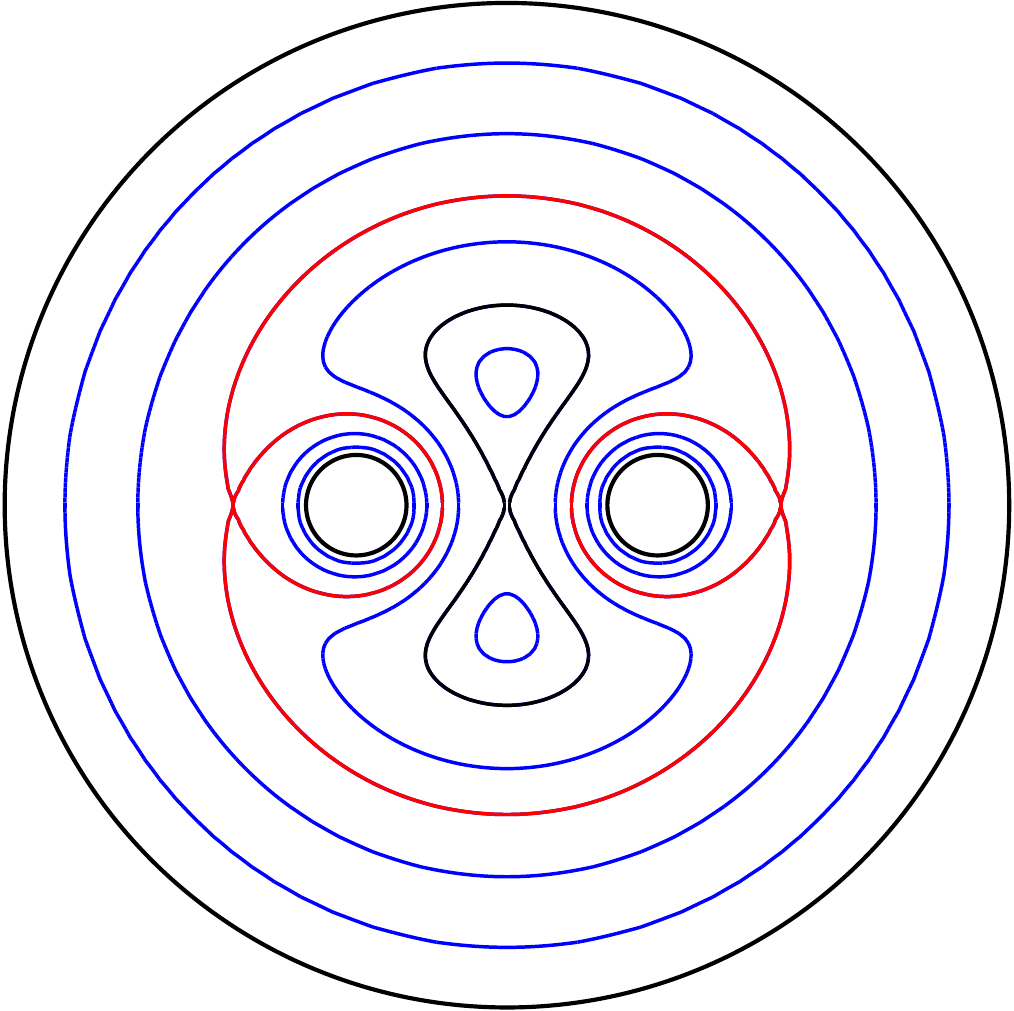}
		\caption{ $p=0.3$}
		\label{fig:2c3}
	\end{subfigure}%
	~
	\begin{subfigure}[b]{0.24\textwidth}
		\centering
		\includegraphics[width=\linewidth,trim=0 0 0 0,clip]{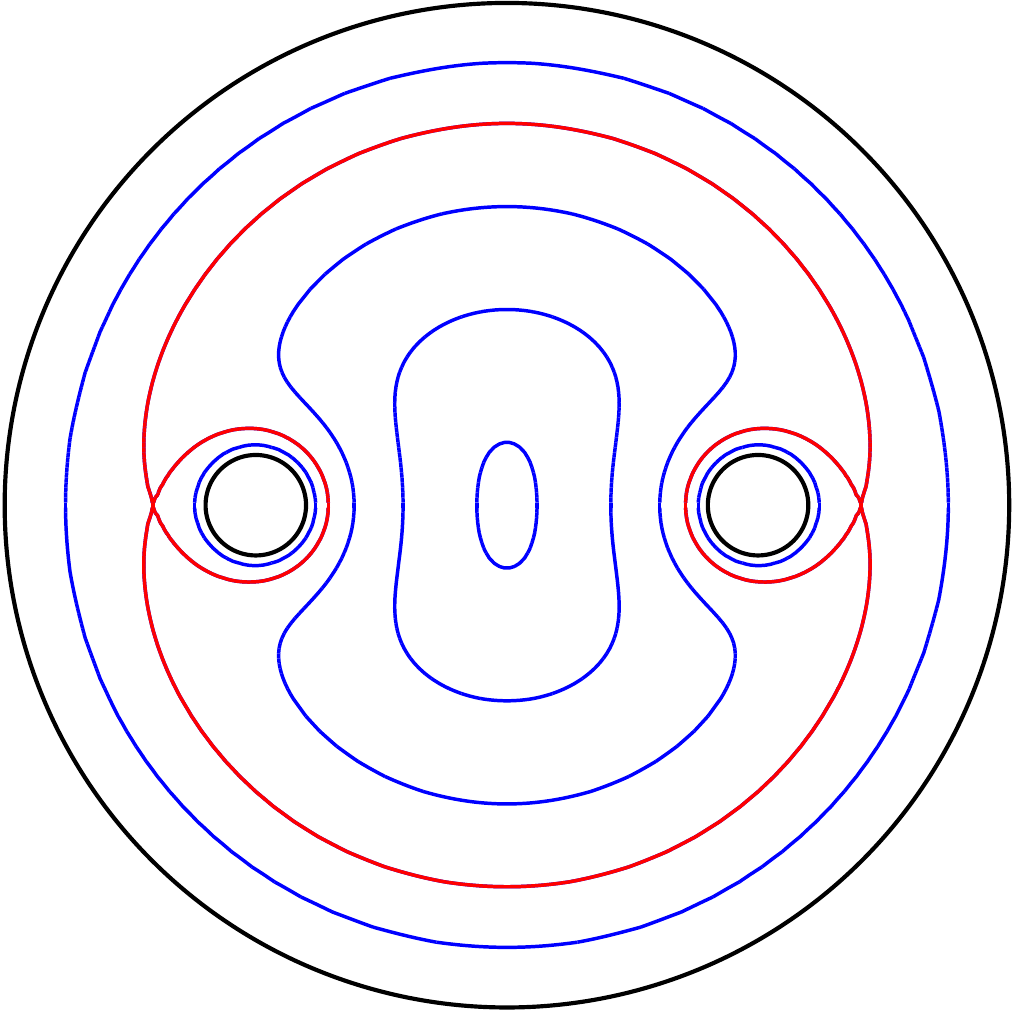}
		\caption{ $p=0.5$}
		\label{fig:2c5}
	\end{subfigure}%
	~ 
	\begin{subfigure}[b]{0.24\textwidth}
		\centering
		\includegraphics[width=\linewidth,trim=0 0 0 0,clip]{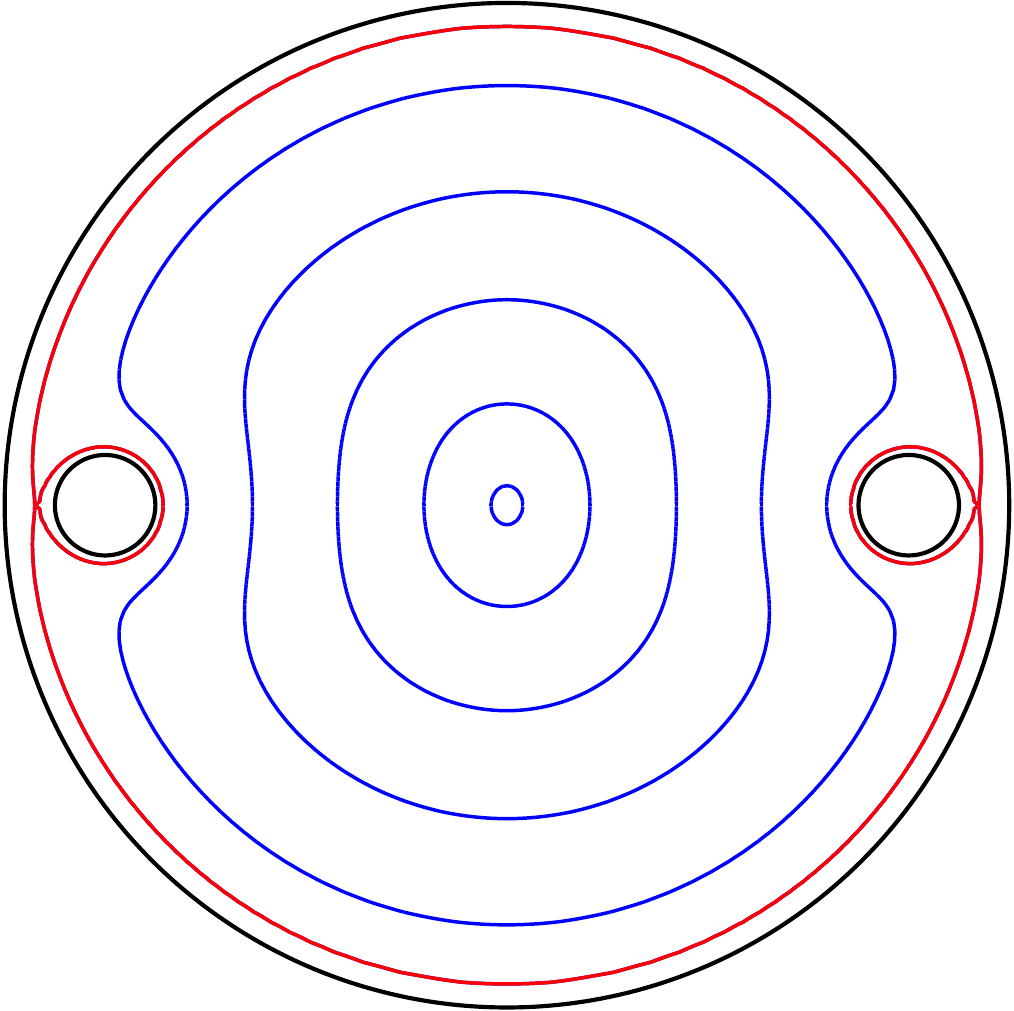}
		\caption{$p=0.8$}
		\label{fig:2c8}
	\end{subfigure}%
	\caption{Vortex trajectories in triply connected circular domains. The outer circle is the unit circle $|\zeta|=1$ and the inner circles are $|\zeta\pm p|=0.1$.}
	%the vertices of the square obstacle are at different positions $p+0.1+0.1\i,p+0.1-0.1\i,p-0.1-0.1\i,p-0.1+0.1\i$.}
	\label{fig:2c}
\end{figure}

In his paper~\cite{Tak-eq}, Sakajo studied also the motion of a single point vortex in multiply connected circular domains. The centers of the inner circles are assumed to be on the real axis. The contour plot of the Hamiltonian for these domains is given in Figure~\ref{fig:2c}. 
Crowdy and Marshal~\cite{cm-ana} have also considered several circular domains. 

\begin{figure}[!htb] %
	\centering
	\begin{subfigure}[b]{0.24\textwidth}
		\centering
		\includegraphics[width=\linewidth,trim=0 0 0 0,clip]{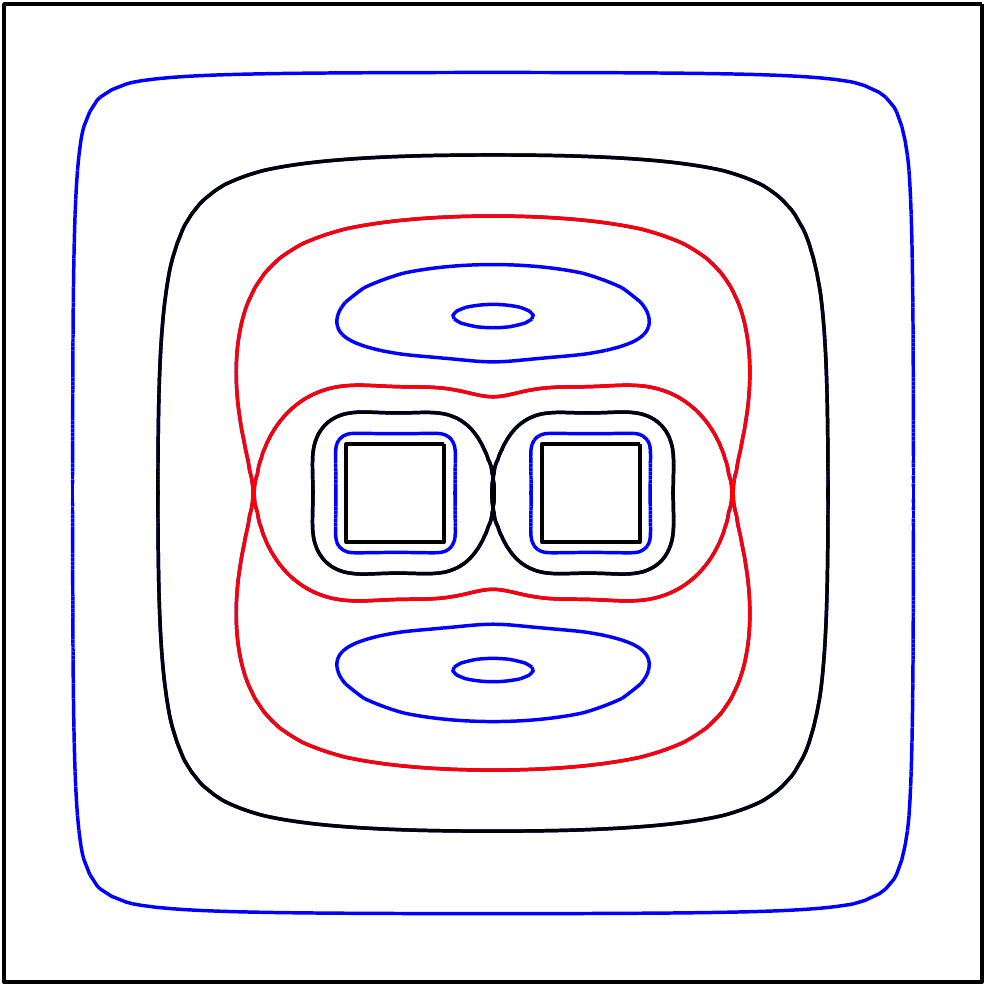}
		\caption{$p=0.2$}
		\label{fig:2s2h}
	\end{subfigure}%
	~ 
	\begin{subfigure}[b]{0.24\textwidth}
		\centering
		\includegraphics[width=\linewidth,trim=0 0 0 0,clip]{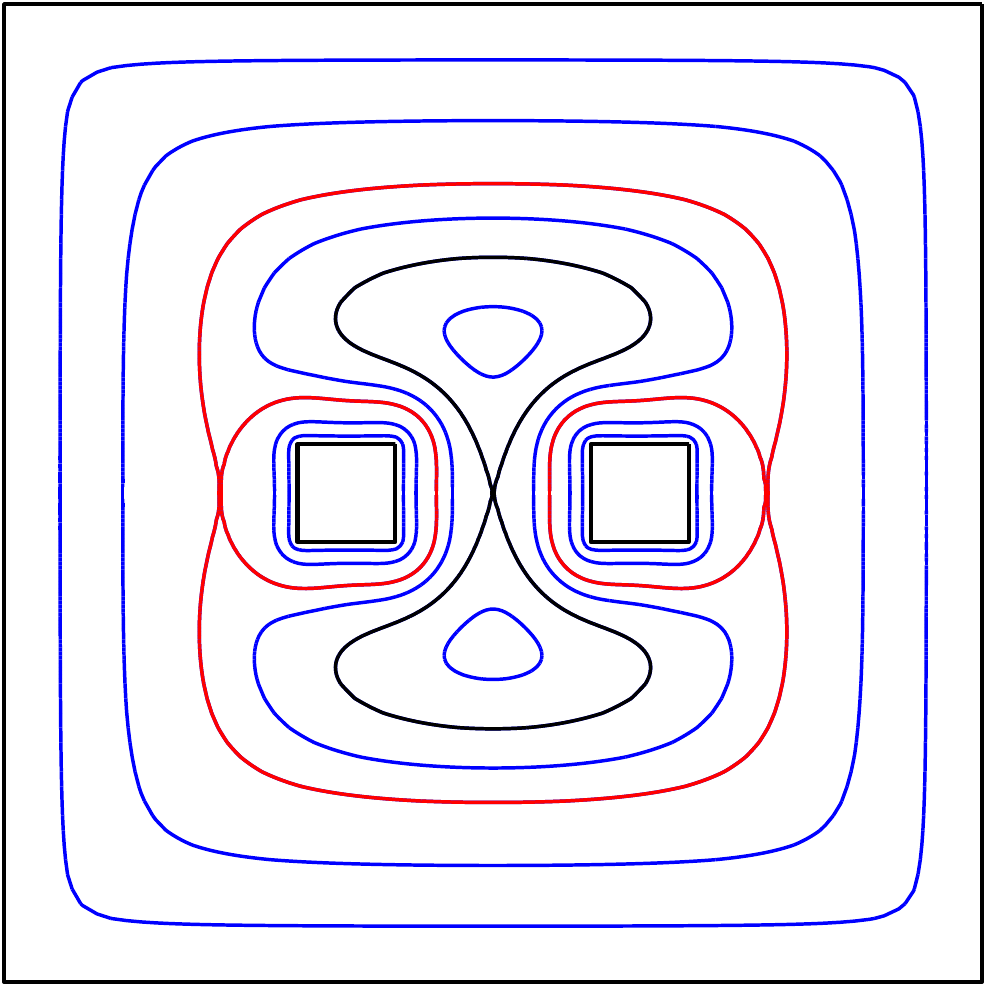}
		\caption{ $p=0.3$}
		\label{fig:2s3h}
	\end{subfigure}%
	~
	\begin{subfigure}[b]{0.24\textwidth}
		\centering
		\includegraphics[width=\linewidth,trim=0 0 0 0,clip]{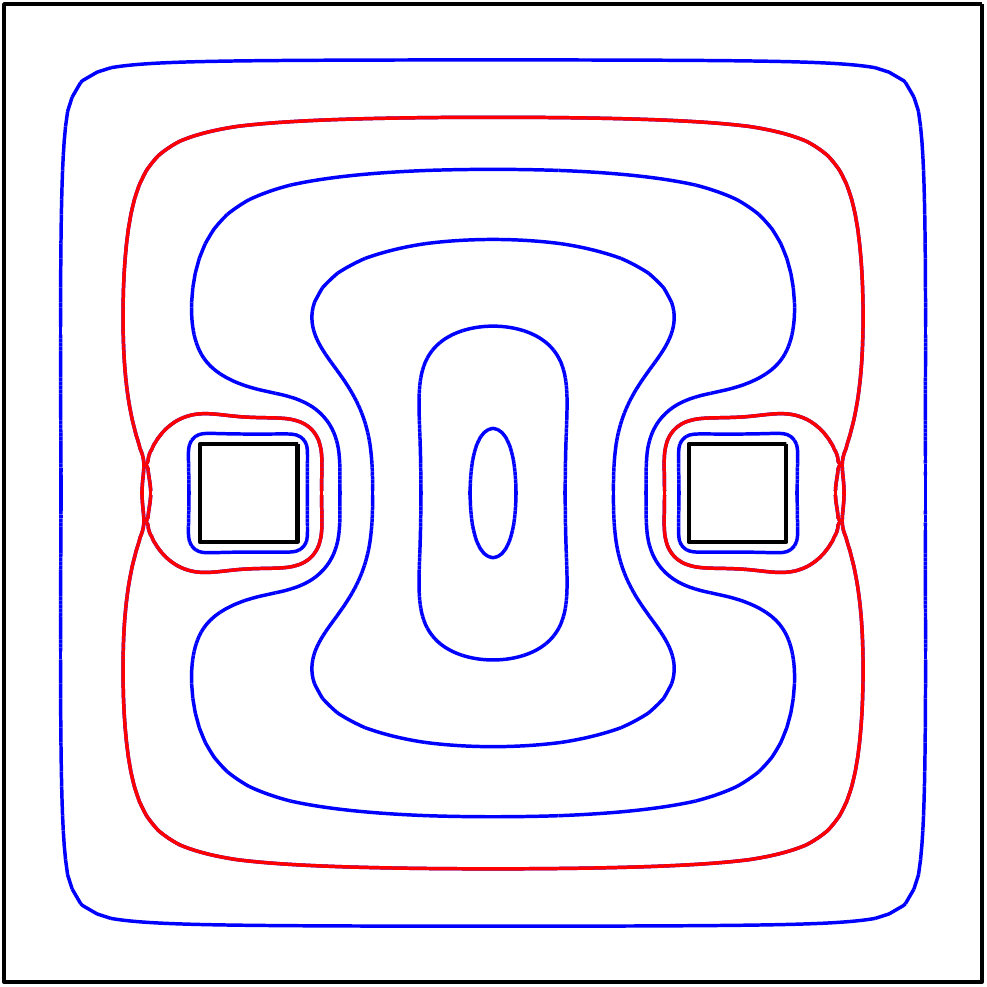}
		\caption{ $p=0.5$}
		\label{fig:2s5h}
	\end{subfigure}%
	~ 
	\begin{subfigure}[b]{0.24\textwidth}
		\centering
		\includegraphics[width=\linewidth,trim=0 0 0 0,clip]{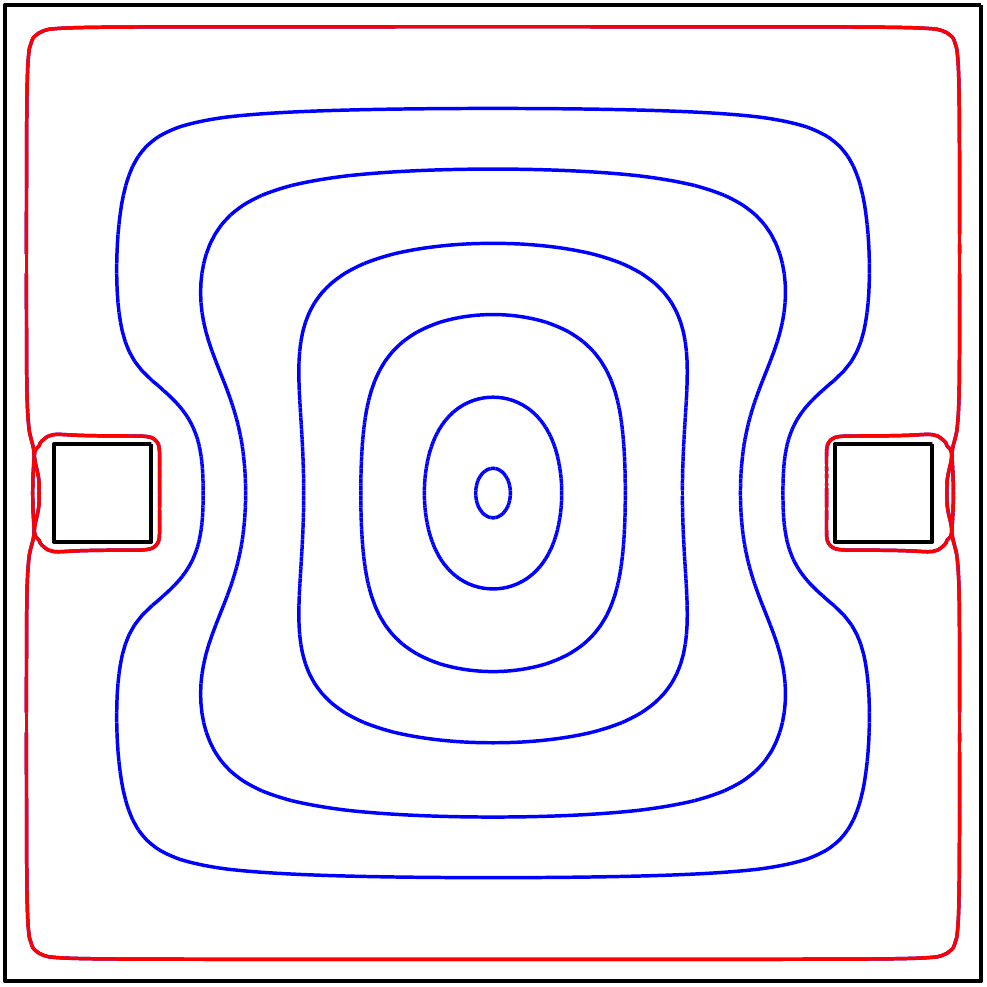}
		\caption{$p=0.8$}
		\label{fig:2s8h}
	\end{subfigure}
	\vskip\baselineskip
	\begin{subfigure}[b]{0.24\textwidth}
		\centering
		\includegraphics[width=\linewidth,trim=0 0 0 0,clip]{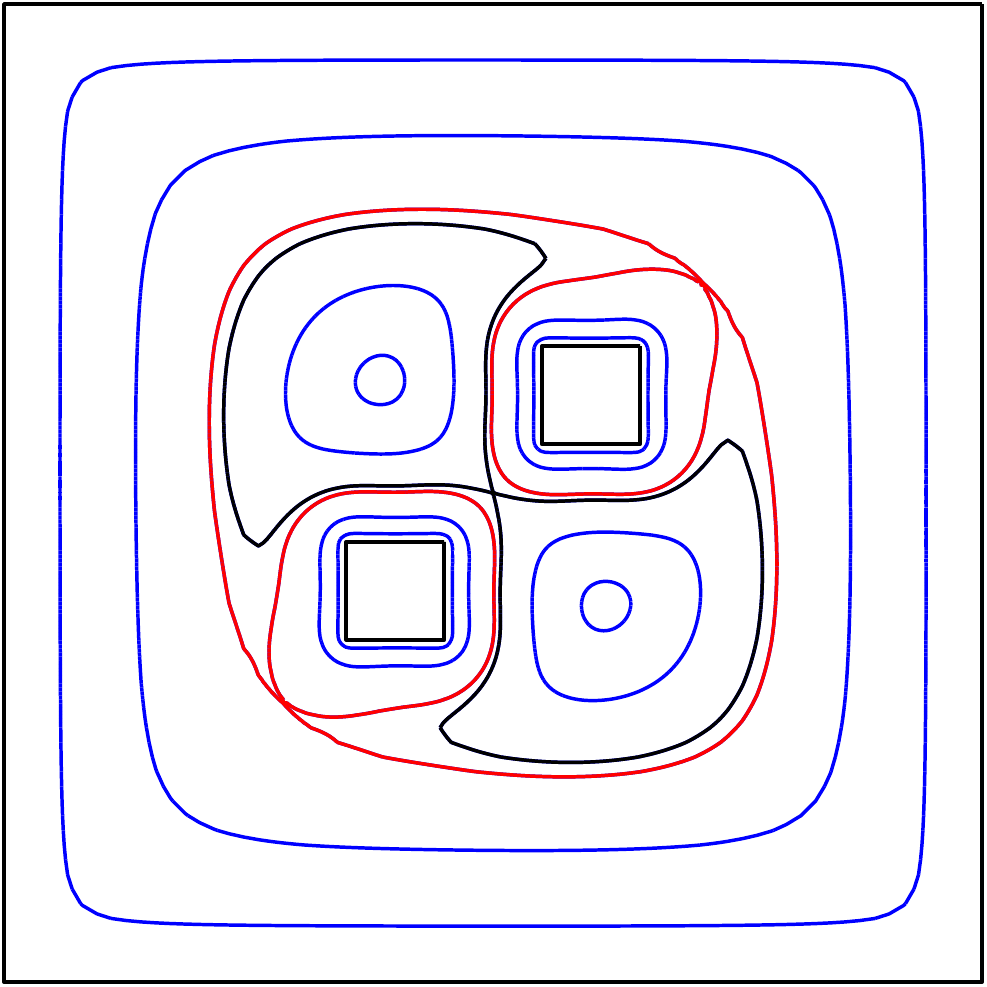}
		\caption{$p=0.2+0.2\i$}
		\label{fig:2s2d}
	\end{subfigure}%
	~ 
	\begin{subfigure}[b]{0.24\textwidth}
		\centering
		\includegraphics[width=\linewidth,trim=0 0 0 0,clip]{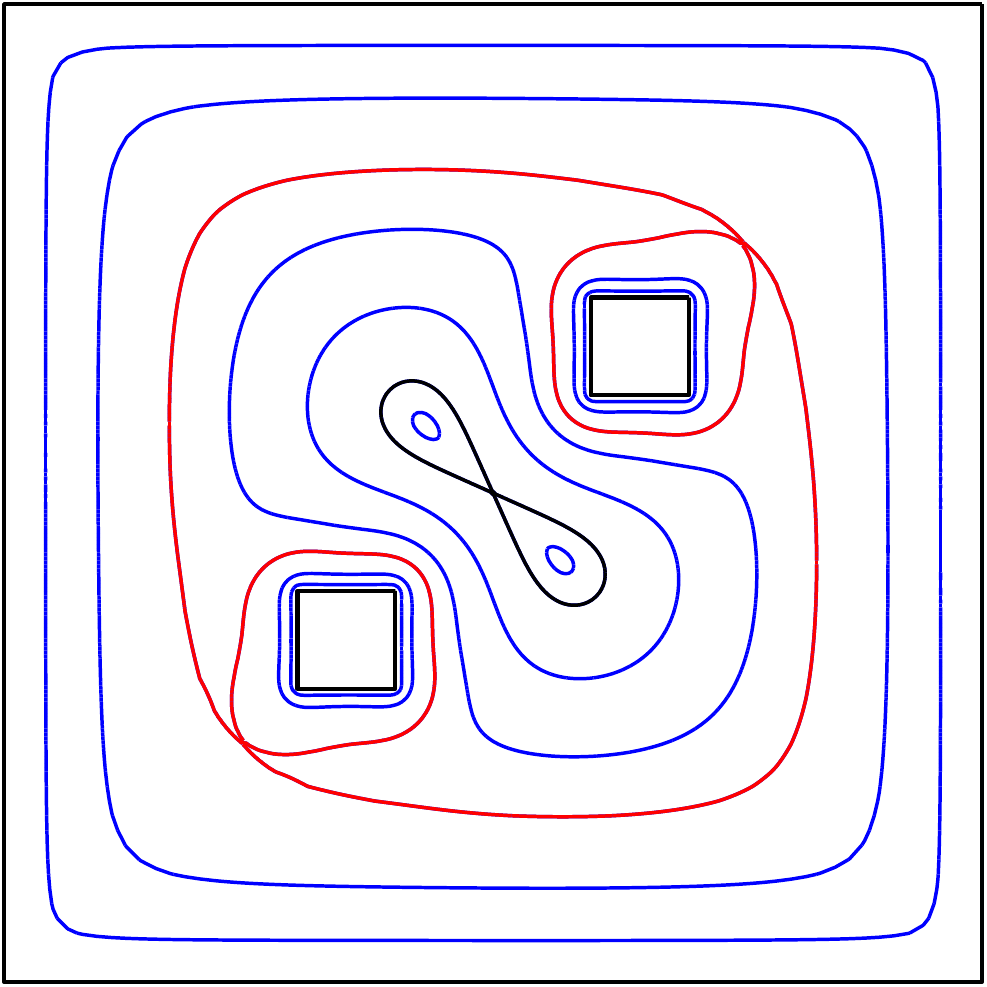}
		\caption{$p=0.3+0.3\i$}
		\label{fig:2s3d}
	\end{subfigure}%
	~
	\begin{subfigure}[b]{0.24\textwidth}
		\centering
		\includegraphics[width=\linewidth,trim=0 0 0 0,clip]{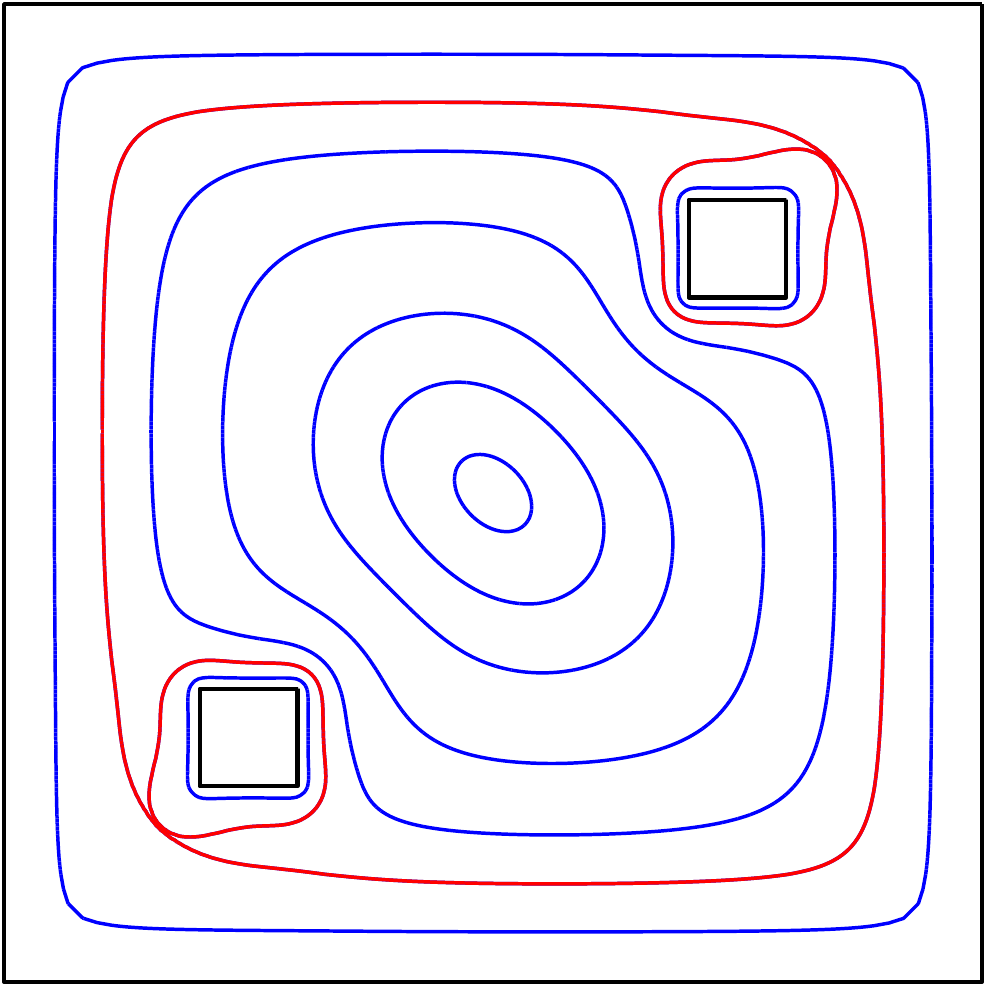}
		\caption{$p=0.5+0.5\i$}
		\label{fig:2s5d}
	\end{subfigure}%
	~ 
	\begin{subfigure}[b]{0.24\textwidth}
		\centering
		\includegraphics[width=\linewidth,trim=0 0 0 0,clip]{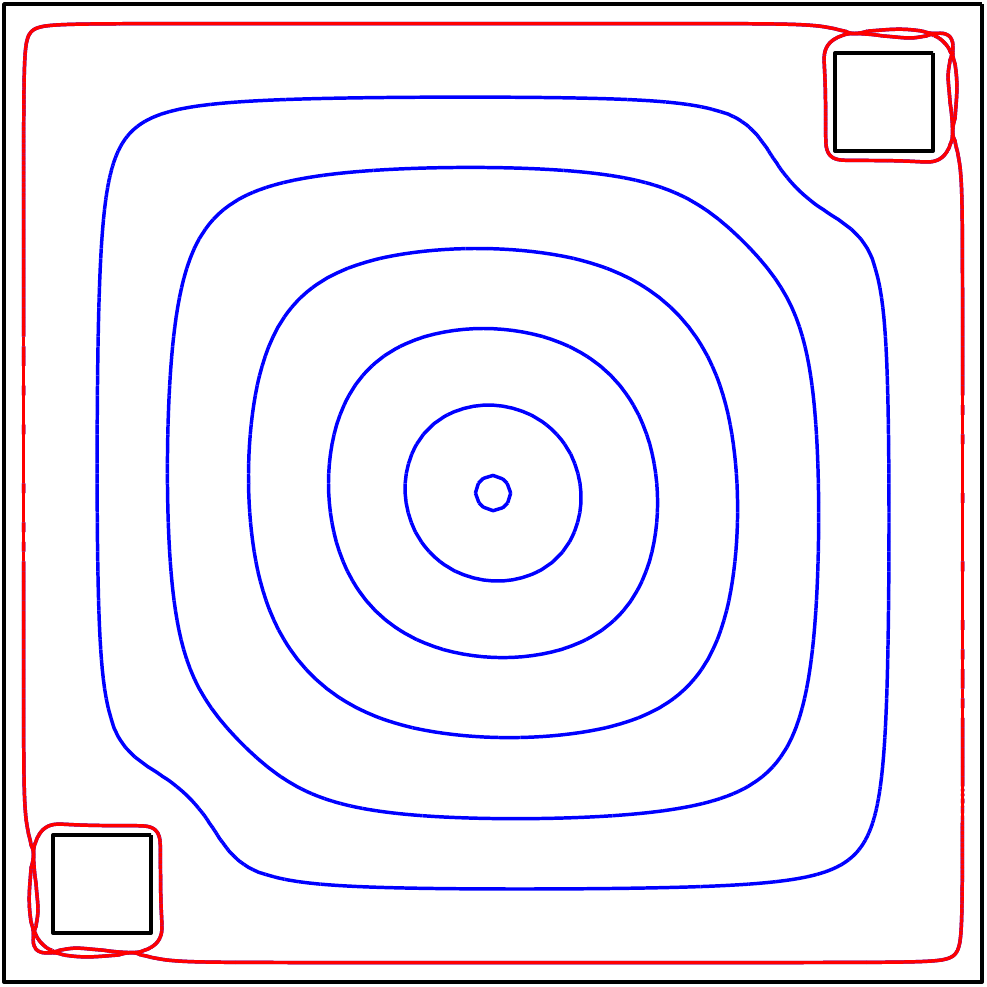}
		\caption{$p=0.8+0.8\i$}
		\label{fig:2s8d}
	\end{subfigure}%
	\caption{Vortex trajectories in triply connected polygonal domains where the inner squares have side length equal to $0.2$ and centers at $p$ and $-p$.}
	\label{fig:2s}
\end{figure}

Here, we investigate vortex motion in several multiply connected polygonal domains. First, we replace the circles in Figure~\ref{fig:2c} by squares where the outer one has vertices at $\pm 1\pm\i$.   The contour plot of the Hamiltonian in this case is then shown in the top row of Figure~\ref{fig:2s}. The topology pattern of vortex motion looks similar to the circular domain case.
Looking at how the point vortex trajectories interact with the polygonal obstacles, we observe that this interaction depends on the distance between these obstacles. When they are close to each other, as in Figure~\ref{fig:2s2h}, then we have two saddle points between them connected by a heteroclinic loop, compared to one saddle for the circular case. As the separation between the two square obstacles increases, we get a vortex merging from two saddles and one center to one saddle at which two homoclinic orbits intersect and surround the previous upper and lower centers separately; see Figure~\ref{fig:2s3h}.
Increasing further the separation distance, like for example to $0.8$ unit in Figure~\ref{fig:2s5h}, we get another vortex merging from the last homoclinic structure to a vortex center located at zero. This configuration continues to hold when each of the two square obstacles approaches the external eastern and western boundaries, as shown in Figure~\ref{fig:2s8h}. The only difference to the circular domain case is that we have two saddle points to the right and two to the left of the square obstacles instead of one to the right and one to the left.

We consider also the case when the centers of the inner squares are on the diagonal line $y=x$, as shown in the bottom row of Figure~\ref{fig:2s}. When these centers are located at $\pm p$ for $p=0.2+0.2\i$ or $p=0.3+0.3\i$ in Figures~\ref{fig:2s2d} and~\ref{fig:2s3d}, the same dynamical configuration occurs which is topologically equivalent to the pattern in Figure~\ref{fig:2s3h}. Observe that all three saddle points are located on the main diagonal line. 
After displacing the square obstacles toward the top right and bottom left outer corners in Figure~\ref{fig:2s5d}, the homoclinic structure at the center disappears, and a unique vortex center is born in a similar way to the horizontal separation. As expected, when the obstacles gets closer to the corners in Figure~\ref{fig:2s8d}, the phase portrait between each obstacle and the external boundary becomes similar to the case of one square obstacle. The Hamiltonian has four saddles near each side and all eight points are joined by heteroclinic connections.

\begin{figure}[!htb] %
	\centering
	\begin{subfigure}[b]{0.24\textwidth}
		\centering
		\includegraphics[width=\linewidth,trim=0 0 0 0,clip]{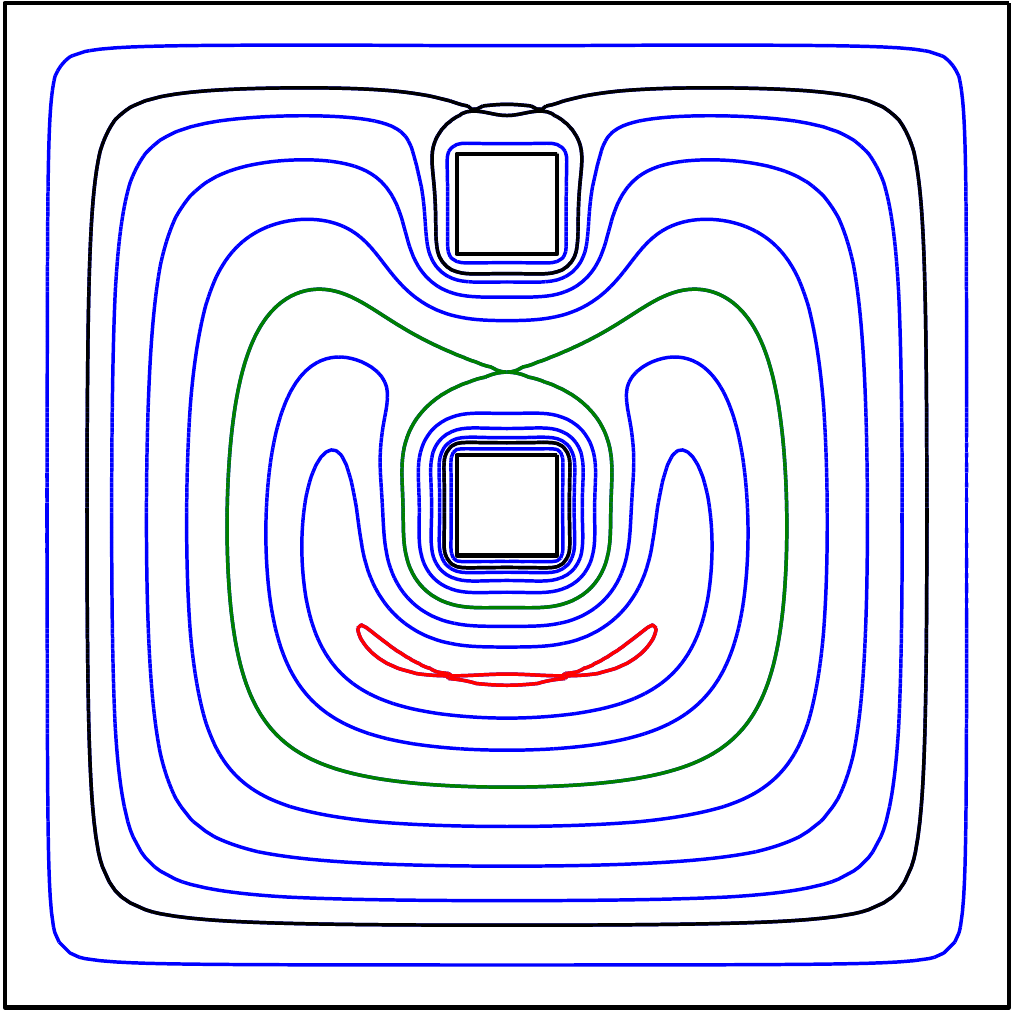}
		\caption{$p=0, q=0.6\i$}
		\label{fig:2s-1s1}
	\end{subfigure}%
	~
	\begin{subfigure}[b]{0.24\textwidth}
		\centering
		\includegraphics[width=\linewidth,trim=0 0 0 0,clip]{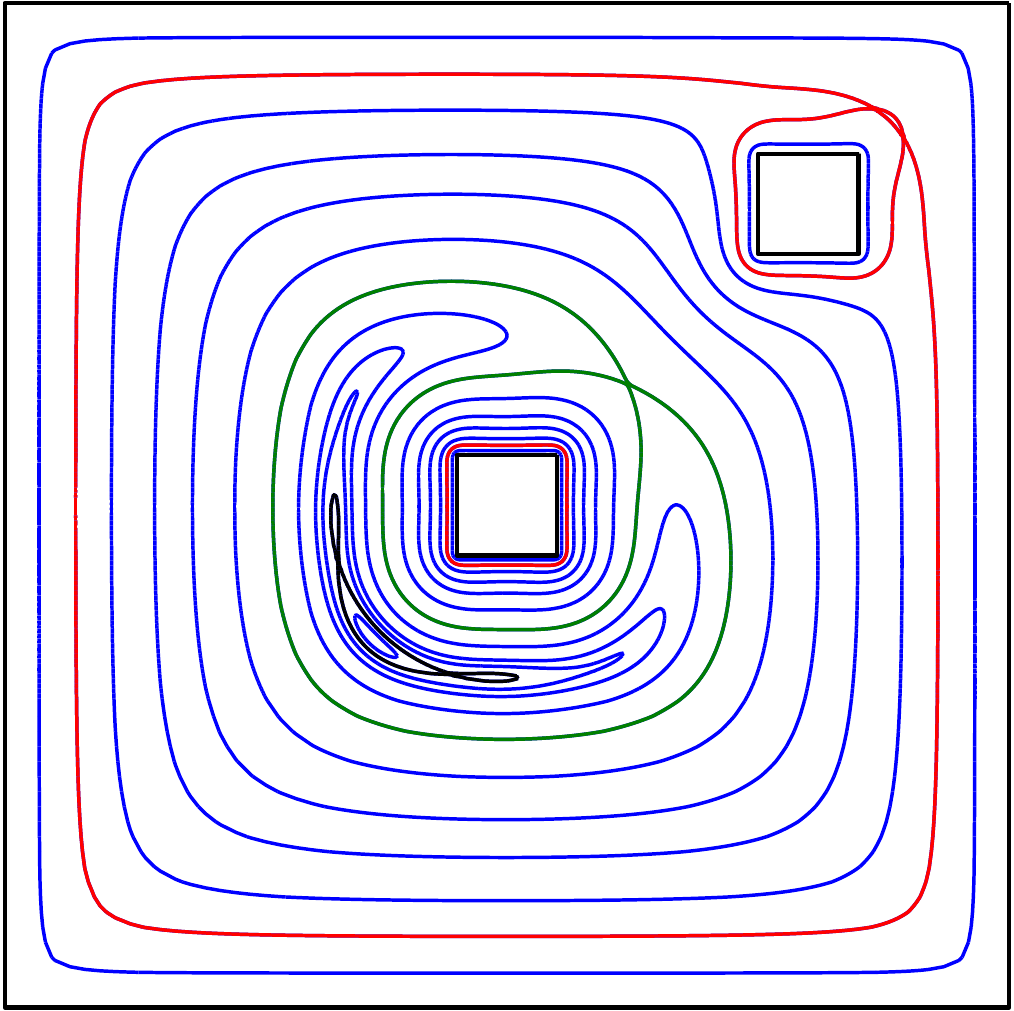}
		\caption{\scalebox{0.9}{$p=0, q=0.6+0.6\i$}}
		\label{fig:2s-1s2}
	\end{subfigure}%
	~ 
	\begin{subfigure}[b]{0.24\textwidth}
		\centering
		\includegraphics[width=\linewidth,trim=0 0 0 0,clip]{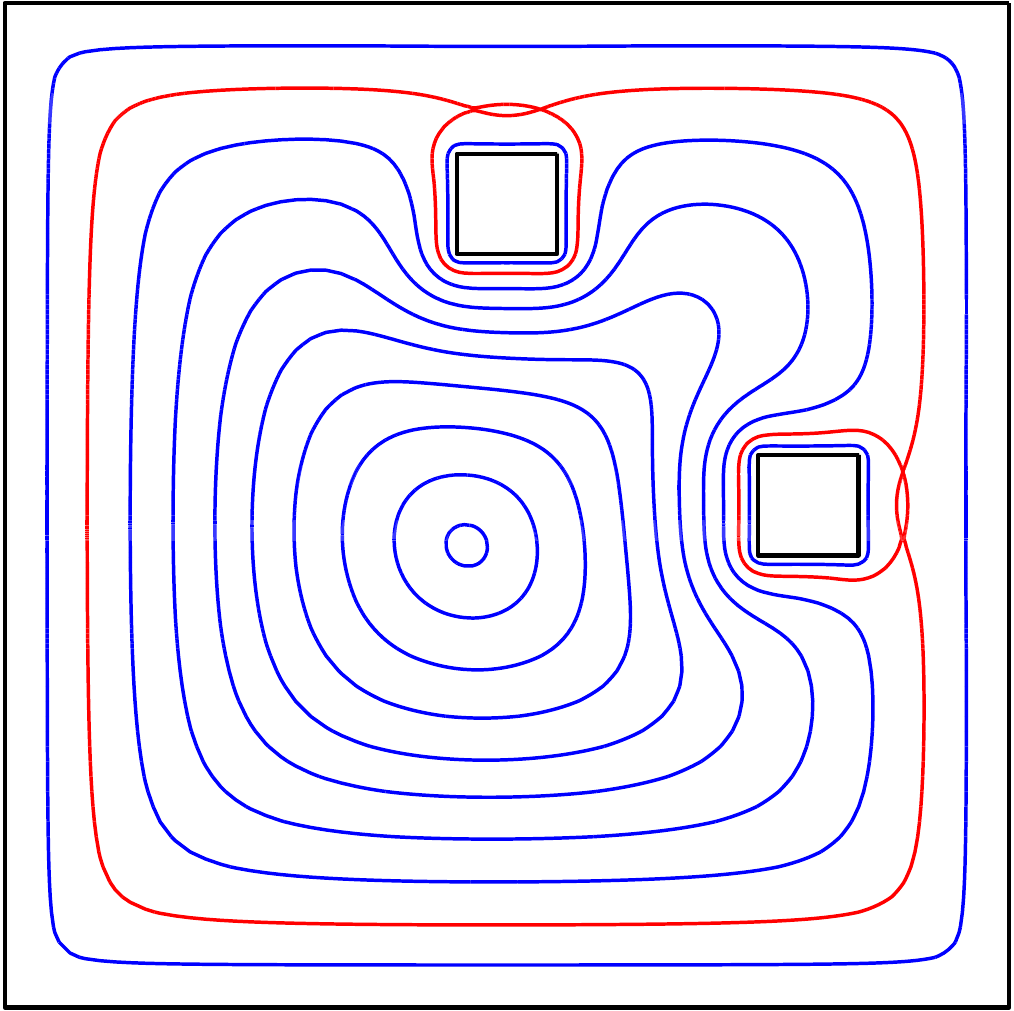}
		\caption{$p=0.6, q=0.6\i$}
		\label{fig:2s-1s3}
	\end{subfigure}%
	~
	\begin{subfigure}[b]{0.24\textwidth}
		\centering
		\includegraphics[width=\linewidth,trim=0 0 0 0,clip]{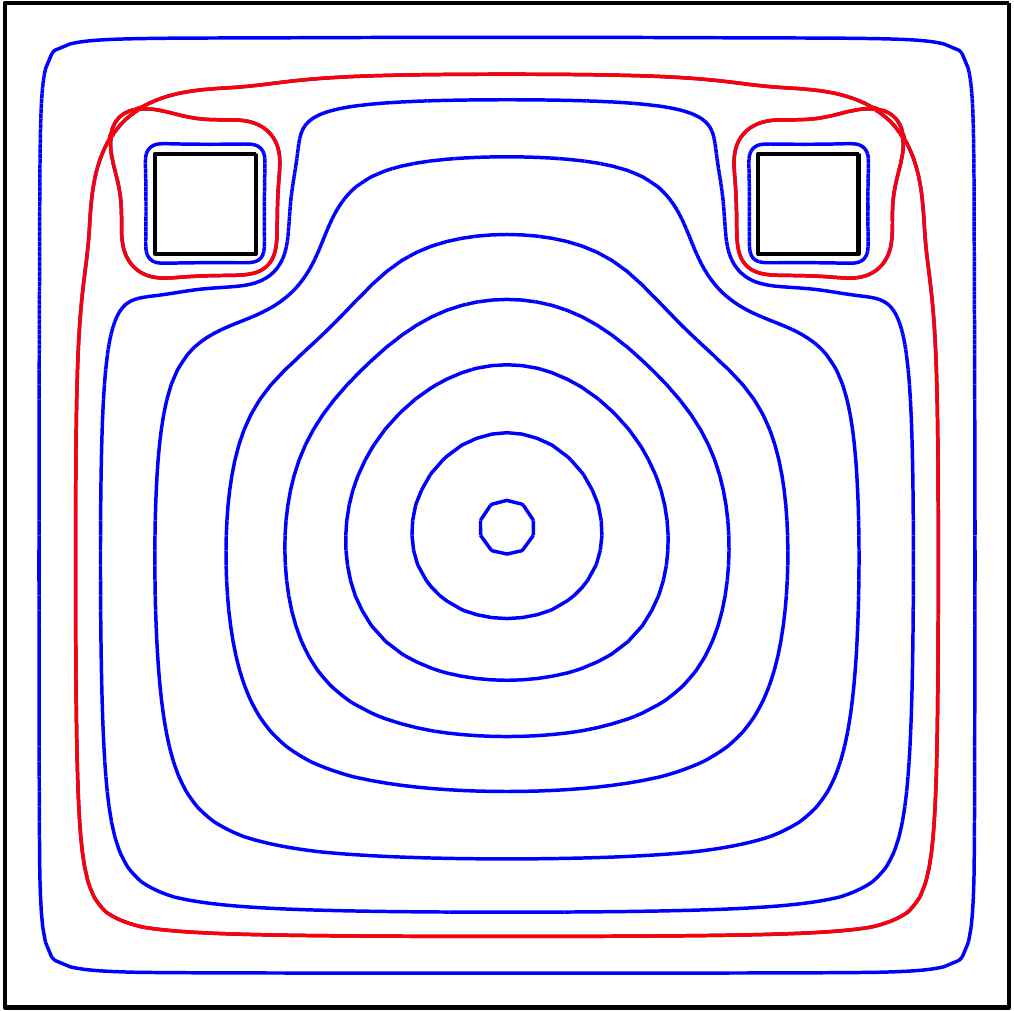}
		\caption{ $p,q=\pm0.6+0.6\i$}
		\label{fig:2s-1s4}
	\end{subfigure}%
	\vskip\baselineskip
	\begin{subfigure}[b]{0.24\textwidth}
		\centering
		\includegraphics[width=\linewidth,trim=0 0 0 0,clip]{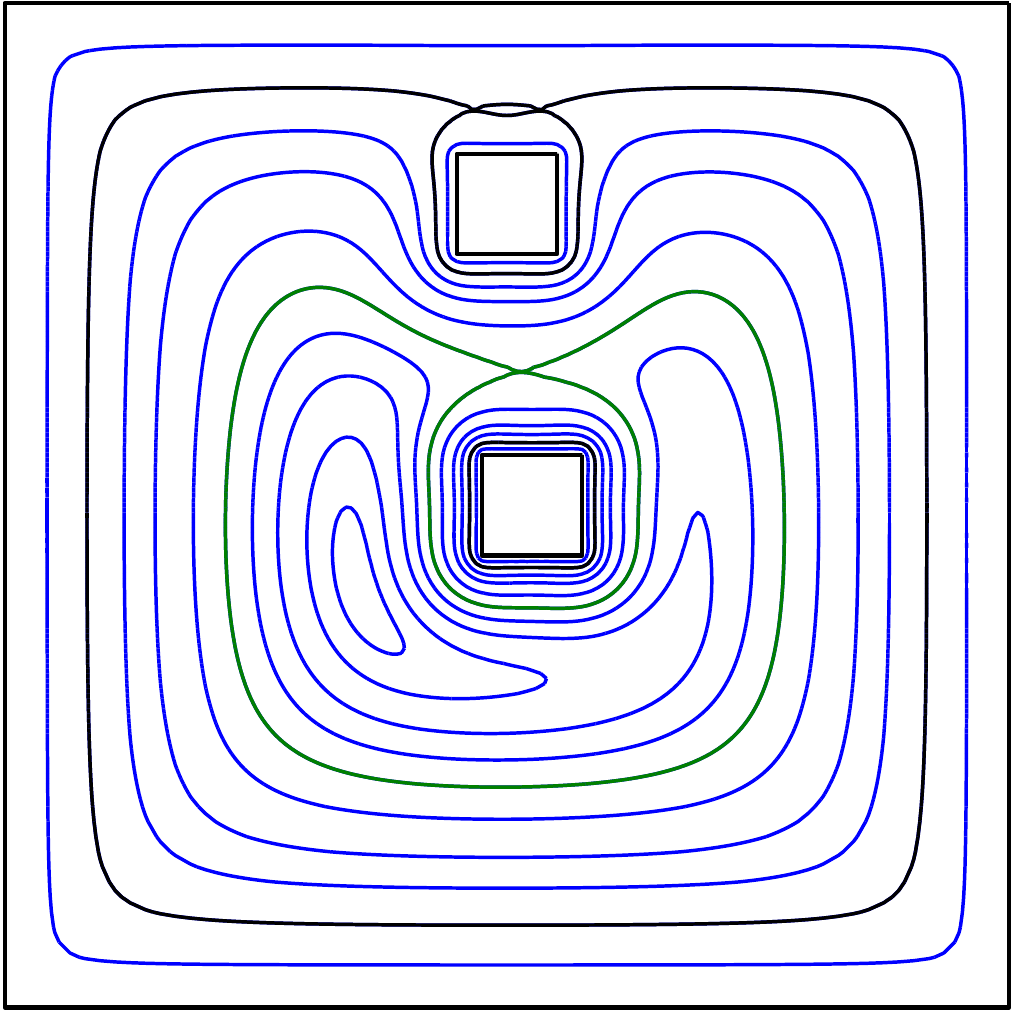}
		\caption{$p=0.05,q=0.6\i$}
		\label{fig:2s-ns1}
	\end{subfigure}%
	~
	\begin{subfigure}[b]{0.24\textwidth}
		\centering
		\includegraphics[width=\linewidth,trim=0 0 0 0,clip]{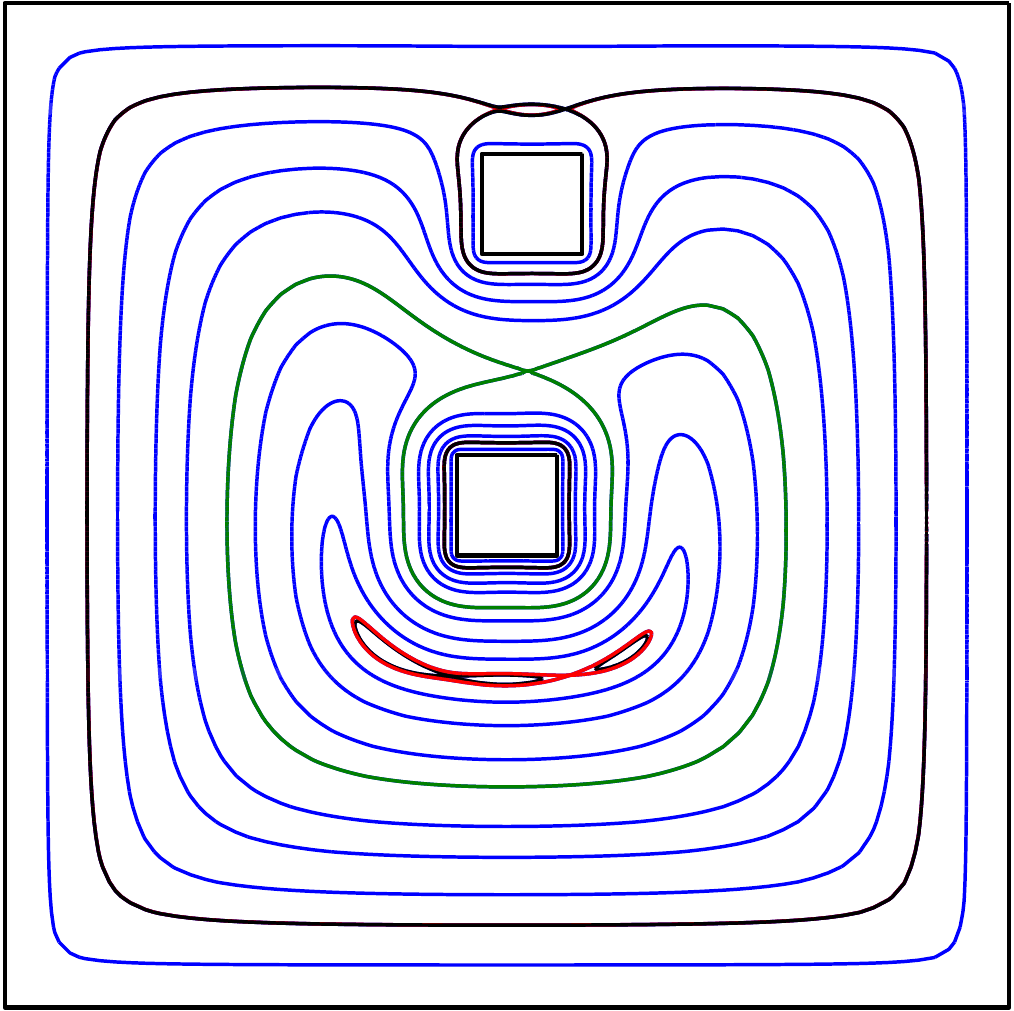}
		\caption{\scalebox{0.85}{$p=0, q=0.05+0.6\i$}}
		\label{fig:2s-ns2}
	\end{subfigure}%
	~ 
	\begin{subfigure}[b]{0.24\textwidth}
		\centering
		\includegraphics[width=\linewidth,trim=0 0 0 0,clip]{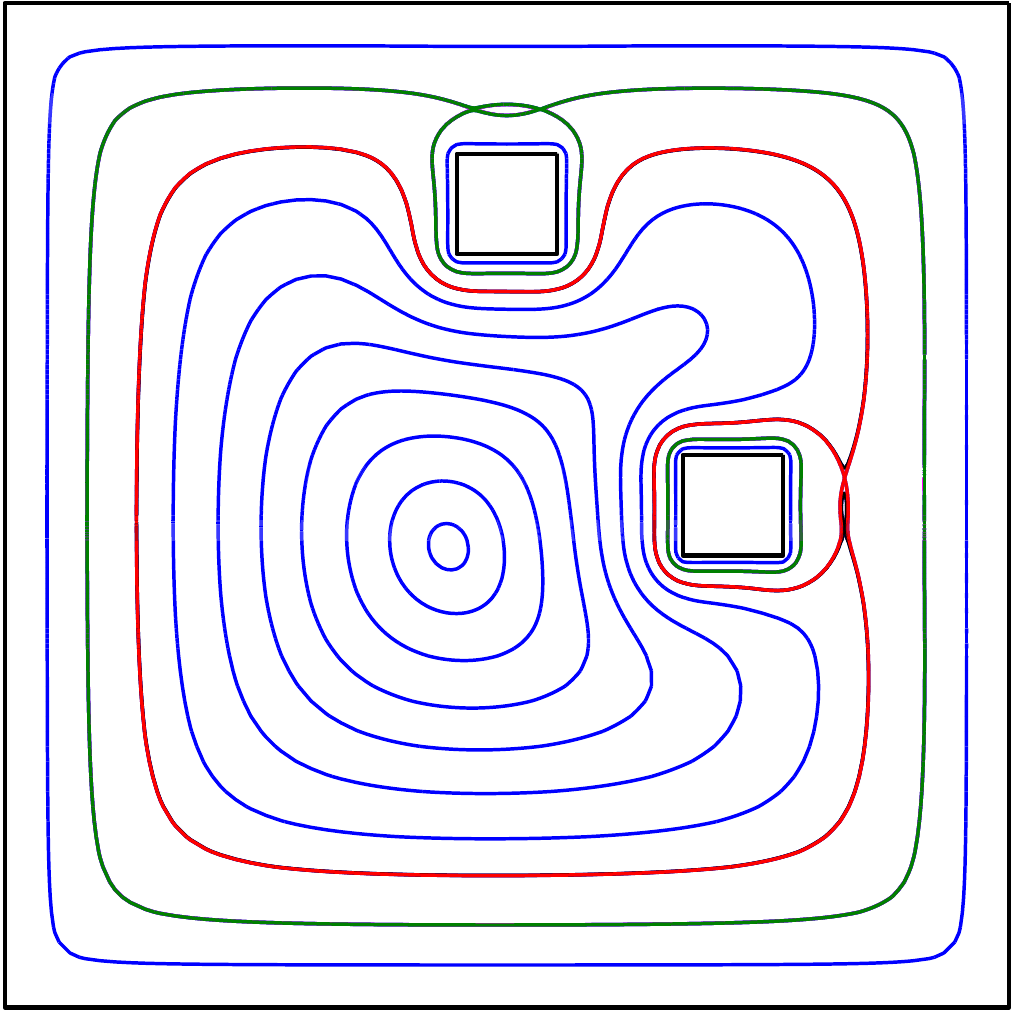}
		\caption{$p=0.45, q=0.6\i$}
		\label{fig:2s-ns3}
	\end{subfigure}%
	~
	\begin{subfigure}[b]{0.24\textwidth}
		\centering
		\includegraphics[width=\linewidth,trim=0 0 0 0,clip]{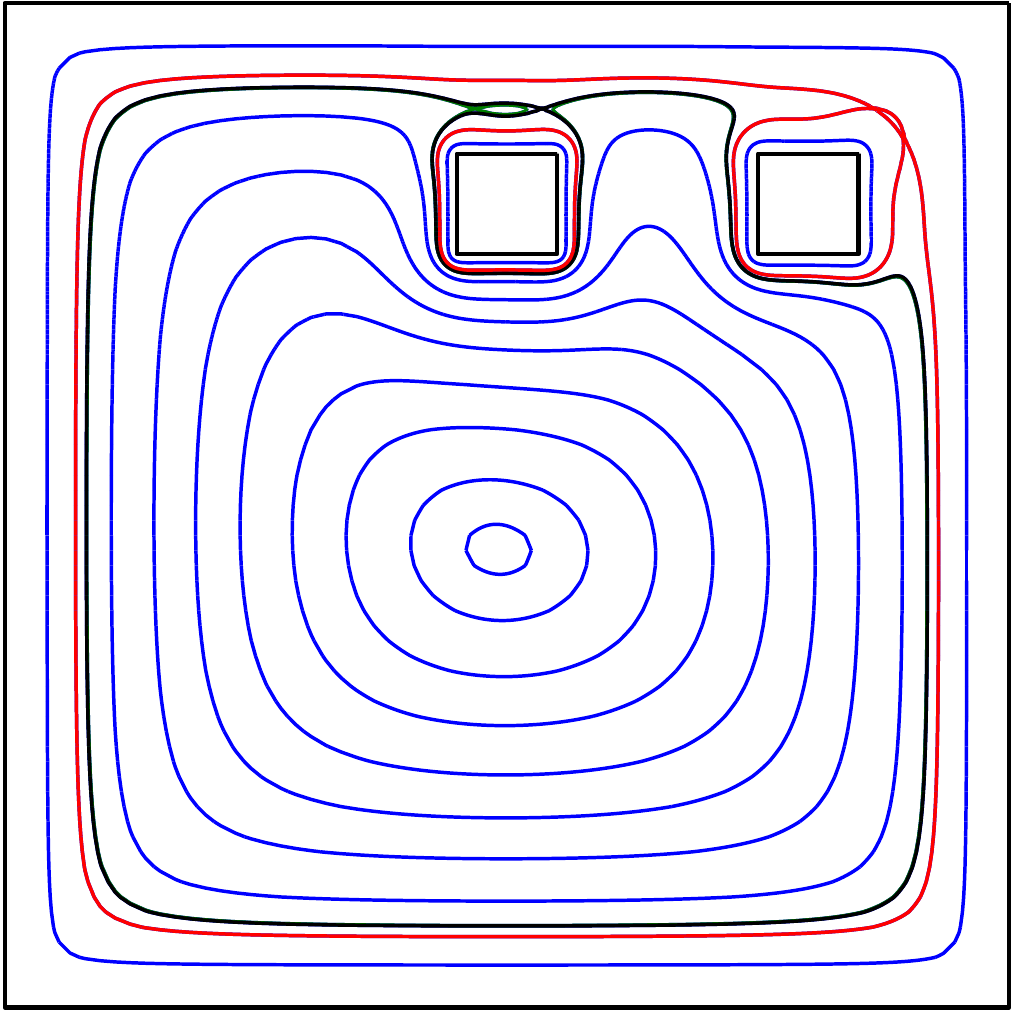}
		\caption{\scalebox{0.8}{$p=0.6\i,q=0.6+0.6\i$}}
		\label{fig:2s-ns4}
	\end{subfigure}%
	\caption{Vortex trajectories in triply connected polygonal domains where the inner squares have side length equal to $0.2$, and centers at $p$ and $q$, respectively.}
	\label{fig:2s-1sns}
\end{figure}

We examine now the effect of  reducing reflectional symmetry to a unique line by placing the obstacle at different locations, as shown in the top row of Figure~\ref{fig:2s-1sns}. We first place one obstacle centered at zero and the other once in the north and then in the right east; see Figures~\ref{fig:2s-1s1} and~\ref{fig:2s-1s2}. In these two examples, the dynamics restricted near the lower side of the line of reflectional symmetry consists of a $(2,1)$ homoclinic-heteroclinic structure similar to the one square obstacle case. The Hamiltonian has also one saddle between the two obstacles and two saddles connected by a heteroclinic loop between the obstacle and the external boundary. In Figures~\ref{fig:2s-1s3} and~\ref{fig:2s-1s4}, we consider two cases where the two obstacles are placed at symmetric locations with respect to the line $y=x$ and the imaginary axis, respectively. The behavior of vortex trajectories in these two examples is the same. There is a vortex center located almost at zero and two saddles between each obstacle and the external boundary. Each two points of the four saddles are connected by a heteroclinic loop.

In the bottom row of Figure~\ref{fig:2s-1sns} we demonstrate the effect of symmetry breaking on vortex motion. As the patterns in Figures~\ref{fig:2s-1s1} and~\ref{fig:2s-1s2} are topologically equivalent, we break the vertical line symmetry only in the first example. First, after a small horizontal perturbation of $0.01$ unit on the center of the lower obstacle, as shown in Figure~\ref{fig:2s-ns1}, the $(2,1)$ homoclinic-heteroclinic pattern in the lower side of this obstacle is no longer present, and a new vortex center is born near the lower left corner of the same obstacle. The other center and three saddles are still present in the phase space with the same patterns. However, a similar horizontal perturbation of the upper obstacle does not cause the two lower saddles to disappear, but it does split their heteroclinic connections. Observe also that the upper critical points still occur; see Figure~\ref{fig:2s-ns2}. Looking at Figures~\ref{fig:2s-ns3} and~\ref{fig:2s-ns4}, we see the effect of breaking the line symmetry in the two examples shown in Figures \ref{fig:2s-1s3} and~\ref{fig:2s-1s4}. Indeed, if we move one obstacle and keep the other at the same location, then the two saddle points near one obstacle get disconnected from the two saddles near the other.

\begin{figure}[!htb] %
	\centering
	\begin{subfigure}[b]{0.33\textwidth}
		\centering
		\includegraphics[width=\linewidth,trim=0 0 0 0,clip]{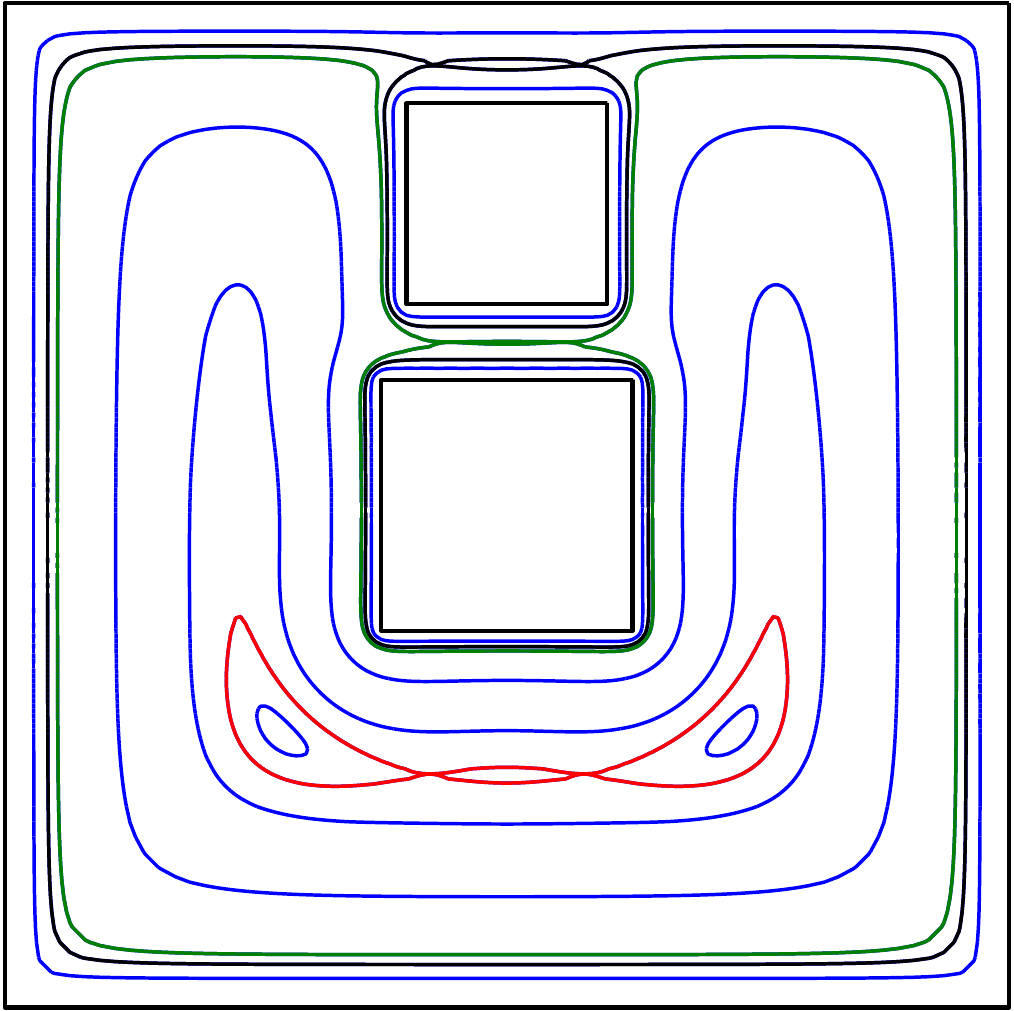}
		\caption{$p=0$ and $q=0.6\i$}
		\label{fig:2sg1}
	\end{subfigure}%
	~
	\begin{subfigure}[b]{0.33\textwidth}
		\centering
		\includegraphics[width=\linewidth,trim=0 0 0 0,clip]{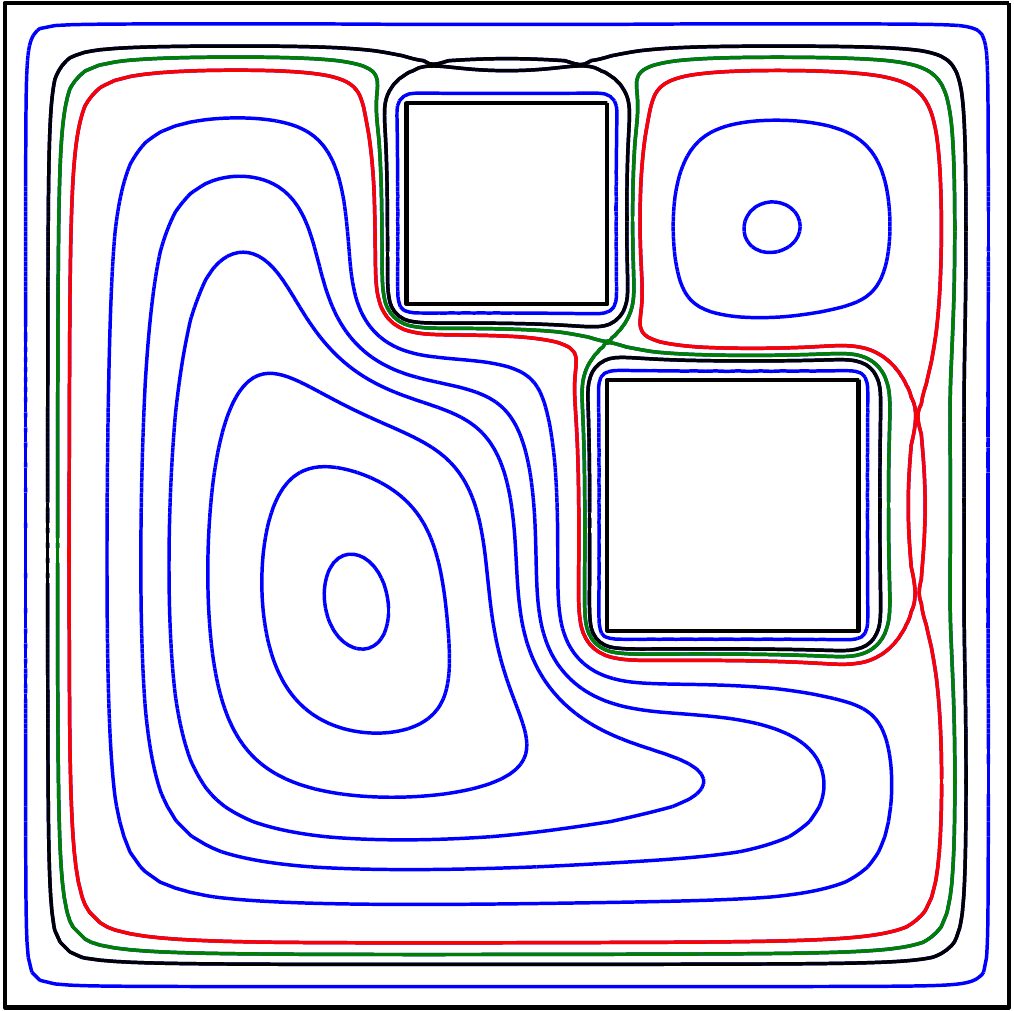}
		\caption{ $p=0.45$ and $q=0.6\i$}
		\label{fig:2sg2}
	\end{subfigure}%
	~ 
	\begin{subfigure}[b]{0.33\textwidth}
		\centering
		\includegraphics[width=\linewidth,trim=0 0 0 0,clip]{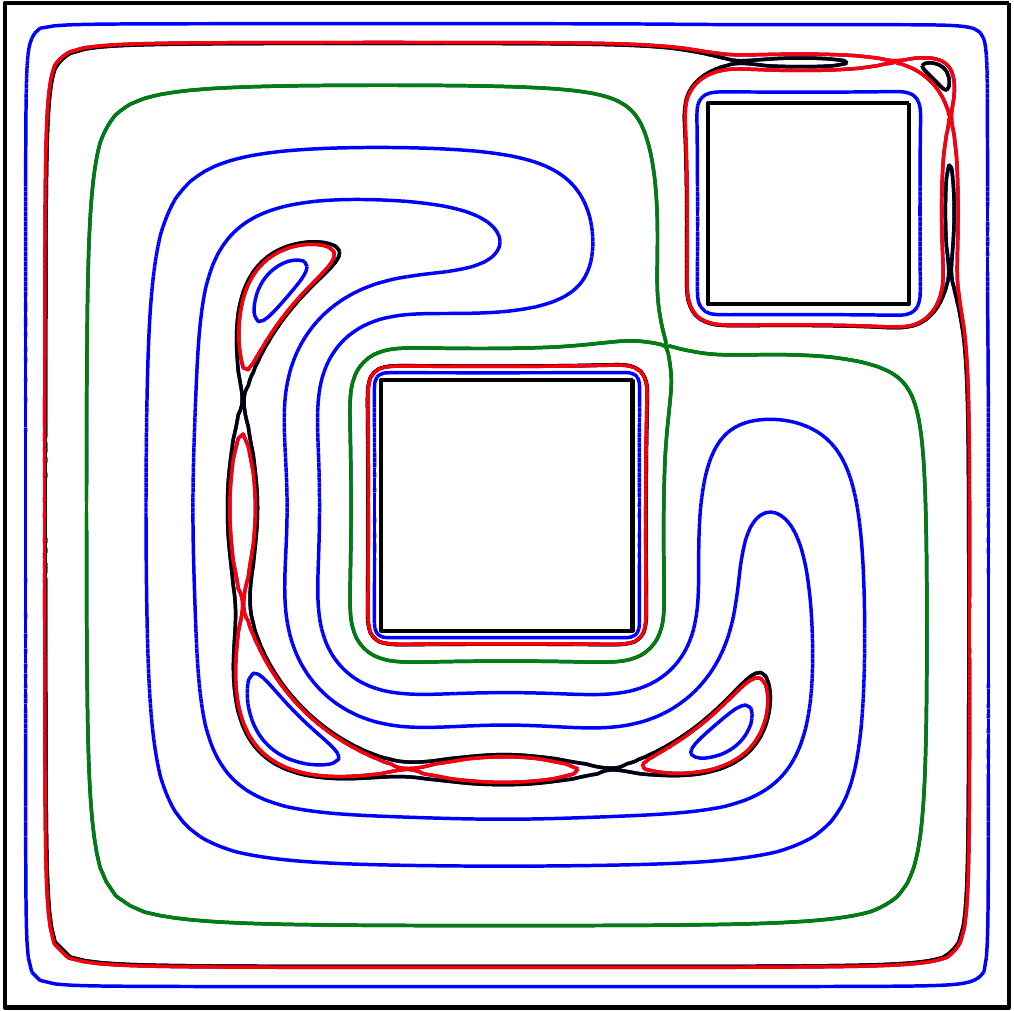}
		\caption{ $p=0$ and $q=0.6+0.6\i$}
		\label{fig:2sg3}
	\end{subfigure}%
	\caption{Vortex trajectories in triply connected polygonal domains where the large and small inner squares have side length equal to $0.5$ and $0.4$, and centers at $p$ and $q$, respectively.}
	\label{fig:2sg}
\end{figure}

In Figure~\ref{fig:2sg}, we show the effect of increasing the size of the square obstacles. Since the squares become too close, a new saddle point appears between them in Figures~\ref{fig:2sg1} and~\ref{fig:2sg2}. The dynamics of the point vortex motion is more remarkable in the case of Figure~\ref{fig:2sg3}. Two new saddles connected by a heteroclinic loop appear near the top right corner of the outer square. Moreover, on the other side, the $(2,1)$ homoclinic-heteroclinic pattern in Figure~\ref{fig:2s-1s2} becomes an inner structure within a heteroclinic loop in a new $(2,1)$ pattern.

\begin{figure}[!htb] %
	\centering
	\begin{subfigure}[b]{0.24\textwidth}
		\centering
		\includegraphics[width=\linewidth,trim=0 0 0 0,clip]{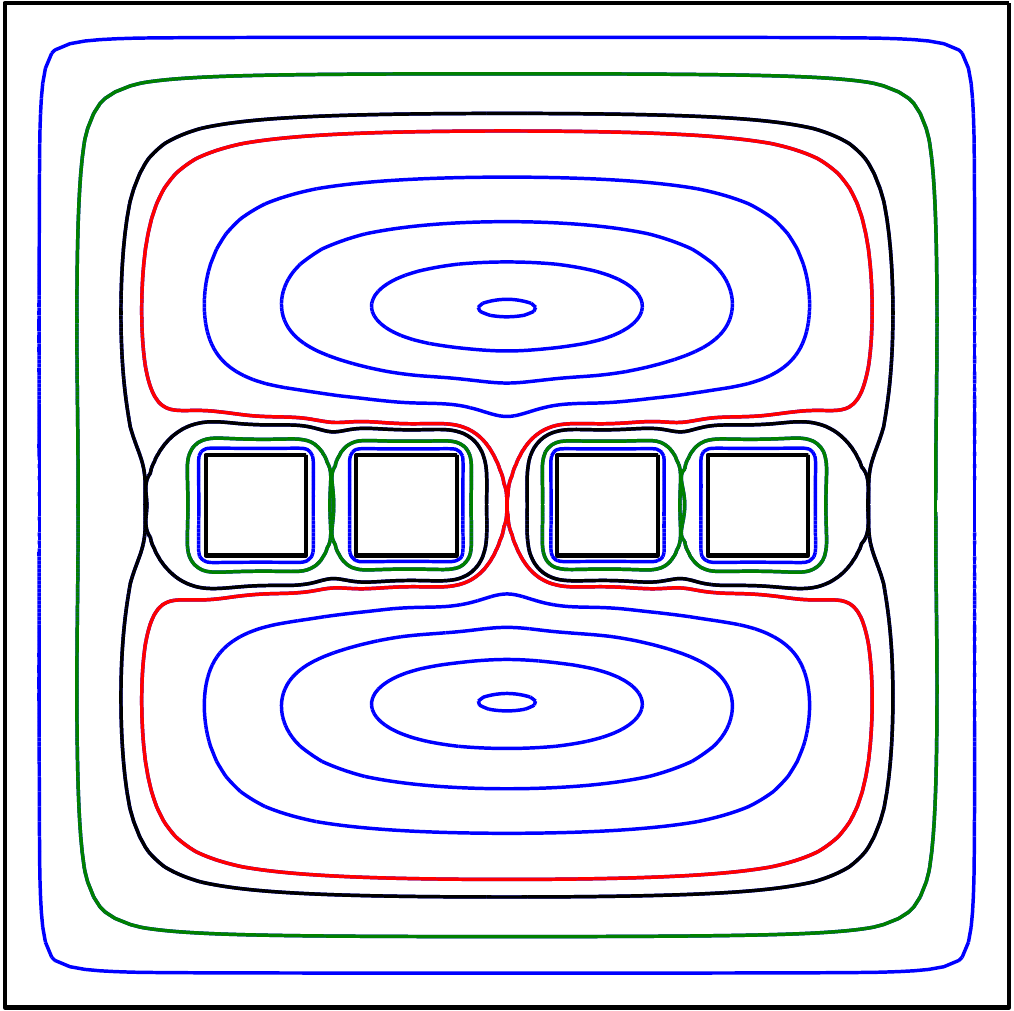}
		\caption{$p=0.2, \, q=0.5$}
		\label{fig:4sr}
	\end{subfigure}%
	~ 
	\begin{subfigure}[b]{0.24\textwidth}
		\centering
		\includegraphics[width=\linewidth,trim=0 0 0 0,clip]{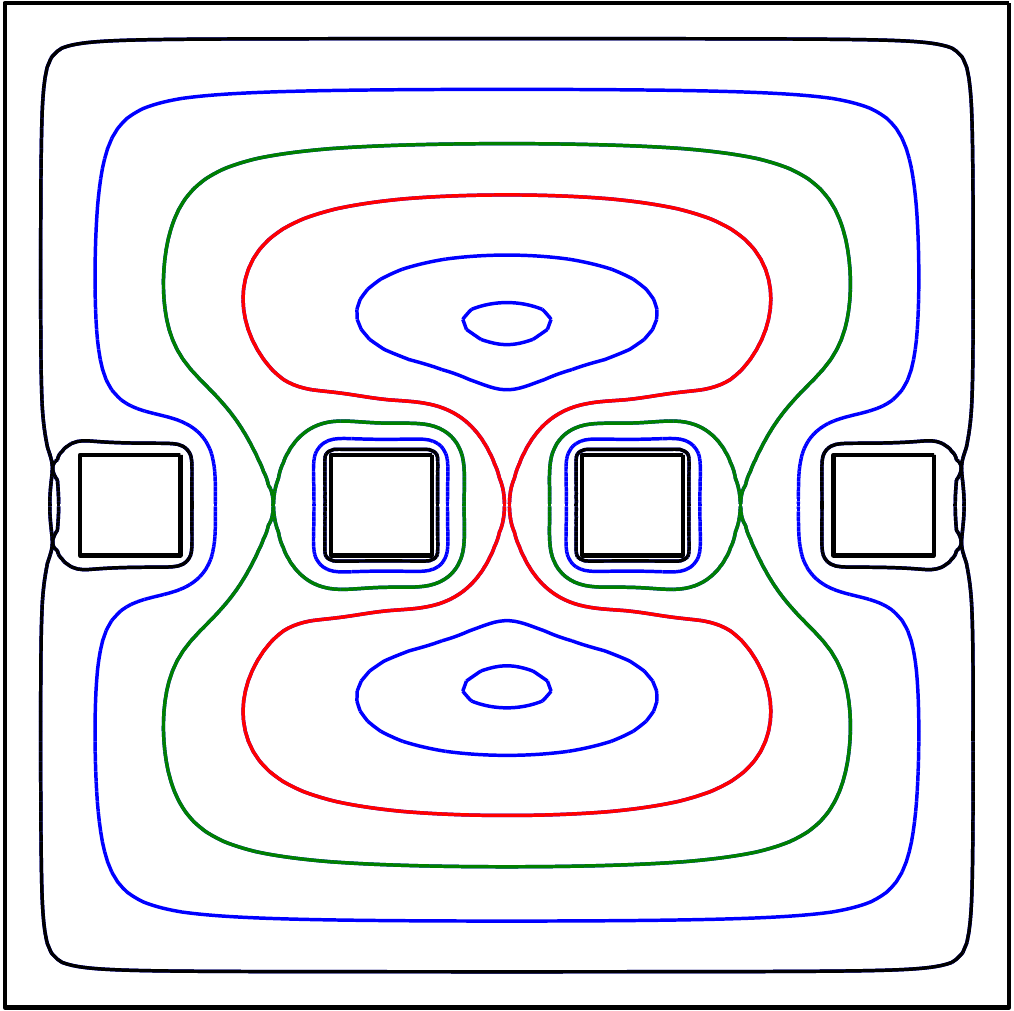}
		\caption{$p=0.25, q=0.75$}
		\label{fig:4sr2}
	\end{subfigure}%
	~
	\begin{subfigure}[b]{0.24\textwidth}
		\centering
		\includegraphics[width=\linewidth,trim=0 0 0 0,clip]{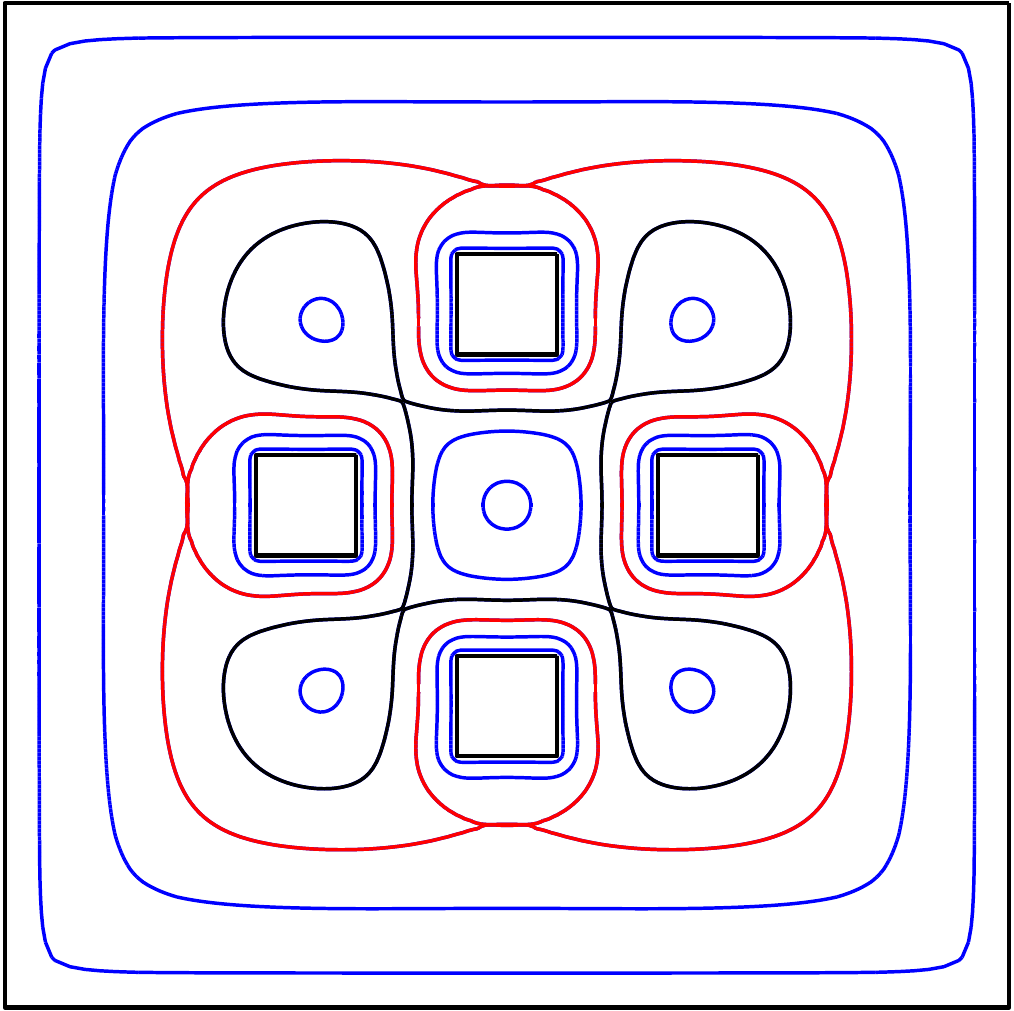}
		\caption{$p=0.4, \, q=0.4\i$}
		\label{fig:4sd}
	\end{subfigure}%
	~
	\begin{subfigure}[b]{0.24\textwidth}
		\centering
		\includegraphics[width=\linewidth,trim=0 0 0 0,clip]{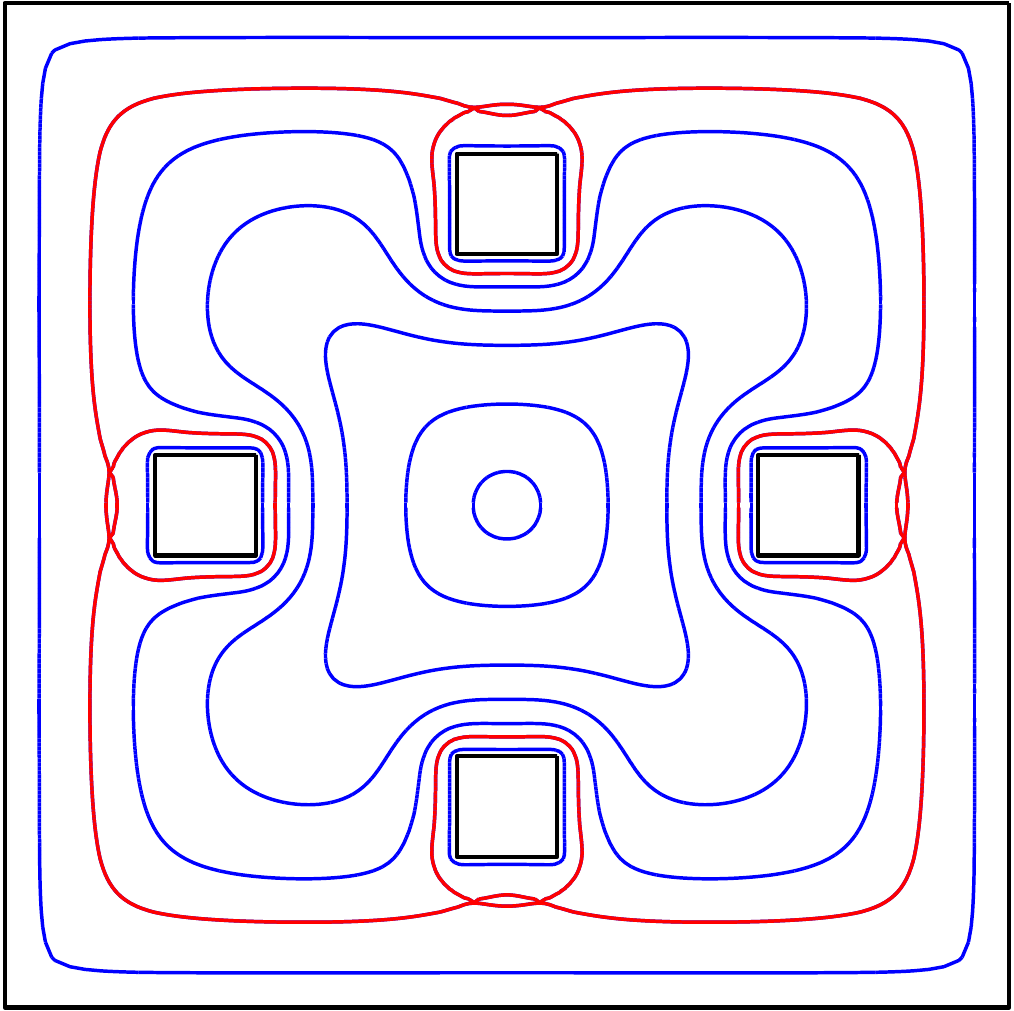}
		\caption{ $p=0.6,\, 0.6\i$}
		\label{fig:4sd2}
	\end{subfigure}%
	\caption{Vortex trajectories in quintuply connected polygonal domains where the inner squares have side length equal to $0.2$ and centers at $\pm p$ and $\pm q$.}
	\label{fig:quint}
\end{figure}

\begin{figure}[!htb] %
	\centerline{
		\scalebox{1}{\includegraphics[trim=0 0 0 0,clip]{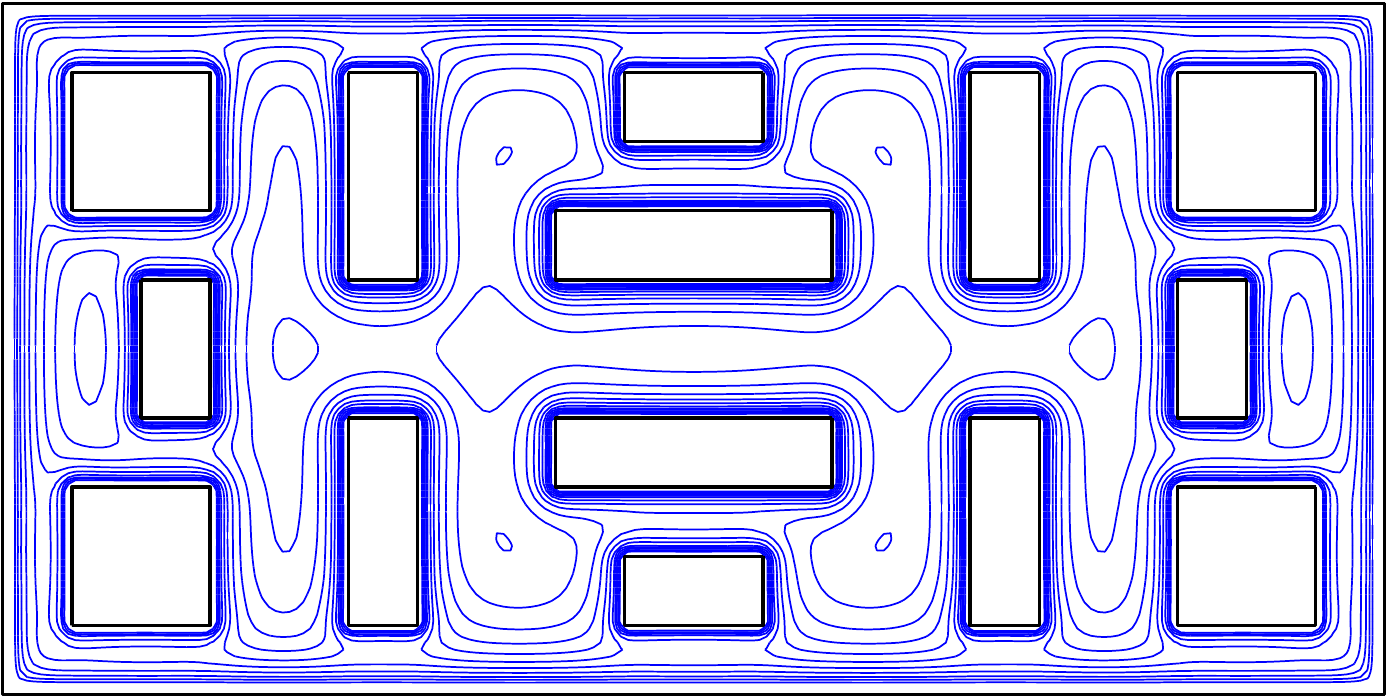}}
	}
	\caption{Vortex trajectories in a multiply connected polygonal domain with $15$ obstacles.}
	\label{fig:m15s}
\end{figure}

Finally, when we consider a quintuply connected polygonal domain with four square obstacles in Figure~\ref{fig:quint}, we notice the same dynamical feature as in the corresponding case of triply connected domains.   It is worth mentioning that the presented method can also be efficiently used to compute vortex trajectories in multiply connected polygonal domains of high connectivity as illustrated in Figure~\ref{fig:m15s}.

%%%%%%%%%%%%%%%%%%%%%%%%%%%%%%%%%%%%%%%%%%
\section{Conclusions}

\label{sc:conc}
%-------------------------------------------------------------------

We have numerically computed the Hamiltonian of a point vortex motion in simply and multiply connected bounded polygonal domains. We have first conformally mapped the polygonal domain onto a circular domain, and then applied the analytic formula presented in~\cite{cm-ana} for the Hamiltonian in multiply connected circular domains.

For simply connected domains, 
we have considered four polygonal shapes that are non convex.
In these four examples, we have studied first the case when the system is topologically equivalent to the unit disk where the Hamiltonian has a unique critical point. Then keeping the same general shape and changing the distance between the center point location and boundaries has yielded a lost of this uniqueness property. In all cases, the number of critical points satisfies the Morse Theorem; that is the number of centers is equal to that of saddles plus one.

We have then examined the case of doubly connected polygonal domains with four examples of an obstacle in a square. For concentric domains, we have noticed a pure heteroclinic structure of vortex trajectories around the standard and rotated squares obstacle, which was not the case for triangle and hexagon obstacles. Afterwards, we have concerned ourselves with the manner the vortex motion changes due to small perturbations of the obstacle location and its displacement toward the external boundary.
Through various examples, the effect of symmetry breaking, and obstacle shape and location on vortex trajectories has been well demonstrated. 

In triply connected polygonal domains, we have looked at how vortex
trajectories interact with the square obstacles and the outer square boundaries. We have observed that this interaction
depends on the distance between these obstacles, and between each obstacle and the external boundary. The common dynamical feature at close boundaries interaction involves two saddles for side to side and one saddle for corner to side or corner to corner. As the separation between
the two obstacles or between one obstacle and the outer boundaries increases, a vortex merging has been noticed. Another worth commenting dynamical feature is that heteroclinic connections between saddle points are quickly broken by lost of symmetry, which confirms their structural instability.

We have also demonstrated that our method can also be efficiently used to study single point vortex motion in multiply connected polygonal domains of high connectivity. Indeed, we have shown vortex trajectories in a square domain with four square obstacles, and a rectangle with fourteen rectangular obstacles.

Finally, as an outlook on future perspectives, we mention three directions. We have seen in this paper that 
heteroclinic loops is generally associated with the presence of symmetries in the model. Breaking such symmetries  often break heteroclinicity and produce new homoclinic orbits.  The formation of homoclinic orbits might be important for the mixing of passive tracers~\cite{scott2006local}.
Future work could also focus on the case in which the Hamiltonian is not a log function but follows a power law, which are an important new frontier for applied mathematics and physics~\cite{badin2018collapse}.
On the other hand, in this paper, we have considered the multiply connected polygonal domains. With the help of the method presented in~\cite{Nas-ETNA,Nas-CMFT,nasser2018fast}, our method can be extended to multiply connected domains other than polygonal domains such as domains exterior to rectilinear slits.

%%%%%%%%%%%%%%%%%%%%%%%%%%%%%%%%%%%%%%%%%%

%% optional
%\supplementary{The following are available online at \linksupplementary{s1}, Figure S1: title, Table S1: title, Video S1: title.}


%% for journal Sci
%\reviewreports{\\
%Reviewer 1 comments and authors’ response\\
%Reviewer 2 comments and authors’ response\\
%Reviewer 3 comments and authors’ response
%}

%%%%%%%%%%%%%%%%%%%%%%%%%%%%%%%%%%%%%%%%%%

\begin{thebibliography}{99}
	
\bibitem{aref2007point}
Aref, H. Point vortex dynamics: a classical mathematics playground. 
{\em Journal of mathematical Physics} {\bf 2007}, {\em 48(6)}, 065401.

\bibitem{badin2018collapse}
Badin, G.; Barry, A.M. Collapse of generalized Euler and surface quasigeostrophic point vortices. {\em Physical Review E} {\bf 2018}, {\em 98(2)}, 023110.

\bibitem{boatto2006point}
Boatto, S.; Crowdy, D.G. Point vortex dynamics. In {\em Encyclopedia of Mathematical Physics}; Springer–Verlag, 2006.

\bibitem{c-book}
Crowdy, D.
\newblock {Solving problems in multiply connected domains}.
{\em SIAM, CBMS-NSF Regional Conference Series in Applied Mathematics}, to appear.

\bibitem{cm-ana}
Crowdy, D.; Marshall, J. 
\newblock{Analytical formulae for the Kirchhoff-Routh path function in multiply connected domains}. 
\newblock{\em Proc. R. Soc. A} {\bf 2005}, {\em 461}, 2477--2501.

\bibitem{cm-mot1}
Crowdy, D.; Marshall, J.  
\newblock{The motion of a point vortex around multiple circular islands}. 
\newblock{\em Phys. Fluids} {\bf 2005}, {\em 17}, 056602.

\bibitem{cm-mot2}
Crowdy, D.; Marshall, J. 
\newblock{The motion of a point vortex through gaps in walls}. 
\newblock{\em J. Fluid Mech.} {\bf 2006}, {\em 551}, 31--48.

\bibitem{ckgn}
Crowdy, D.G.;  Kropf, E.H.; Green, C.C.; Nasser, M.M.S. 
\newblock{The Schottky-Klein prime function: A theoretical and computational tool for applications}. 
\newblock{\em IMA J. Appl. Math.} {\bf 2016}, {\em 81}, 589--628.

\bibitem{flucher1997vortex}
Flucher, M.; Gustafsson, B. Vortex motion in two-dimensional hydrodynamics.
{\em Royal Inst. Techn. Stockholm} {\bf 1997}, TRITA-MAT-97-MA-02.	

\bibitem{Gre-Gim12}
Greengard, L.; Gimbutas, Z.
\newblock {{FMMLIB2D}: A {MATLAB} toolbox for fast multipole method in two dimensions}.
\newblock Version 1.2, \url{http://www.cims.nyu.edu/cmcl/fmm2dlib/fmm2dlib.html}. Accessed 1 Jan 2018.

\bibitem{gustafsson1979motion}
Gustafsson, B. On the motion of a vortex in two-dimensional flow of an ideal fluid in simply and multiply connected domains. 
{\em Royal Inst. Techn. Stockholm} {\bf 1979}, RITA-MAT-1979-7.

\bibitem{gustafsson1990convexity}
Gustafsson, B. On the convexity of a solution of Liouville's equation. 
{\em Duke Mathematical Journal} {\bf 1990}, {\em  60(2)}, 303--311.

\bibitem{helmholtz}
Helmholtz, H.
\"Uber integrale der hydrodynamischen gleichungen, welche den Wirbelbewegungen entsprechen. 
{\em J. Reine Angew. Math.} {\bf 1858}, {\em 55}, 25--55.

\bibitem{johnson2004motion}
Johnson, E.R.; McDonald, N.R.  The motion of a vortex near two circular cylinders. 
{\em Proc. R. Soc. A} {\bf 2004}, {\em 460}, 939–954.

\bibitem{kirchhoff}
Kirchhoff, G. Vorlesungen \"uber mathematische Physik. {\em Teubner}: Leipzig, 1876.

\bibitem{Krantz}
Krantz, S.G.
\newblock {Geometric Function Theory: Explorations in Complex Analysis}.
{\em Birkh\"auser}: Boston, 2006.

\bibitem{lin1941motion}
Lin, C.C. On the motion of vortices in two dimensions. 
{\em Proc. Nat. Acad. Sci. USA} {\bf 1941}, {\em 27}, 570–577. 

%\bibitem{Nas-log}
%J.~{Liesen}, O.~{S\'ete} and M.M.S.~{Nasser}.
%\newblock {A fast and accurate computation of the logarithmic capacity of compact sets}. 
%\newblock {\em Comput. Methods Funct. Theory},  17:689--713, 2017.


%\bibitem{Nas-jsc19}
%M.M.S.~{Nasser}.
%\newblock {Numerical computing of preimage domains for bounded multiply connected slit domains}.
%\newblock {\em J. Sci. Comput.}, 78:582--606, 2019.

\bibitem{Nas-ETNA}
Nasser, M.M.S.
\newblock {Fast solution of boundary integral equations with the generalized Neumann kernel}.
{\em Electron. Trans. Numer. Anal.} {\bf 2015}, {\em 44}, 189--229.

\bibitem{Nas-CMFT}
Nasser, M.M.S.
\newblock {Fast Computation of the Circular Map}.
\newblock {\em Comput. Methods Funct. Theory} {\bf 2015}, {\em 15}, 187--223.

\bibitem{Nas-plg}
Nasser, M.M.S.
\newblock {PlgCirMap: A MATLAB toolbox for computing the conformal mapping from polygonal multiply connected domains onto circular domains}.
\newblock {\em arXiv:1911.01787}, 2019.

\bibitem{nasser2018fast} Nasser, M.M.S.; Green, C.C. A fast numerical method for ideal fluid flow in domains with multiple stirrers.
{\em Nonlinearity}  {\bf 2018}, {\em 31}, 815--837.

%\bibitem{Nas-Gre}
%M.M.S.~{Nasser} and C.C.~{Green}.
%\newblock {A fast numerical method for ideal fluid flow in domains with multiple stirrers}.
%\newblock {\em Nonlinearity}, 31:815--837, 2018.

\bibitem{newton2002n-vortex}
P. Newton. The N–Vortex Problem: Analytical Techniques. {\em Springer}: New York, 2002.

\bibitem{routh}
Routh, E.J. Some applications of conjugate functions. {\em Proc. London Mat. Soc.} {\bf 1881}, {\em 12}, 73–89.

\bibitem{saffman1992vortex}
Saffman, P. Vortex Dynamics. {\em Cambridge University Press}: Cambridge, 1992.

\bibitem{Tak-eq}
Sakajo, T. 
\newblock{Equation of motion for point vortices in multiply connected circular domains}. 
{\em Proc. R. Soc. A} {\bf 2009}, {\em 465},  2589--2611.

\bibitem{scott2006local} Scott, R.K. Local and nonlocal advection of a passive scalar. {\em Physics of Fluids} {\bf 2006}, {\em 18(11)}, 116601.

\end{thebibliography}
\end{document}